\documentclass[%
preprint,
 amsmath,amssymb,
 aps,
longbibliography
]{revtex4-1}

\usepackage{comment} 
\usepackage{graphicx}
\usepackage{multirow}
\usepackage{subcaption}
\usepackage{tikz}
\usepackage{dcolumn}
\usepackage{bm}
\usepackage{color}
\usepackage{upgreek}
\usepackage{multirow}
\usepackage{lipsum}
\usepackage{float}
\usepackage{textcomp}
\usepackage[colorinlistoftodos, color=blue!20!white,bordercolor=gray,textsize=tiny,textwidth=0.8in]{todonotes}
\usepackage{fancyhdr}
\usepackage{mleftright}

\newcommand{\Learned}{\raisebox{2pt}{\tikz{\draw[red,solid,line width=1pt](0,0) -- (5mm,0);}}}
\newcommand{\LEVM}{\raisebox{2pt}{\tikz{\draw[blue,dashed,line width=1pt](0,0) -- (5mm,0);}}}
\newcommand{\DNS}{\raisebox{2pt}{\tikz{\draw[black,solid,line width=2pt](0,0) -- (5mm,0);}}}

\newcommand{\NoisyLearned}{\raisebox{2pt}{\tikz{\draw[red,dashed,line width=1pt](0,0) -- (5mm,0);}}}
\newcommand{\Sindy}{\raisebox{2pt}{\tikz{\draw[red,solid,line width=1pt](0,0) -- (5mm,0);}}}
\newcommand{\LRRIP}{\raisebox{2pt}{\tikz{\draw[black,solid,line width=1pt](0,0) -- (5mm,0);}}}

\newcommand{\uu}{\raisebox{2pt}{\tikz{\draw[black,dashed,line width=1pt](0,0) -- (5mm,0);}}}
\newcommand{\uv}{\raisebox{2pt}{\tikz{\draw[black,dotted,line width=1pt](0,0) -- (5mm,0);}}}
\newcommand{\vv}{\raisebox{2pt}{\tikz{\draw[black,solid,line width=1pt](0,0) -- (5mm,0);}}}
\newcommand{\ww}{\raisebox{2pt}{\tikz{\draw[black,dashdotted,line width=1pt](0,0) -- (5mm,0);}}}

\begin{document}

\preprint{APS/Physical Review Fluids}

\title{Formulating turbulence closures using sparse regression with embedded form invariance}

\author{S. Beetham}
\email{snverner@umich.edu}
\author{J. Capecelatro}
 \affiliation{Department of Mechanical Engineering, University of Michigan, Ann Arbor, MI 48105}%


\begin{abstract}
A data-driven framework for formulation of closures of the Reynolds-Average Navier--Stokes (RANS) equations is presented. In recent years, the scientific community has turned to machine learning techniques to distill a wealth of highly resolved data into improved RANS closures. While the body of work in this area has primarily leveraged Neural Networks (NNs), we alternately leverage a sparse regression framework. This methodology has two important properties: (1) The resultant model is in a closed, algebraic form, allowing for direct physical inferences to be drawn and naive integration into existing computational fluid dynamics solvers, and (2) Galilean invariance can be guaranteed by thoughtful tailoring of the feature space. Our approach is demonstrated for two classes of flows: homogeneous free shear turbulence and turbulent flow over a wavy wall. This work demonstrates equivalent performance to that of modern NNs but with the added benefits of interpretability, increased ease-of-use and dissemination, and robustness to sparse training datasets.   
\end{abstract}


\maketitle

\section{Introduction}\label{sec:Background}
Simulation frameworks based on the Reynolds-Averaged Navier--Stokes (RANS) equations~\cite{Reynolds1895, Tennekes1992, Menter1994, Menter2003} have been the most widely-used tool in industrial and large-scale applications of turbulent flows for the last several decades \cite{Moin1997} and will remain to be a central tool for guiding design decisions well into the coming decades \cite{Slotnick2014, Bush2019}. This is primarily driven by the wide range of length- and time-scales associated with turbulent flows of interest. Because of this, direct numerical simulations (DNS) that fully resolve all relevant scales are prohibitively costly. Instead, the RANS equations solve for mean flow quantities that are then used to assess global flow features of interest. A principal challenge associated with RANS is accurate modeling of the unresolved terms, which are denoted `unclosed' because they are not completely specified in terms of the unknowns (e.g., mean velocity, pressure, etc.).  

With the rise of computational power and the accessibility of large, highly resolved datasets, the community has turned to machine learning techniques in recent years to distill this wealth of information into improved RANS models. As a consequence of the interest and prevalence of the use of machine learning in the turbulence modeling community, several thoughtful and thorough reviews have been published and the authors refer the interested reader to several of these works, including \citet{Brenner2019},  \citet{Karthik2019_AgeofData}, \citet{Karthik2019Towards} and \citet{ML_2014Duraisamy_new}.     

Numerous studies in recent years approach the RANS closure problem by leveraging a Neural Network (NN)-based framework. \citet{ML_2016Ling} used an invariant tensor basis integrated into a NN to model the Reynolds stress anisotropy tensor for turbulent duct flow as well as flow over a wavy wall. Galilean invariance, a critical model property, was ensured by nature of the invariant tensor basis as the inputs to the NN and demonstrated excellent agreement with DNS data as compared to traditional (linear and quadratic eddy viscosity) models. Following this work, many others have implemented similar strategies and employed a similar basis technique for ensuring invariance. Of these works, many have used flow through a periodically constricted channel or backward facing step as challenging tests of new modeling methodologies as these flows exhibit massive separation. This complication is notoriously difficult to accurately capture with traditional RANS closures \cite{Rumsey2008}. A large body of works using NNs as the data-driven methodology to formulate closure models have demonstrated promising success \cite{Xiao2020,Kohler2020,Parmar2020,ML_2016Ling,Liu2019,ML_2012Rajabi}.

Despite the demonstration of improved model performance, models based upon NNs have an important drawback. Due to the nature of the algorithm at the heart of NNs, the resultant model acts as a `black box' and cannot be expressed in a compact, algebraic form. This compromises interpretability, introduces difficulty in disseminating the learned model with end users and industries, and increases computational cost of the model in the context of a RANS solver (as compared with traditional algebraic closures). Further, a large number of NN approaches attempt to augment or correct existing models. However, this approach breaks down for more complex turbulent flows, such as disperse two-phase flows \cite{fox2014, capecelatro2014cit, capecelatro2015on, capecelatro2016channel2, beetham2019} or turbulent combustion \cite{Veynante2002, Pitsch2006}, in which  the fundamental assumption of an energy cascade breaks down due to production at the smallest scales. In these cases, existing closures adopted from single-phase flows are not appropriate, which precludes an augmentation modeling approach. For these reasons, the present study proposes an alternate method that allows for the development of physics-based, compact algebraic closures, thus affording interpretability, transportability and efficiency. 

Several studies have taken alternate approaches to NNs using symbolic methods in order to arrive at closed form, algebraic models. Gene Expression Programming \cite{Schmelzer2018, Schmelzer2019, Weatheritt2019, Lav2019, zhao2019_2} and random forest regression \cite{Wang2017_2, Wu2018} have become increasingly popular methodologies. The early success of these works serves as motivation for the present work in which we present a methodology based upon sparse regression as an alternative to NNs for developing new RANS closures, with emphasis on the following key benefits,  
\begin{itemize}
\item \emph{Interpretability}: Sparse regression produces an algebraic model with a limited number of terms, resulting in improved interpretability of underlying physics and better prediction of model behavior and stability outside the scope of training.
\item \emph{Galilean invariance}: By careful construction of the feature space and structuring of the optimization cost functional, Galilean invariance of the resultant model is ensured. 
\item \emph{Efficiency}: Sparse regression models are built using physics-based, functional terms and identifies a subset of these terms that are most important for capturing physics. This is fundamentally different from a naive curve fit in which all possible terms are included in the model. Thus, it is possible to use less training data to develop predictive models than has been demonstrated with other machine learning techniques, such as NNs. This attribute allows for decreased training time and decreased cost for producing training datasets for equally predictive models. Further, the resultant model is algebraic, making for a lighter and more efficient integration with existing solvers.  
\end{itemize}
Beyond developing the methodology, its utility is demonstrated on two canonical cases: homogeneous free shear turbulence and turbulence through a periodically constricted channel. Within the context of homogeneous free shear turbulence, the sparse regression methodology is validated using a `toy' problem in which the training data set is synthetically generated using an existing model. Then, sparse regression is used to recover this existing model. Subsequent cases are based upon DNS data and seek to uncover improved models in comparison with existing closures. 

\section{Methodology \label{sec:Method}}
The sparse regression approach expands upon the data-driven technique presented in \citet{ML_2016Brunton} for using temporally evolving data to `discover' nonlinear, dynamical systems. Rather than uncovering governing equations, this method is employed to identify robust, data-driven closure models.  In this section, the Sparse Identification of Nonlinear Dynamics (SINDy) framework \cite{ML_2016Brunton} is built upon by adapting it for the RANS closure problem and embedding invariance--a key property of any candidate RANS model. 

It is first postulated that a tensor quantity of interest, $\mathbb{D}$, can be characterized by the linear combination of an invariant tensor basis, represented as $\mathbb{T}$, premultiplied by optimal coefficients, represented as $\hat{\boldsymbol{\beta}}$,
\begin{equation}
\mathbb{D} = \mathbb{T} \hat{\boldsymbol{\beta}}. 
\end{equation}
Using this postulated form of the model, the following objective function is minimized in order to determine the optimal coefficient vector, according to
\begin{equation}
\hat{\boldsymbol{\beta}} = \min_{\boldsymbol{\beta}} \vert \vert \mathbb{D} - \mathbb{T}\boldsymbol{\beta} \vert \vert_2^2 + \lambda \vert \vert \bm{\beta} \vert \vert_1, \\ \label{Eq.CostFunc}
\end{equation}
where $\beta$ represents intermediary realizations of the coefficient vector which may not necessarily be the optimal coefficient vector, $\hat{\beta}$. Here, the \textit{L}-2 and the \textit{L}-1 norms are denoted by $\vert \vert \cdot \vert \vert_2^2$ and $\vert \vert \cdot \vert \vert_1$, respectively. The first term in the objective function is ordinary least squares, which regresses the coefficient vector to the trusted data, and the second term is a sparsity-inducing penalty on the coefficients. By choice of the \textit{L}-1 penalty, the minimization of the objective function performs model selection by inducing sparsity (e.g., several of the terms of $\hat{\bm{\beta}}$ are identically zero, indicating that the associated term in the invariant basis, $\mathbb{T}$, is not important. The interested reader can refer to \cite{Tibshirani1996, Zou2005, Bishop2006} for further information.) Minimization of the cost function is performed using the open source iterative algorithm presented in \citet{ML_2016Brunton}.  

In order to obtain a model that is both compact and frame-invariant, consideration must be given to the construction of the trusted data vector, $\mathbb{D}$, and the invariant basis, $\mathbb{T}$. For compactness, $\mathbb{D}$, and as a consequence $\bm{\beta}$, are restricted to column vectors. This ensures that the coefficients for each term in the model is a scalar, which guarantees the same model form regardless of orientation (e.g., if the coefficients were vectors or tensors, this would embed directionality into the coefficients and thereby enslave the model to the orientation in which it was learned). 

In this work, $\mathbb{D}$ is assembled by first assessing the symmetry of the problem. All nonzero, unique entries in the trusted data tensor are concatenated into a column vector. For example, as seen in Fig.~\ref{fig:SIMRsetup}, if $\mathcal{D}$ is symmetric in the $y$-- and $z$-- directions and the only anisotropic contribution is in the $x$--$y$ direction, then the full tensor is represented as $\lbrack \mathcal{D}_{11},  \mathcal{D}_{12}, \mathcal{D}_{22} \rbrack^{\text{T}}$. For each realization (e.g., in time) and for each configuration under consideration, these column vectors are vertically concatenated.    

\begin{figure}
\begin{equation*}
\underbrace{\begin{bmatrix} \begin{Bmatrix} \mathcal{D}_{11}^{t = 1} \\ \mathcal{D}_{12}^{t =1} \\ \mathcal{D}_{22}^{t = 1} \\ \vdots \\ \mathcal{D}_{11}^{t = s} \\ \mathcal{D}_{12}^{t = s} \\ \mathcal{D}_{22}^{t = s}   \end{Bmatrix}^{\text{case } 1} \\ \vdots \\ 
\begin{Bmatrix} \mathcal{D}_{11}^{t = 1} \\ \mathcal{D}_{12}^{t =1} \\ \mathcal{D}_{22}^{t = 1} \\ \vdots \\ \mathcal{D}_{11}^{t = s} \\ \mathcal{D}_{12}^{t = s} \\ \mathcal{D}_{22}^{t = s}   \end{Bmatrix}^{\text{case } c} \end{bmatrix}}_{\mathbb{D} \in \mathbb{R}^{(n\cdot s\cdot c)\times 1}}  = 
\underbrace{\begin{bmatrix} 
\begin{Bmatrix} 
 \mathcal{T}_{11}^{(1),\; t = 1} &  \mathcal{T}_{11}^{(2),\; t = 1} & \cdots &  \mathcal{T}_{11}^{(g),\; t = 1}\\
 \mathcal{T}_{12}^{(1),\; t =1} &  \mathcal{T}_{12}^{(2),\; t = 1} & \cdots &  \mathcal{T}_{12}^{(g),\; t = 1}\\ 
 \mathcal{T}_{22}^{(1),\; t = 1} & \mathcal{T}_{22}^{(2),\; t = 1} & \cdots &  \mathcal{T}_{22}^{(g),\; t = 1}\\ 
 \vdots & \vdots & & \vdots \\ 
 \mathcal{T}_{11}^{(1),\; t = s} &  \mathcal{T}_{11}^{(2),\; t = s} & \cdots &  \mathcal{T}_{11}^{(g),\; t = s}\\
 \mathcal{T}_{12}^{(1),\; t =s} &  \mathcal{T}_{12}^{(2),\; t = s} & \cdots &  \mathcal{T}_{12}^{(g),\; t = s}\\ 
 \mathcal{T}_{22}^{(1),\; t = s} & \mathcal{T}_{22}^{(2),\; t = s} & \cdots &  \mathcal{T}_{22}^{(g),\; t = s}\\ 
\end{Bmatrix}^{\text{case } 1} \\
\vdots \\
\begin{Bmatrix} 
 \mathcal{T}_{11}^{(1),\; t = 1} &  \mathcal{T}_{11}^{(2),\; t = 1} & \cdots &  \mathcal{T}_{11}^{(g),\; t = 1}\\
 \mathcal{T}_{12}^{(1),\; t =1} &  \mathcal{T}_{12}^{(2),\; t = 1} & \cdots &  \mathcal{T}_{12}^{(g),\; t = 1}\\ 
 \mathcal{T}_{22}^{(1),\; t = 1} & \mathcal{T}_{22}^{(2),\; t = 1} & \cdots &  \mathcal{T}_{22}^{(g),\; t = 1}\\ 
 \vdots & \vdots & & \vdots \\ 
 \mathcal{T}_{11}^{(1),\; t = s} &  \mathcal{T}_{11}^{(2),\; t = s} & \cdots &  \mathcal{T}_{11}^{(g),\; t = s}\\
 \mathcal{T}_{12}^{(1),\; t =s} &  \mathcal{T}_{12}^{(2),\; t = s} & \cdots &  \mathcal{T}_{12}^{(g),\; t = s}\\ 
 \mathcal{T}_{22}^{(1),\; t = s} & \mathcal{T}_{22}^{(2),\; t = s} & \cdots &  \mathcal{T}_{22}^{(g),\; t = s}\\ 
\end{Bmatrix}^{\text{case } c}
\end{bmatrix}}_{\mathbb{T} \in \mathbb{R}^{(n\cdot s\cdot c) \times g}}
\underbrace{\begin{Bmatrix} \beta_1 \\ \beta_2 \\ \vdots \\ \beta_g \end{Bmatrix}}_{\boldsymbol{\beta}\in \mathbb{R}^{g\times1}}
\end{equation*}
\caption{The postulated model takes the form $\mathbb{D}_i = \mathbb{T}_{ij} \beta_j$, where $\mathbb{D}$ contains the observed data spanning over $c$ cases, each with $s$ realizations in time. Further, $\mathbb{T}$ spans an invariant tensor basis of dimension $g$ evaluated at each of the samples $s$ for each case $c$.} 
\label{fig:SIMRsetup}
\end{figure}

Finally, form (Galilean) invariance in the resultant model is guaranteed by assembling $\mathbb{T}$ from an invariant tensor basis. The basis is crafted by using dimensional analysis to determine the relevant known tensor quantities that fully describe the physics under study. These tensors are then used to assemble a minimal integrity basis (see, e.g., \citet{ML_1975Pope}, \citet{ML_1991Speziale}, \citet{ML_2016Ling}), using the following arguments:
\begin{itemize} 
\item[1.] Any tensor can be represented by an infinite tensor sum of the form: $$\mathcal{D}_{ij} = \sum_{n=1}^{\infty} G^{(n)} \mathcal{T}_{ij}^{(n)},$$ where $G$ are coefficients that in the general sense may be functions of the invariants of the tensor basis $\mathcal{T}_{ij}^{(n)}$. 
\item[2.] In some cases, the Cayley-Hamilton theorem can be leveraged to reduce the infinite tensor sum to a finite sum that still exactly represents the infinite sum. In cases where this is not possible, the basis is truncated once model improvement stagnates. 
\end{itemize} 

Once the invariant basis is determined, the matrix $\mathbb{T}$ is assembled using the same convention as for $\mathbb{D}$. Then, the sparsity parameter $\lambda$ is adjusted until acceptable model error and sparsity are reached, noting that $\lambda = 0$ is Ordinary Least Squares, and increasing $\lambda$ results in an increasingly sparse coefficient vector, $\hat{\bm{\beta}}$. 

In the following sections, the sparse regression methodology is applied to two canonical cases of increasing complexity. First, homogeneous free shear turbulence is considered. As an initial proof-of-concept, a synthetic dataset is generated using a known model and the sparse regression methodology is used to recover that model. Next, DNS is used to generate the trusted datasets for the same physical configuration and sparse regression is used to uncover alternate models to those traditionally used. Last, turbulence through a periodically constricted channel is considered. This test case has the additional complexities of being a statistically two-dimensional flow (as compared to homogeneous free shear turbulence being statistically one-dimensional in time) as well as giving rise to flow separation. The model learned by sparse regression is compared with the standard Linear Eddy Viscosity Model (LEVM) as well as with the performance reported by previous studies employing NNs. 

\section{Case studies}\label{sec:CaseStudies}
Herein, Reynolds decomposition is denoted by angled brackets, $\langle \cdot \rangle$, given for the velocity vector by $u_i(x_i,t) = \langle u_i(x_i,t) \rangle + {u_i}^{\prime}(x_i,t)$, where $u_i(x_i,t)$ is the field quantity for velocity written in Einstein notation and $\langle u_i(x_i,t) \rangle$ and $u_i^{\prime}(x_i,t)$ denote the mean (which may be spatial, temporal or both) and the fluctuating portions of the velocity, respectively.   

Applying Reynolds averaging to the incompressible Navier--Stokes equations yields the RANS equations,
\begin{align}
&\frac{\partial \langle u_i \rangle}{\partial x_i} = 0 \label{Eq:RANScontinuity} \\
&\frac{\partial \langle u_i \rangle}{\partial t} + \langle u_k \rangle \frac{\partial \langle u_i \rangle}{\partial x_k}  = - \frac{1}{\rho} \frac{\partial \langle p \rangle}{\partial x_i} + \frac{\partial}{\partial x_j} \left \lbrack \nu \left( \frac{\partial \langle u_i \rangle}{\partial x_j} + \frac{\partial \langle u_j \rangle}{\partial x_i} \right) - \langle u_i^{\prime} u_j^{\prime} \rangle \right \rbrack. \label{Eq:RANSmom}
\end{align}
It is notable that the Reynolds averaging process yields a Reynolds stress term, $\langle u_i^{\prime}u_j^{\prime}\rangle$, which requires closure. 

The strategy for closure of the Reynolds stress term generally falls into two categories: (1) an algebraic closure or (2) the inclusion of a transport equation for the Reynolds stresses. In this work, two flows serve as case studies for the implementation of the methodology described in Sec.~\ref{sec:Method}. The first case study (homogeneous free shear turbulence) will develop closures in the form of transport of the Reynolds stresses and the second (turbulence in a periodically constricted channel) will consider algebraic closure. 

\subsection{Homogeneous free shear turbulence}\label{sec:FreeShear}
\subsubsection{Problem statement}
The flow configuration under consideration in this section is homogeneous free shear turbulence, in which an unbounded, three-dimensional fluid volume is subjected to a mean-velocity gradient that generates and sustains turbulence. After sufficient time, the Reynolds stresses reach a `self-similar' state, characterized by the anisotropy of the Reynolds stresses reaching stationarity in time (e.g., $\frac{d}{d t} \left(\langle u'_i u'_j \rangle/k \right)$ with $k = \langle u'_k u'_k\rangle$). Consequently, Reynolds-averaged quantities are statistically one-dimensional (i.e., they depend only on time). It is this `self-similar' behavior that is of specific interest in formulating an improved RANS closure.  

As previously described, the Reynolds stresses in the RANS equations (Eq.~\eqref{Eq:RANSmom}) require closure. In this example, we consider the transport of the Reynolds stresses, which are given exactly as 
\begin{align}
\frac{D \langle u'_i u'_j \rangle}{Dt} = &\underbrace{- \left \lbrack \langle u'_j u'_k \rangle \frac{\partial \langle u_i\rangle}{\partial x_k} + \langle u'_i u'_k \rangle \frac{\partial \langle u_j\rangle}{\partial x_k}\right \rbrack}_{\text{production, } \mathcal{P}_{ij}}- \underbrace{2\nu \left \langle \frac{\partial u'_i}{\partial x_k}\frac{\partial u'_j}{\partial x_k} \right \rangle}_{\text{dissipation, } \varepsilon_{ij}}+\underbrace{\left \langle \frac{p}{\rho} \left( \frac{\partial u'_i}{\partial x_j}+ \frac{\partial u'_j}{\partial x_i} \right)\right \rangle}_{\text{redistribution, } \mathcal{R}_{ij}} \\ \nonumber
& - \frac{\partial}{\partial x_k} \underbrace{ \left \lbrack \langle u'_i u'_j u'_k \rangle - \nu \frac{\partial \langle u'_i u'_j\rangle}{\partial x_k} + \left \langle \frac{p}{\rho} \left(\delta_{ik} u'_i + \delta_{ik}u'_j \right)\right \rangle \right \rbrack},_{\text{Reynolds stress flux } }
\end{align}
where $D/Dt$ denotes the material derivative, $\nu$ is the kinematic viscosity, and $\rho$ and $p$ denote fluid density and pressure, respectively.
In the case of homogeneous free shear turbulence, the domain is spatially homogeneous and consequently, spatial gradients of mean quantities are null. Thus, the transport of Reynolds stresses is reduced to 
\begin{equation}
\frac{d \langle u'_i u'_j \rangle}{dt} = \underbrace{- \left \lbrack \langle u'_j u'_k \rangle \Gamma_{ij} + \langle u'_i u'_k \rangle \Gamma_{ij} \right \rbrack}_{\text{production, } \mathcal{P}_{ij}} - \underbrace{2\nu \left \langle \frac{\partial u'_i}{\partial x_k}\frac{\partial u'_j}{\partial x_k} \right \rangle}_{\text{dissipation, } \varepsilon_{ij}}+\underbrace{\left \langle \frac{p}{\rho} \left( \frac{\partial u'_i}{\partial x'_j}+ \frac{\partial u'_j}{\partial x_i} \right)\right \rangle}_{\text{redistribution, } \mathcal{R}_{ij}} \label{Eq:RSTransport_shear}, 
\end{equation} 
where the shear rate tensor is given as $\Gamma_{ij} =  {\partial \langle u_i\rangle}/{\partial x_j}$,
Here, the production term is closed, however the dissipation and redistribution tensors both require closure. In this work, new modeling efforts are directed toward the redistribution tensor and the dissipation tensor is closed using the standard transport equation proposed by \citet{Hanjalic1972}, 
\begin{equation}
\frac{\partial \varepsilon}{\partial t} =  C_{\varepsilon 1} \frac{\mathcal{P} \varepsilon}{k} - C_{\varepsilon 2} \frac{\varepsilon^2}{k},  \label{Eq.DispModel} 
\end{equation}
where $\mathcal{P} = \text{tr}\left(\mathcal{P}_{ij}\right)/2$ and model constants are given by $\lbrack C_{\varepsilon 1}, C_{\varepsilon 2} \rbrack = \lbrack 1.44, 1.92 \rbrack$ \cite{Launder1990}.

\subsubsection{Proof-of-concept: a synthetic dataset} 
As an initial proof-of-concept for the sparse regression methodology described in Sec.~\ref{sec:Method}, a set of data is generated using a well-established closure for the redistribution tensor with the goal of recovering the known model. The closure utilized to generate the synthetic dataset was proposed by \citet{Launder1975} and is known as the LRR-IP model,
\begin{equation}
\mathcal{R}_{ij} =  -C_R \frac{\varepsilon}{k} \left(\langle u'_i u'_j \rangle - \frac{2}{3} k \delta_{ij} \right) - C_2 \left( \mathcal{P}_{ij} - \frac{2}{3} \mathcal{P} \delta_{ij}\right), \label{Eq:LRRIP}
\end{equation}
where the constants are given as $\lbrack C_R, C_2 \rbrack = \lbrack 1.8, 0.6 \rbrack$ \cite{Launder1975}. This closure, embedded in the transport equation for the Reynolds stresses in Eq.~\eqref{Eq:RSTransport_shear} and the transport equation for dissipation given in Eq.~\eqref{Eq.DispModel} are solved for three shear rates ($\Gamma = \Gamma_{12} = [2.25, 11.24, 20.23]$). This results in one-dimensional (time-dependent) data for the Reynolds stresses for each shear rate. 

Given the simple flow configuration, the redistribution tensor can be normalized by the viscous dissipation rate, $\varepsilon$, and characterized by a linear combination of the the following non-dimensionalized, mean flow quantities: 
\begin{center}
\begin{tabular}{l r l} 
(1) Anisotropic stress tensor & $b_{ij}=$ & $\frac{\langle u^{\prime}_i u^{\prime}_j \rangle}{2k} - \frac{1}{3} \delta_{ij}$ \\
(2) Mean rotation rate tensor & $\hat{R}_{ij}=$ & $\frac{1}{2}\frac{k}{\varepsilon}\left(\frac{\partial \langle u_i \rangle}{\partial x_j}-\frac{\partial \langle u_j \rangle}{\partial x_i} \right)$ \\
(3) Mean shear rate tensor & $\hat{S}_{ij}=$ & $\frac{1}{2}\frac{k}{\varepsilon}\left(\frac{\partial \langle u_i \rangle}{\partial x_j}+\frac{\partial \langle u_j \rangle}{\partial x_i} \right)$ 
\end{tabular} 
\end{center}
such that, 
\begin{equation}
\mathcal{R}_{ij} = \varepsilon \bm{f}(b_{ij}, \hat{R}_{ij}, \hat{S}_{ij}),
\end{equation}
where $\bm{f}$ is a \emph{form invariant} tensor-valued function, which, due to the linearity in $b_{ij}, \hat{R}_{ij},$ and $\hat{S}_{ij}$ automatically satisfies,
\begin{equation}
\bm{Q} \bm{f}(b_{ij}, \hat{R}_{ij}, \hat{S}_{ij})\bm{Q}^{\text{T}} = \bm{f} (\bm{Q}b_{ij}\bm{Q}^{\text{T}}, \bm{Q}\hat{R}_{ij}\bm{Q}^{\text{T}}, \bm{Q}\hat{S}_{ij}\bm{Q}^{\text{T}}).
\end{equation}
Here, $\bm{Q}$ is a Galilean rotation matrix, e.g., $\bm{Q}\bm{Q}^{\text{T}} = \bm{Q}^{\text{T}} \bm{Q} = \bm{I}$ (where $\bm{I}$ is the identity tensor) and $\det{\bm{Q}} = \pm 1$.

These bases have been extensively used in the literature (see e.g, \cite{ML_1991Speziale, Gatski1993}), however it must be clarified that unlike the closure for the anisotropic stress tensor (which will be discussed in greater detail in Sec.~\ref{sec:WavyWall}), the infinite basis cannot be reduced exactly. Because the redistribution tensor, $\mathcal{R}_{ij}$ is symmetric and deviatoric, and its dependence on each of the bases is linear, it is implied that each basis tensor must also satisfy these same properties. The anisotropic stress tensor is symmetric (but \emph{not} deviatoric), the mean rotation rate tensor is deviatoric (but \emph{not} symmetric), and the mean shear rate tensor is symmetric (but \emph{not} deviatoric). Because of this, there is no minimal set of bases beyond which subsequent terms can be expressed as a linear combination of a minimal set. 

For this reason, it is common practice to truncate the basis by assessing the reduction in model error while increasing the terms present in the basis. Point of truncation is typically classified by leading order in the anisotropy tensor and nearly all contemporary models do not include terms beyond the third order (e.g., $b_{ij}^3$).  In this study, it is also found that terms beyond third order in anisotropy are unnecessary for improving model accuracy. These terms are shown in the leftmost column of Table \ref{NoiseTable}. 

Using the data generated by solving Eqs.~\eqref{Eq:RSTransport_shear}--\eqref{Eq:LRRIP}, the basis tensors are computed and the redistribution tensor is populated by taking the time derivative of the Reynolds stresses (using a 6th-order central difference scheme) and solving Eq.~\eqref{Eq:RSTransport_shear} for the redistribution tensor. Then, these quantities are assembled into $\mathbb{D}$ and $\mathbb{T}$ as described in Sec.~\ref{sec:Method}. Note that since we are interested in modeling the self-similar regime, only data from this region is used for training. After $\mathbb{D}$ and $\mathbb{T}$ are assembled, the cost functional (Eq.~\eqref{Eq.CostFunc}) is optimized for decreasing values of $\lambda$ until the model error is minimized.  

\begin{figure} 
  \centering
      \subcaptionbox{$\mathcal{N}(\mu, 0.1 \mu)$\label{NoiseTen}}
        {\includegraphics[height = 0.2\textwidth]{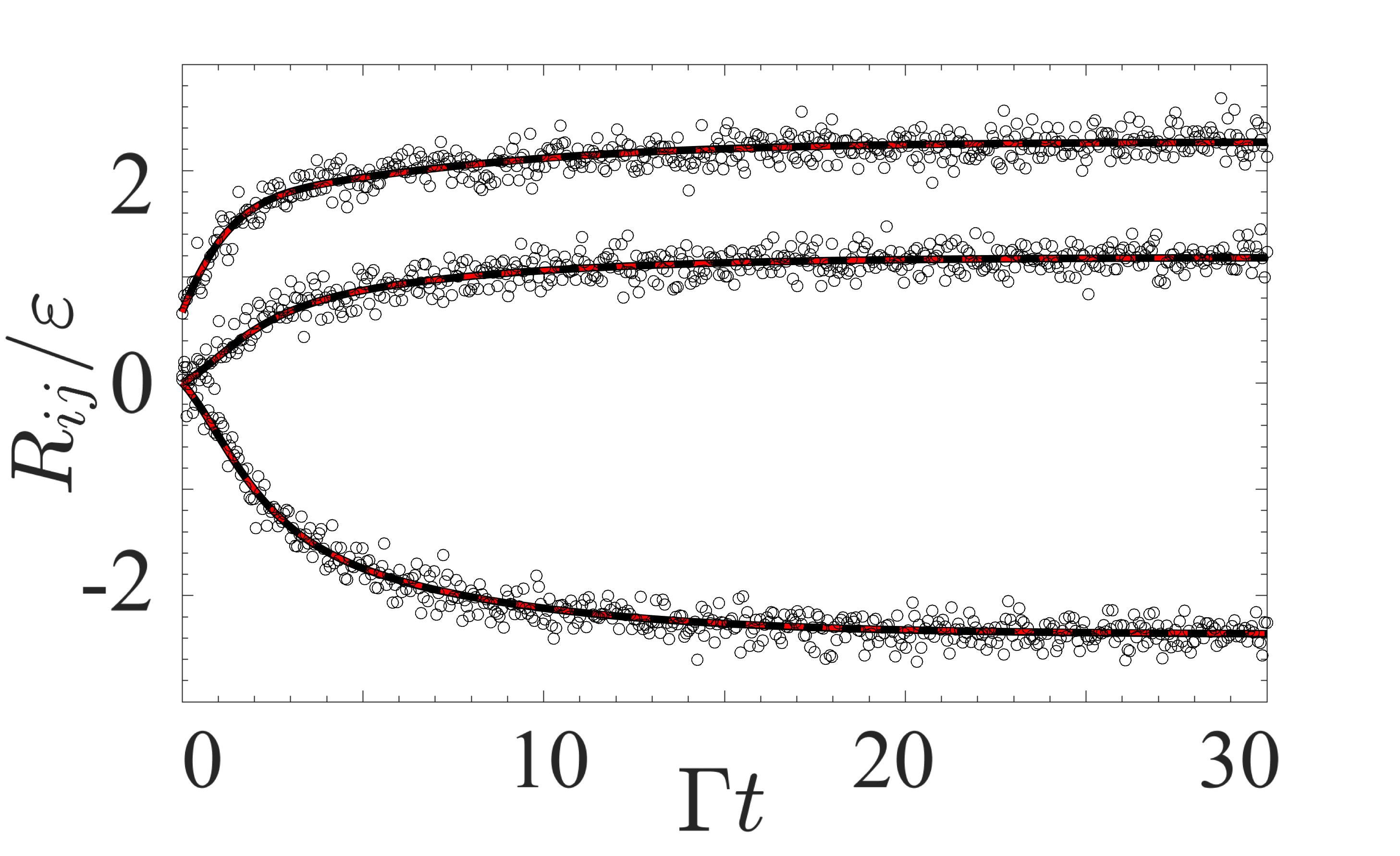} }
      \subcaptionbox{$\mathcal{N}(\mu, 0.2 \mu)$\label{NoiseTwenty}}
        {\includegraphics[height = 0.2\textwidth]{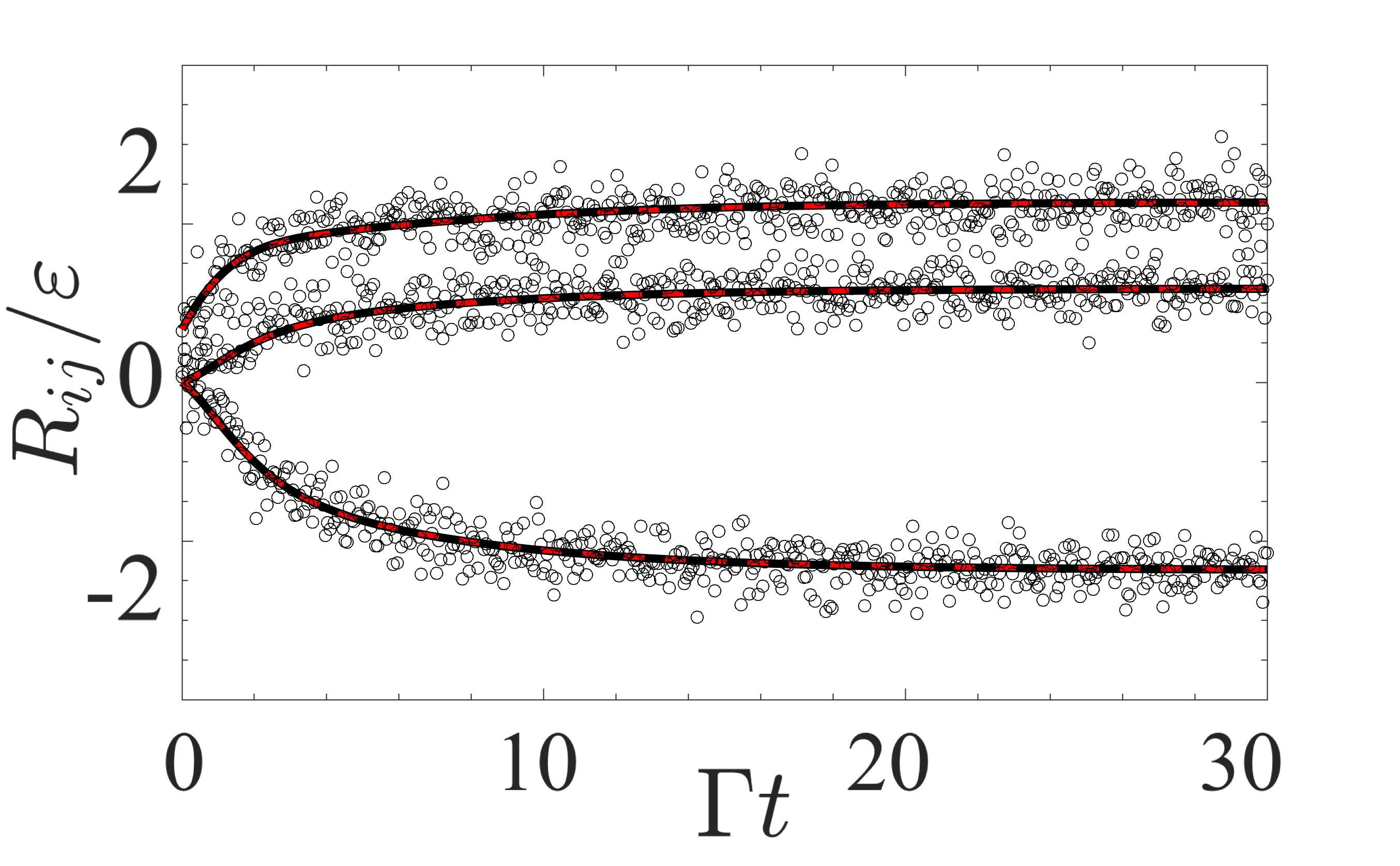}}
       \subcaptionbox{$\mathcal{N}(\mu, 0.3 \mu)$\label{NoiseThirty}}
        {\includegraphics[height = 0.2\textwidth]{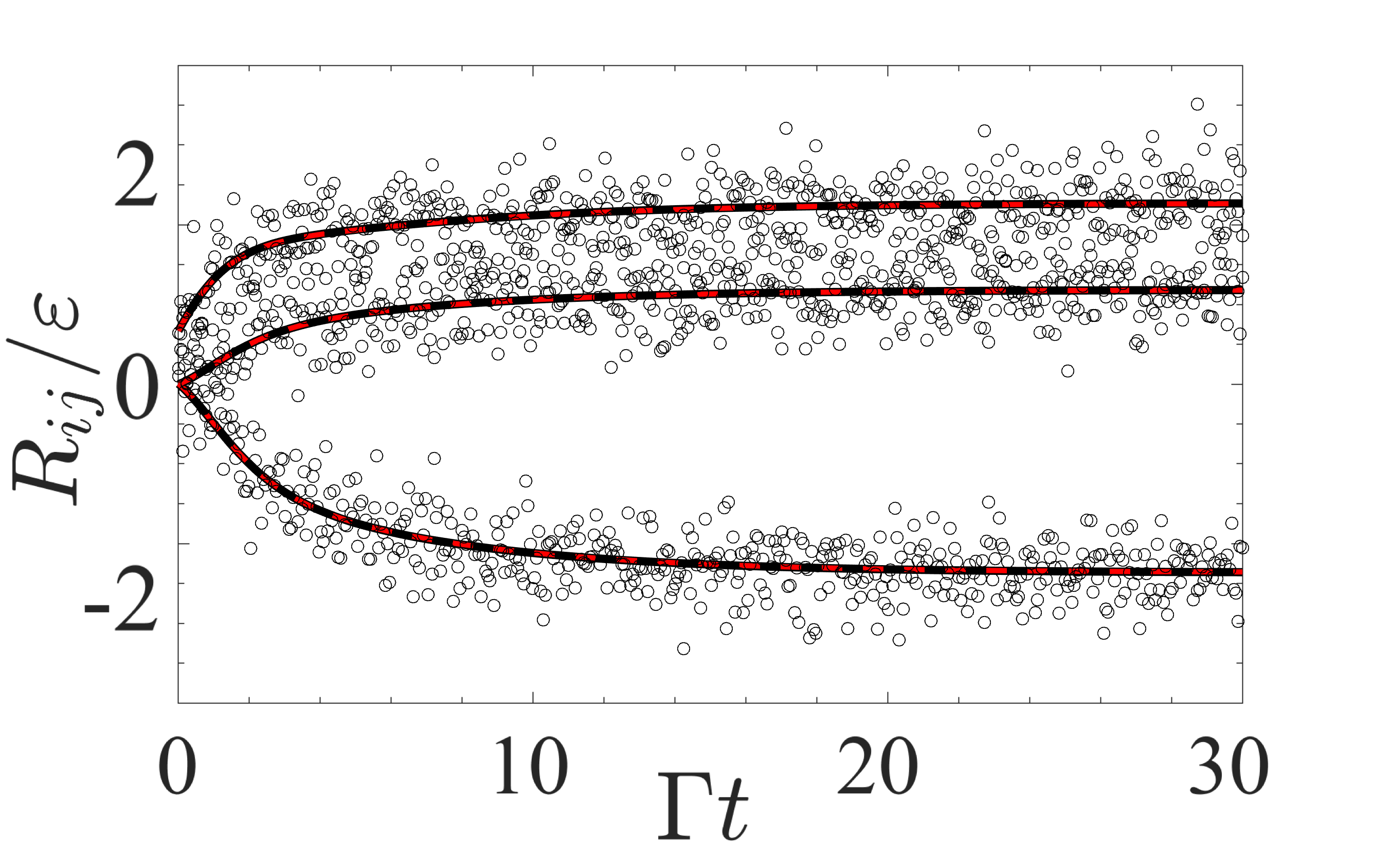}}
     \caption{Comparison between prescribed LRR-IP model (\protect \LRRIP) with the learned models (\protect \NoisyLearned) and artificially noisy data ($\circ$), for $\Gamma = 2.25$.} \label{fig:Noise}
\end{figure}

As shown in Table \ref{NoiseTable}, the methodology exactly returns the LRR-IP model used to generate the dataset. Artificial noise was added to the raw synthetic data set to test the robustness of the method. The noise was normally distributed about the mean of the synthetic data, denoted as $\mathcal{N}(\mu, \sigma)$, where $\sigma$ is the standard deviation that is prescribed in terms of percentage of the mean value, $\mu$. We consider $\sigma =0.1, 0.2,$ and $0.3\mu$. In each case, $\lambda$ was reduced until the model error plateaued. Even in the case of the noisiest data provided, the learned model deviated from the expected LRR-IP model by only 2.3\%, where error is defined by the\textit{L}-2 norm  
\begin{equation}
\epsilon = \frac{ \vert \vert \mathbb{D} - \mathbb{T}\beta \vert \vert_2}{\vert \vert \mathbb{D} \vert \vert_2}.
\end{equation} 
This level of performance indicates the sparse regression methodology is robust to substantial noise in the training data without compromising the accuracy in learning the underlying physics. This is further demonstrated in Figs.~\ref{NoiseTen}--\ref{NoiseThirty} where the models learned from the noisy data are shown against the LRR-IP model and the artificially noisy data. In all three cases, the learned model accurately describes the behavior of the LRR-IP model despite small amounts of error in the coefficients. 

\begin{table}
\begin{tabular}{c  @{\hskip 0.15in}  l c @{\hskip 0.15in} c c c c}
\hline
\hline
& \multirow{3}{*}{$\mathcal{T}^{(n)}$} & \multirow{3}{*}{LRR-IP} & \multicolumn{4}{c}{Sparse Regression} \\ [-2ex]
order & & & $\lambda = 0.1$ & $\lambda = 0.1$ & $\lambda = 0.1$ & $\lambda = 0.5$ \\ [-1ex]
in $b_{ij}$& & & $\mathcal{N} = 0 $ & $\mathcal{N}(\mu, 0.1 \mu)$ & $\mathcal{N}(\mu, 0.2 \mu)$ & $\mathcal{N}(\mu, 0.3 \mu) $ \\
\hline
{ 0} & $S_{ij}$ & 0.8 & 0.8 & 0.8010 & 0.8020 & 0.803\\[1ex]
\multirow{3}{*}{1} & $b_{ij}$ & -3.6  & -3.6 & -3.5761 & -3.5522 & -3.5282\\ [-1ex]
&$R_{il}b_{lj}+R_{jl}b_{li}$ &1.2  & 1.2 & 1.2003 & 1.2007 & 1.2010\\[-1ex]
&$S_{il}b_{lj}+S_{jl}b_{li}-\frac{2}{3}S_{lm}b_{ml}\delta_{ij}$ &1.2  & 1.2 & 1.2018 & 1.2036 & 1.2054\\ [1ex]
\multirow{3}{*}{ 2}  & $b_{ij}^2-\frac{1}{3}b_{ll}^2\delta_{ij}$ & 0 &0 & 0 & 0 & 0\\[-1ex]
&$S_{il}b_{lj}^2+S_{jl}b_{li}^2 - \frac{2}{3} S_{lm}b_{ml}^2\delta_{ij}$ &0& 0 & 0 & 0 & 0\\[-1ex]
&$R_{il}b_{lj}^2+R_{jl}b_{li}^2$& 0 & 0 & 0 & 0 &0\\[1ex]
{3 } & $b_{ik}^2R_{kp}b_{pj} -b_{il}R_{lk}b_{kj}^2$& 0& 0 & 0 & 0 & 0\\
\hline
\multicolumn{2}{c}{$\epsilon^{b_{ij}}$} & -- &\textit{0.0} & \textit{0.0076} & \textit{0.015} & \textit{0.023} \\
\hline
\hline
\end{tabular}
 \caption{Summary of model forms and associated error in the self-similar region of homogeneous free shear turbulence, with increasing amounts of artificial noise added to the synthetic dataset.\label{NoiseTable}}
\end{table}

\subsubsection{DNS-generated data: Can sparse regression improve upon existing models?}

\begin{figure}
\centering
\includegraphics[width=0.65\textwidth]{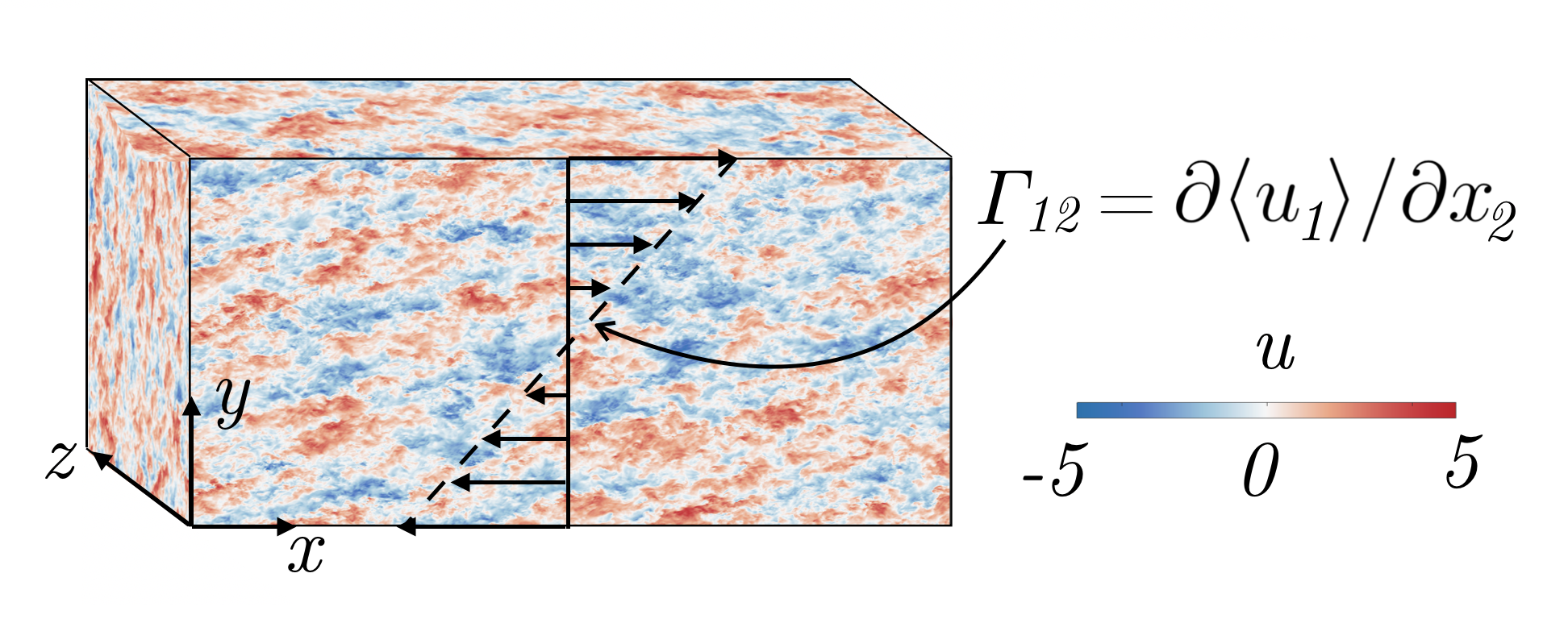}
\caption{Snapshot of the instantaneous velocity field in DNS homogeneous free shear turbulence at $\mathcal{S} = 11.2$.}
\label{Fig:Shear}
\end{figure}

Here, the same physical configuration is considered, albeit with the trusted data generated using DNS (see Fig.~\ref{Fig:Shear}). 

To generate the DNS datasets, NGA~\cite{desjardins2008high}, a fully conservative, low-Mach number finite volume solver is used. A pressure Poisson equation is solved to enforce continuity via fast Fourier transforms in all three periodic directions. The Navier--Stokes equations are solved on a staggered grid with second order spatial accuracy and time is advanced with second order accuracy using the semi-implicit Crank-Nicolson scheme of \citet{pierce2001progress}.  Shear periodic boundary conditions are enforced using the recently developed algorithm of \citet{Kasbaoui2017}. Turbulence in the domain is initialized using spectral methods in order to ensure consistency with Kolmogorov's `-5/3' spectrum \cite{Passot1986, Sreenivasan1995}. 

Five cases are simulated for non-dimensional shear rates $\mathcal{S} = 2 \Gamma k_0/\varepsilon_0 = (2.3, 6.6, 11.2, 13.2, 20.2)$ on a grid of size $1024\times512\times512$, corresponding to a domain size of $2 \pi \times \pi \times \pi$. Here, $k_0$ and $\varepsilon_0$ denote the initial values of TKE and dissipation, respectively. The grid resolution ensures that the flow captures the dissipative scales. Each case is simulated to a non-dimensional time of $\Gamma t \approx 25-30$ to ensure sufficient data in the self-similar region is captured. Of the five datasets, three are selected as training sets ($\mathcal{S} = 2.3, 11.2, 20.2$) from which a new model is learned. The remaining two datasets ($\mathcal{S} = 6.6, 13.2$) serve as testing sets in order to assess the accuracy of the learned model. 
 
In the same fashion as was described for the synthetic dataset, the DNS data is organized into $\mathbb{D}$ and the $\mathbb{T}$, and the cost functional is optimized for decreasing values of the sparsity parameter $\lambda$ until model error is minimized. The resulting models from this procedure are shown in Table \ref{Tab:summary}. As $\lambda$ is decreased, additional terms are included in the learned model and the coefficients adjust accordingly. The four learned models are compared against existing models, the Rotta \cite{Rotta1951}, LRR-IP and LRR-QI  models \cite{Launder1975} (Eqs.~\eqref{Eq:Rotta}--\eqref{Eq:LRRQI}), written in terms of the basis tensors as, 
\begin{align}
\mathcal{R}^{\text{Rotta}}_{ij} &= -2 C_R b_{ij}, \label{Eq:Rotta} \\ 
\mathcal{R}^{\text{LRR-IP}}_{ij} &= -2 C_R b_{ij} + \frac{4}{3}C_2 S_{ij} + 2 C_2 \left(R_{il}b_{lj} + R_{jl} b_{li} \right) + \\ \nonumber
&2 C_2 \left(S_{il}b_{lj} + S_{jl} b_{li} - \frac{2}{3} S_{lm}b_{ml}\delta_{ij} \right), \label{Eq:LRRIP} \\ 
\mathcal{R}^{\text{LRR-QI}}_{ij} &= -2 C_R^{\prime} b_{ij} + \frac{4}{5}S_{ij} +  \frac{2}{11}\left(10-7 C_2^{\prime}\right) \left(R_{il}b_{lj} + R_{jl} b_{li} \right) + \\ \nonumber 
 &\frac{6}{11}\left(2+3C_2^{\prime}\right) \left(S_{il}b_{lj} + S_{jl} b_{li} - \frac{2}{3} S_{lm}b_{ml}\delta_{ij} \right) \label{Eq:LRRQI}.
\end{align}
Here, the coefficients are given as $\lbrack C_R, C_2 \rbrack = \lbrack 1.8, 0.6 \rbrack$ and $\lbrack C_R^{\prime}, C_2^{\prime} \rbrack = \lbrack 1.5, 0.4 \rbrack$. The Rotta model assumes a linear relationship with the anisotropy tensor and thus models a linear return to isotropy. In contrast, the LRR-IP model includes nonlinear terms that are important for characterizing homogeneous \emph{anisotropic} turbulence. In comparing these three models with four learned models of increasing complexity, it is observed that the least complex learned model, corresponding to $\lambda = 0.75$, already shows marked improvement over the highest performing existing models and reduces error in the anisotropic stress tensor from 26\% to 9\%. 

\begin{table}
\centering
\begin{tabular}{c @{\hskip 0.15in} l  c @{\hskip 0.15in} c @{\hskip 0.15in} c @{\hskip 0.15in}  c c c c }
\hline
\hline
{order} & \multirow{2}{*}{$\mathcal{T}^{(n)}$} & \multirow{2}{*}{Rotta} & \multirow{2}{*}{LRR-IP} & \multirow{2}{*}{LRR-QI} & \multicolumn{4}{c}{Sparse Regression} \\ [-1ex]
{in $b_{ij}$} & & & & & $\lambda = 0.75$ & $\lambda = 0.6$ & $\lambda = 0.5$ & $\lambda = 0$ \\
\hline
0 &$S_{ij}$ & 0 &0.8 & 0.8 & 1.01 & 1.01 & 0.98 & 0.98\\ [1ex]
\multirow{3}{*}{1} &$b_{ij}$ & -3.6 & -3.6 & -3.0 & 1.27 & 1.31 & 1.45 & 1.46\\[-1ex]
&$R_{il}b_{lj}+R_{jl}b_{li}$ & 0 &1.2 & 1.31 & 1.53 & 1.56 & 1.49 & 1.48\\[-1ex]
&$S_{il}b_{lj}+S_{jl}b_{li}-\frac{2}{3}S_{lm}b_{ml}\delta_{ij}$ & 0 &1.2 & 1.74 & 1.73 & 1.71 & 1.79 & 1.78\\ [1ex]
\multirow{3}{*}{2}& $b_{ij}^2-\frac{1}{3}b_{ll}^2\delta_{ij}$ & 0 & 0 & 0 &5.22 & 4.64 & 7.02 & 6.71\\[-1ex]
&$S_{il}b_{lj}^2+S_{jl}b_{li}^2 - \frac{2}{3} S_{lm}b_{ml}^2\delta_{ij}$ &0 & 0 & 0& 0 & 0 & 0.57 & 0.56\\[-1ex]
&$R_{il}b_{lj}^2+R_{jl}b_{li}^2$& 0 &0 & 0 & 0 & 0 & 0 &0.13\\[1ex]
3 & $b_{ik}^2R_{kp}b_{pj} -b_{il}R_{lk}b_{kj}^2$&0 & 0 & 0& 0 &-0.65 & 2.08 & 2.45 \\
\hline
& training error, $\epsilon^{\bm{b}}_{\text{train}}$ & \multirow{2}{*}{0.68} & \multirow{2}{*}{0.26} & \multirow{2}{*}{0.26} &0.090 & 0.092 & 0.078 & 0.073 \\ [-1ex]
& testing error, $\epsilon^{\bm{b}}_{\text{test}}$ & 			             &				       &                                  & 0.086 & 0.089 & 0.070 & 0.078 \\
\hline
\hline
\end{tabular}
\vspace{1em}
\caption{Summary of learned and existing models and associated error in the self-similar region for homogeneous free shear turbulence.}
\label{Tab:summary}
\end{table}

\begin{figure}[h!]
      \centering
      \subcaptionbox{Up to first order terms \label{fig:Order1Coeff}}
        {\includegraphics[width = 0.3\textwidth]{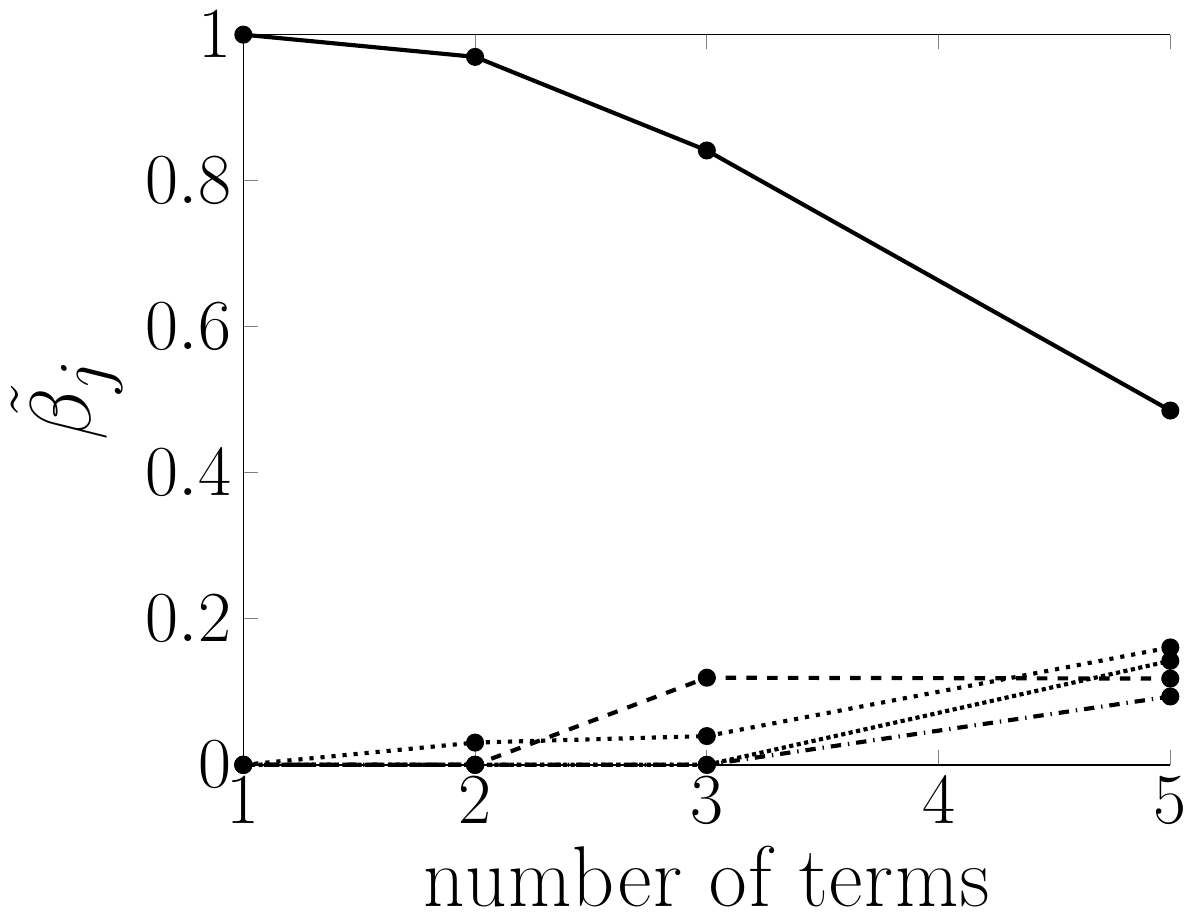} }
      \subcaptionbox{Up to second order terms \label{fig:Order2Coeff}}
        {\includegraphics[width = 0.3\textwidth]{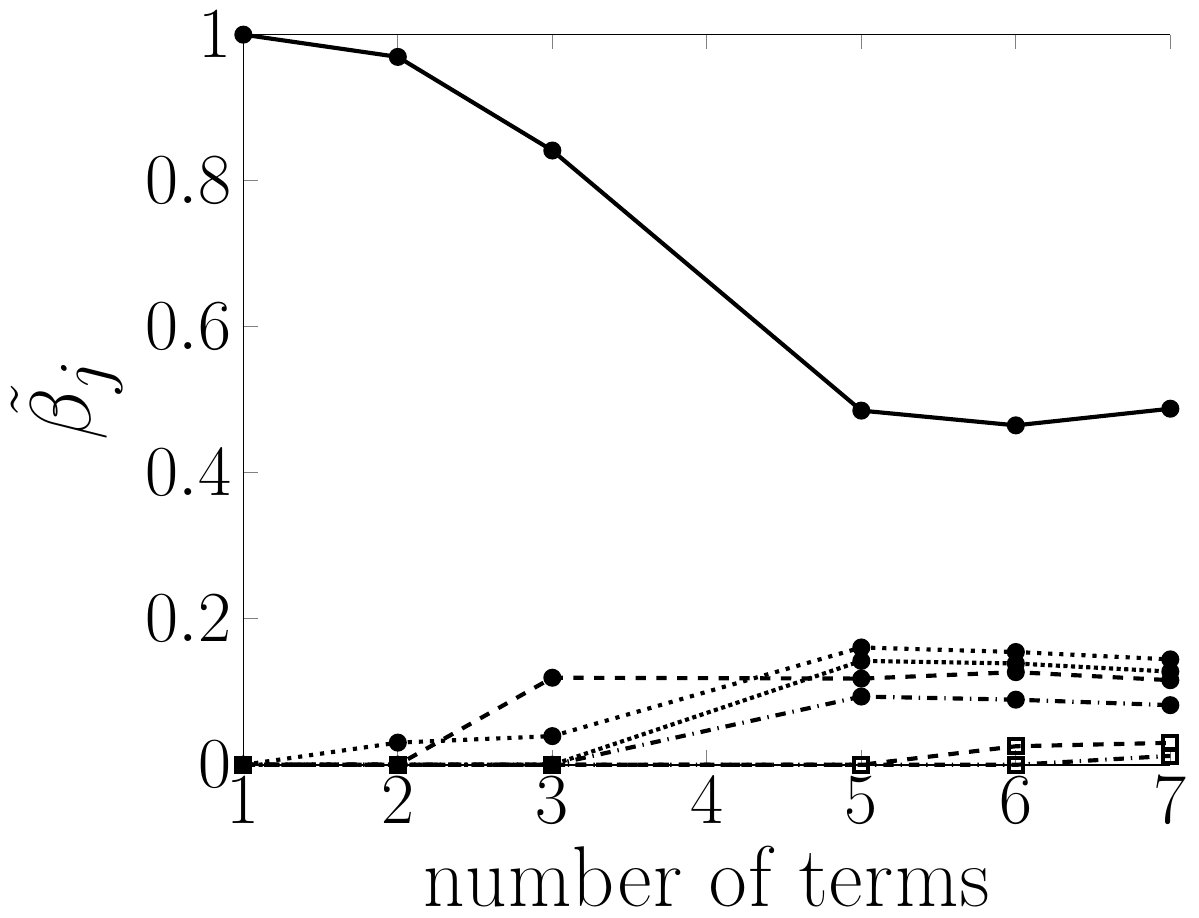}}
       \subcaptionbox{Up to third order terms \label{fig:Order3Coeff}}
        {\includegraphics[width = 0.3\textwidth]{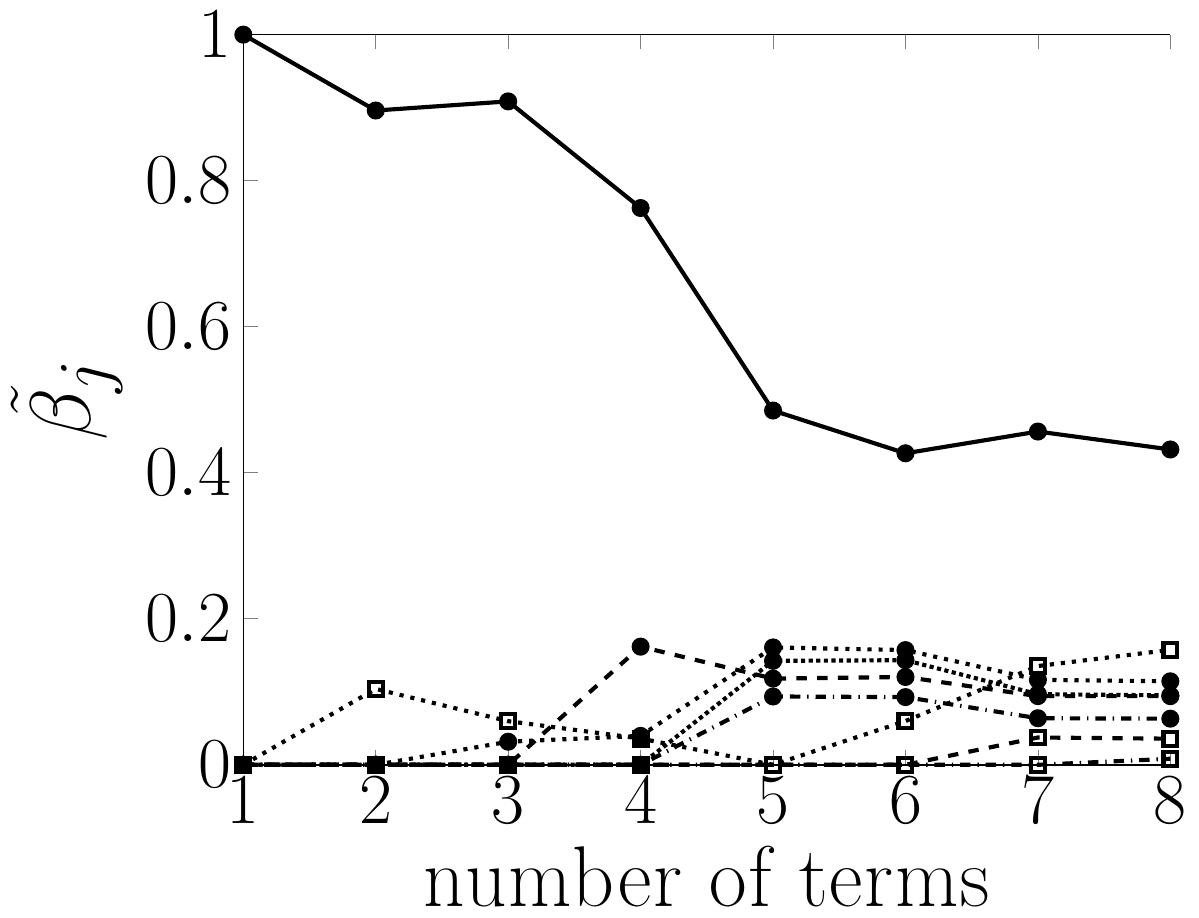}} 
     \caption{As the number of terms in the model increases (by decreasing $\lambda$), terms that are most important for capturing key redistribution physics arise in the sparsest models and persist with prominent coefficients as terms are added.} \label{Fig:Coeffs}
\end{figure}

Shown in Fig.~\ref{Fig:Coeffs}, as $\lambda$ is decreased and terms are added to the learned model, the normalized coefficients, $\tilde{\beta}$, change to accommodate contributions from additional terms. In Fig.~\ref{fig:Order1Coeff}, the sparse regression methodology is employed on a basis that is restricted to up to first order in $b_{ij}$. The basis is then expanded to second and third order terms in $b_{ij}$ in Figs. \ref{fig:Order2Coeff} and \ref{fig:Order3Coeff}, respectively. In each instance, the most prominent coefficient remains the largest contribution to the learned model, though its contribution is decreased as subsequent terms are added. The order of prominence of the lesser contributing terms does not remained fixed once the number of terms in the model grows. This behavior has an insignificant effect on model performance and sensitivity and serves to demonstrate the relative lesser importance of these terms to the overall model performance as compared with the terms with larger contributions. Finally, it is observed in Figs. \ref{fig:Order2Coeff} and \ref{fig:Order3Coeff} that the dominant coefficient stagnates beyond a five term model. This mirrors the reduction in overall model error as shown in Table \ref{Tab:summary}. 

\begin{figure}
\centering
\includegraphics[width = 0.5\textwidth]{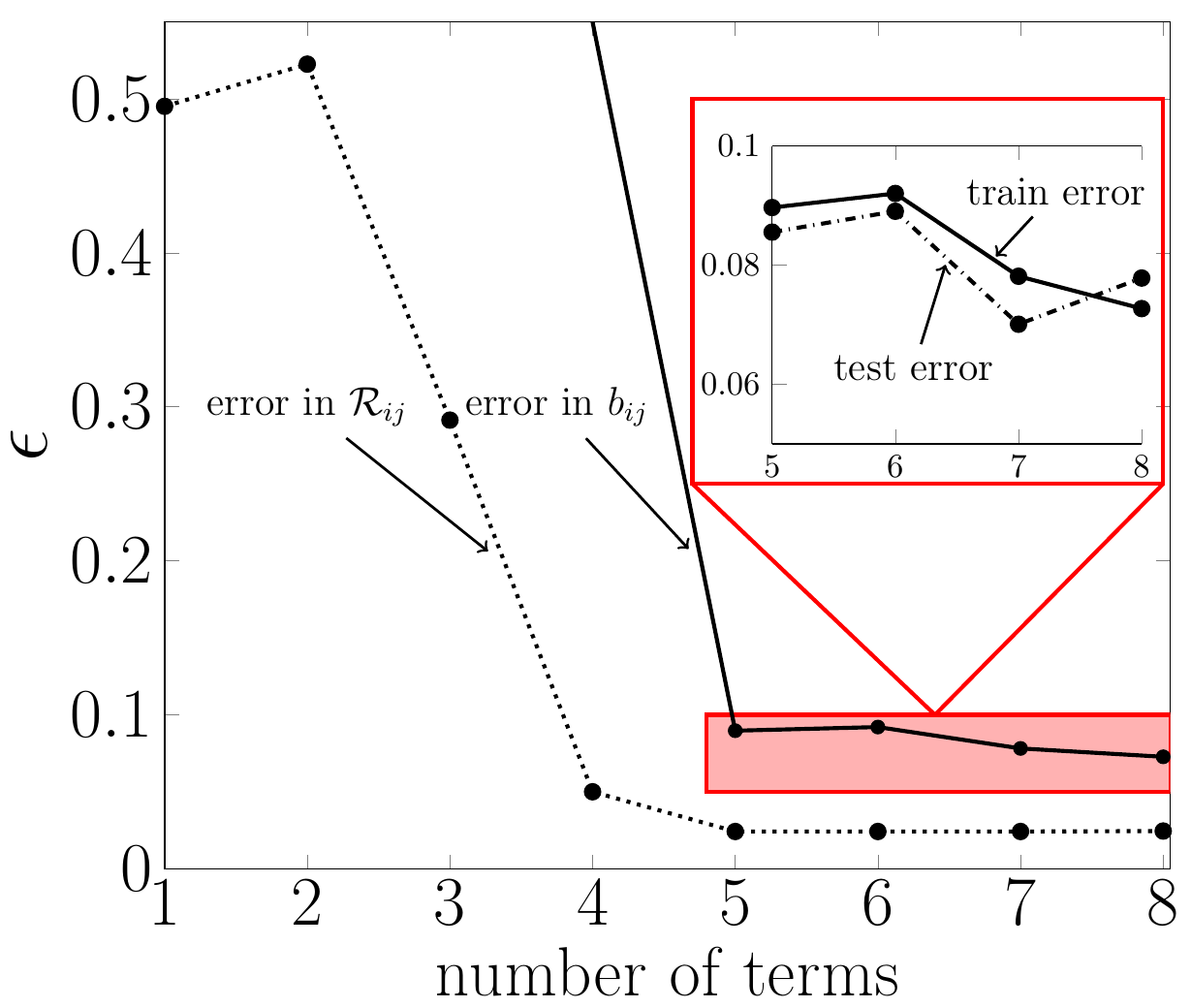}
\caption{Error in $\mathcal{R}_{ij}$ and $b_{ij}$ are shown for models of increasing complexity for homogeneous free shear turbulence. The inset figure delineates testing and training error.}
\label{Fig:error}
\end{figure}

A comparison of the training and testing errors give the clearest indication of when a learned model begins to exhibit symptoms of over-fitting (see Fig.~\ref{Fig:error}). While the learned closure predicts the redistribution tensor, $\mathcal{R}_{ij}$, the ultimate goal is to improve performance in predicting anisotropy in the Reynolds stresses, $b_{ij}$, making both measures of error relevant to assessing learned models. As shown in Fig.~\ref{Fig:error}, the error in $\mathcal{R}_{ij}$ decreases monotonically beyond a two term model, however, we observe that five terms are required for stability in the transport equation for the Reynolds stresses. Testing and training errors are also compared in Fig.~\ref{Fig:error}. As might be expected, the training error generally decreases as terms are added, but beyond seven terms in the learned model, an increase in testing error is observed. This is indicative of over-fitting, thus making the seven term model the ideal model that minimizes model error while maximizing accuracy of the model across different shear rates. As seen in Table \ref{Tab:summary}, the ideal learned model reduces error in predicting self-similar behavior by more than half as compared with the LRR-IP or LRR-QI models and more than eight fold as compared with the Rotta model. 

\begin{figure}[h!]
      \centering
      \subcaptionbox{$\mathcal{S}=3.2$\label{fig:SindyS3}}
        {\includegraphics[width = 0.3\textwidth]{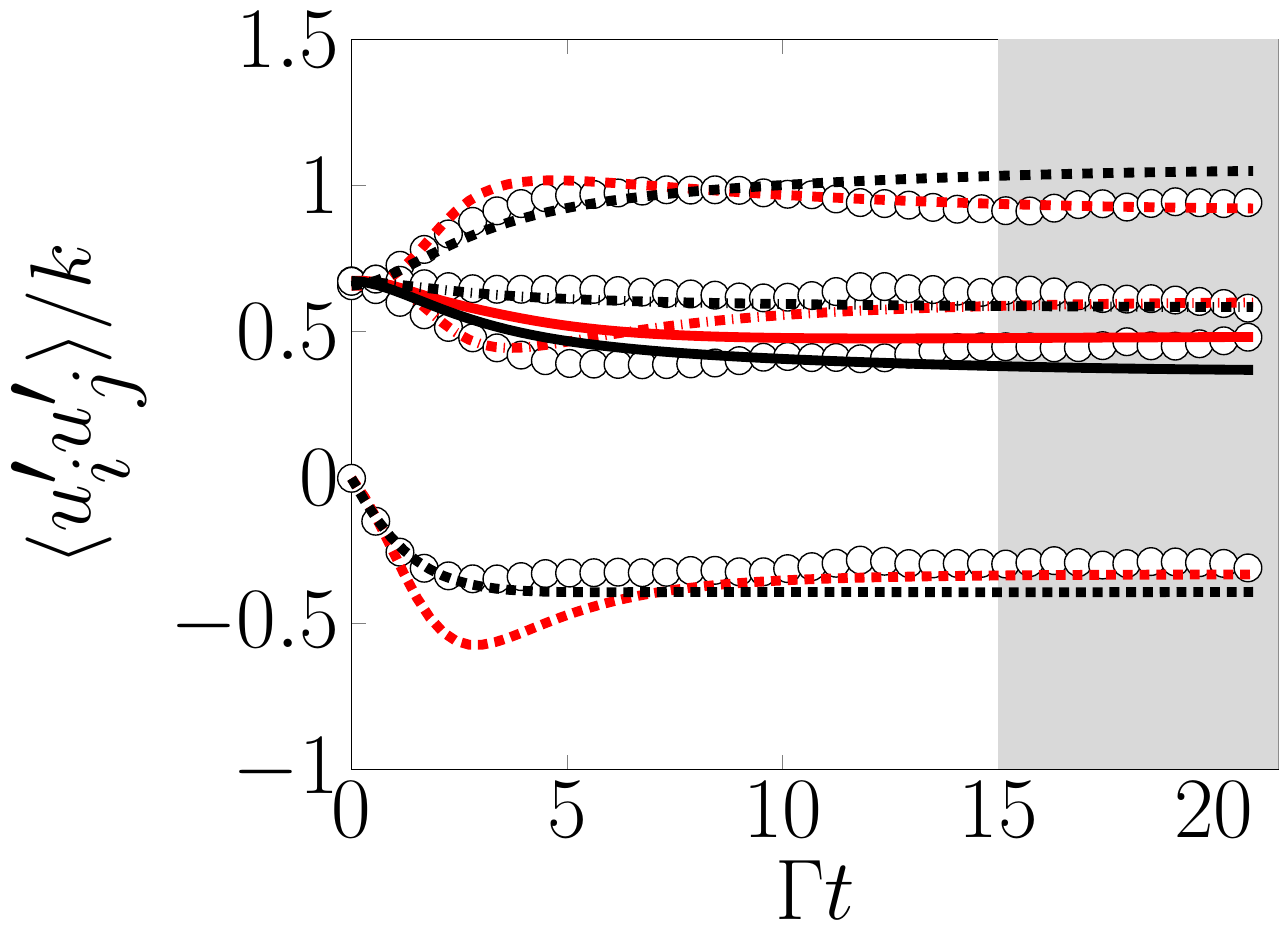} }
      \subcaptionbox{$\mathcal{S}=16.1$\label{fig:SindyS15}}
        {\includegraphics[width = 0.3\textwidth]{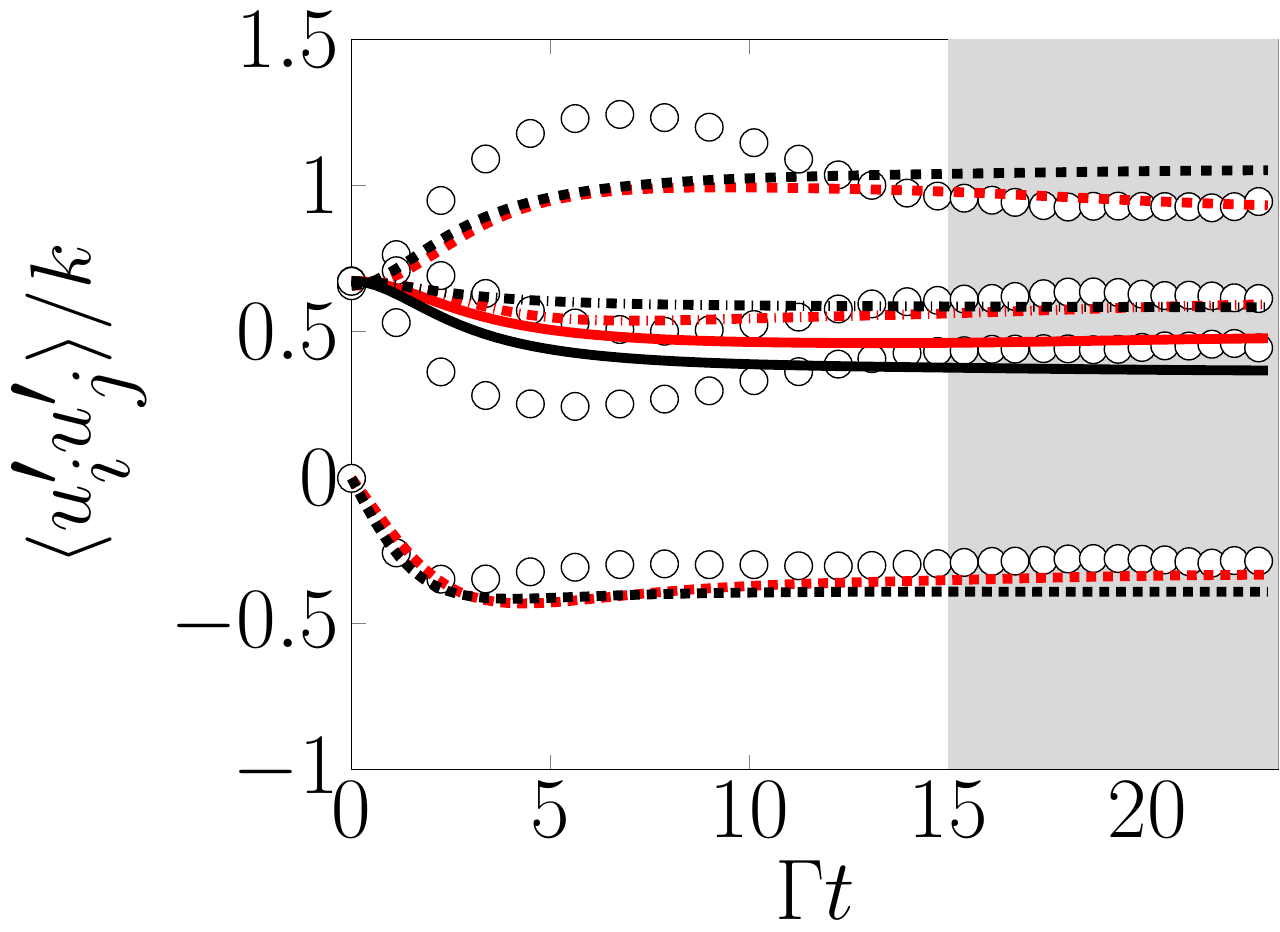}}
       \subcaptionbox{$\mathcal{S}=30.7$\label{fig:SindyS27}}
        {\includegraphics[width = 0.3\textwidth]{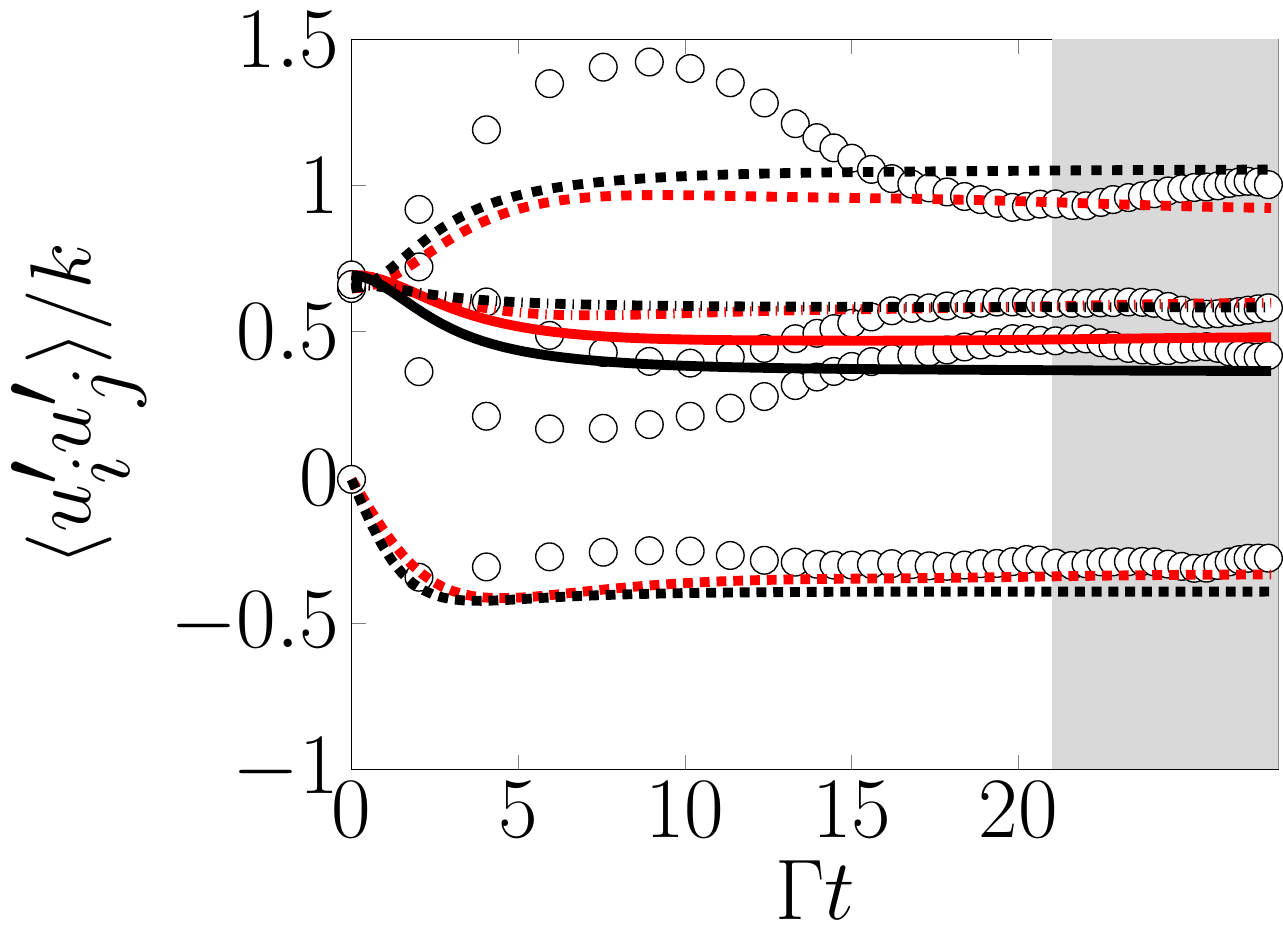}} \\
         \subcaptionbox{$\mathcal{S}=10.0$\label{fig:SindyS10}}
        {\includegraphics[width = 0.3\textwidth]{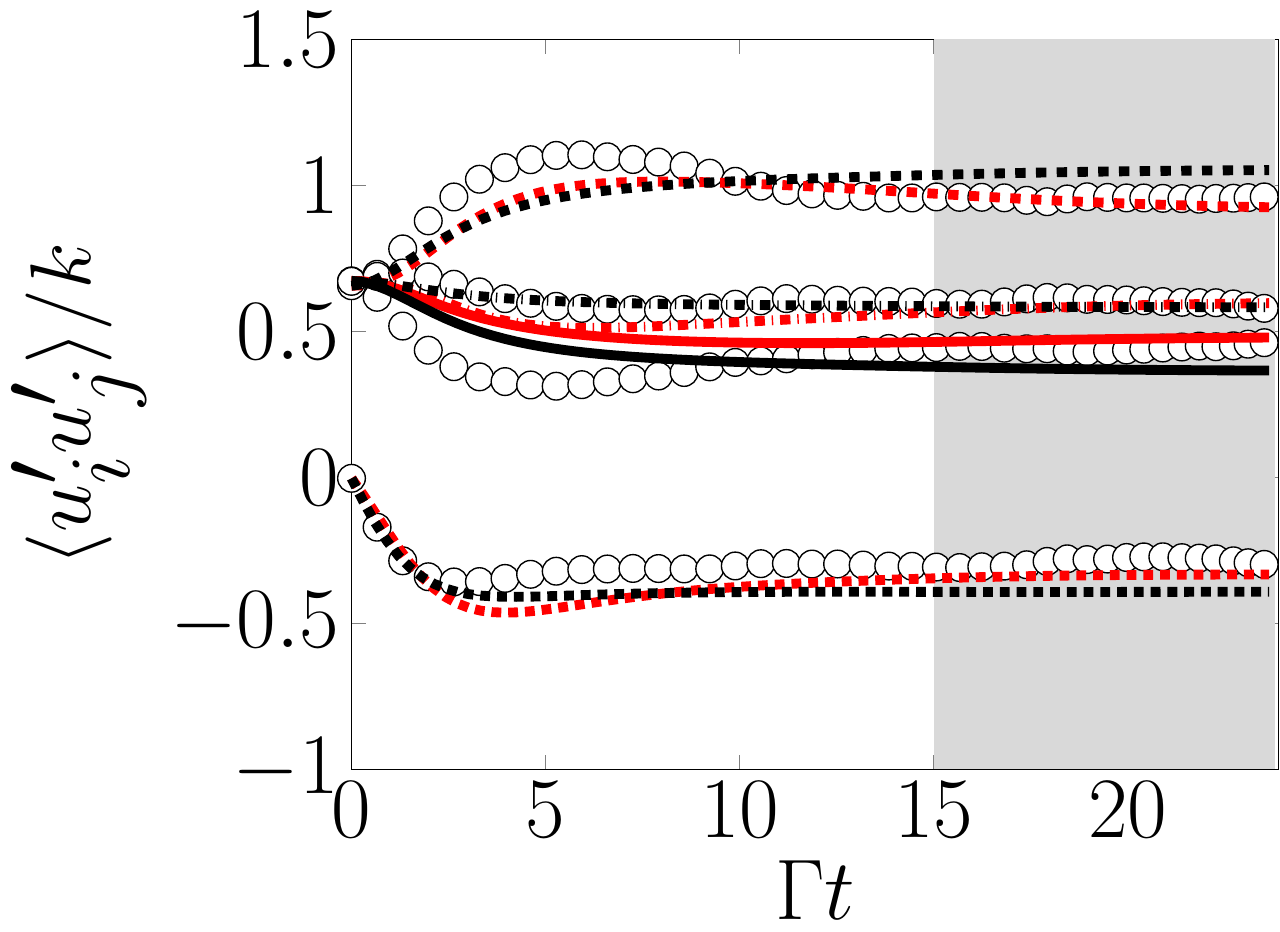}}
       \subcaptionbox{$\mathcal{S}=20.1$\label{fig:SindyS20}}
        {\includegraphics[width = 0.3\textwidth]{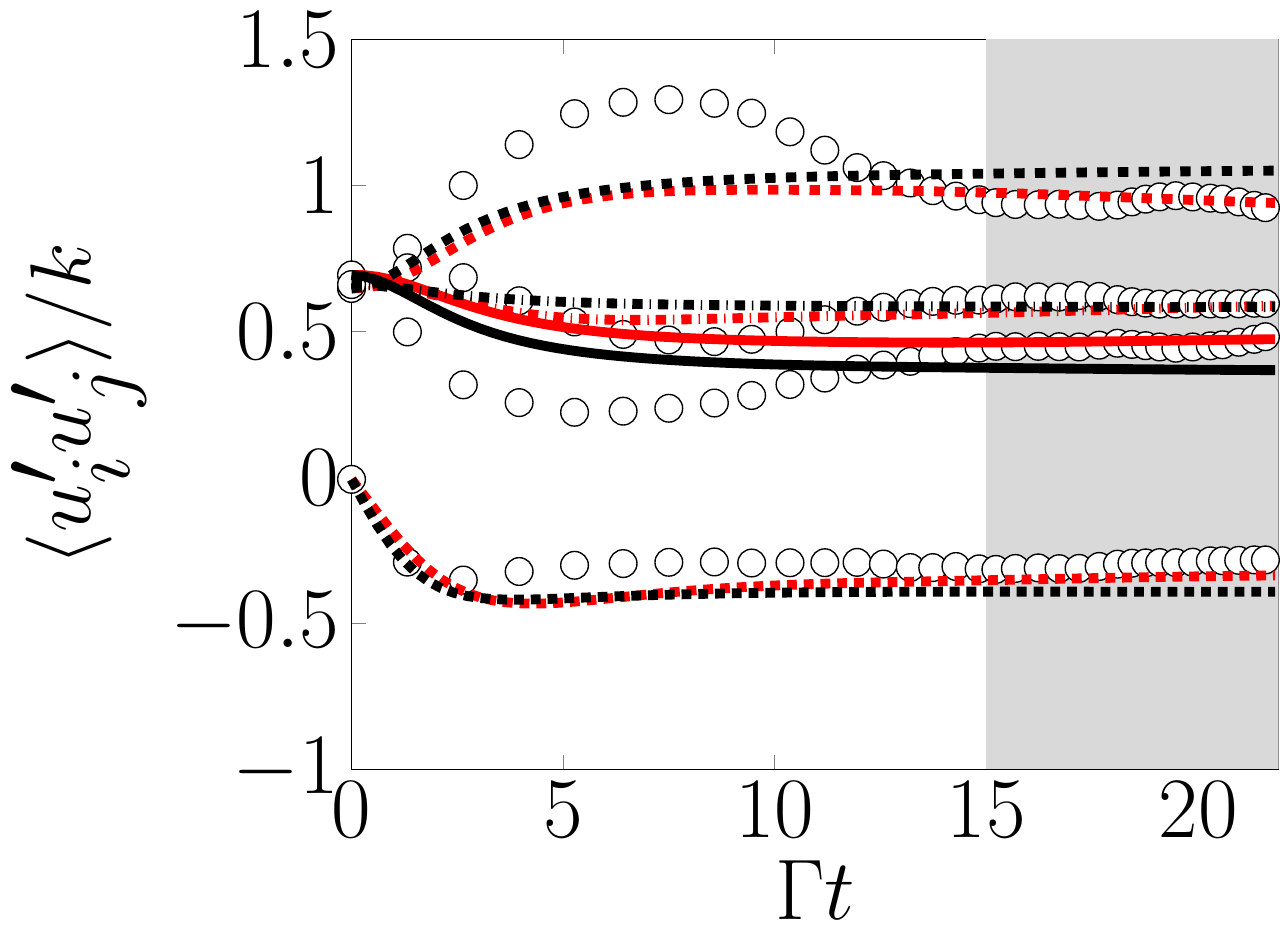}} \\
     \caption{Sparse regression (\protect \Sindy) produces a more accurate model as compared with the most accurate traditional closure available (LRR-QI, \protect \LRRIP). DNS data is denoted by open circles and the four lines correspond to the unique components of the normalized Reynolds stresses: $\langle u^{\prime} u^{\prime} \rangle$: \protect \uu,  $\langle u^{\prime}v ^{\prime}\rangle$: \protect \uv, $\langle v^{\prime}v^{\prime} \rangle$: \protect \vv,   and $\langle w^{\prime}w^{\prime} \rangle $ \protect \ww. The shaded portion denotes the self-similar region of the flow. } \label{Fig:summaryGraphs}
\end{figure}

In Fig.~\ref{Fig:summaryGraphs}, the ideal seven-term model is compared with the highest performing existing model, the LRR-QI model. Both are plotted against the DNS values used for training (\ref{fig:SindyS3}--\ref{fig:SindyS27}) and for testing (\ref{fig:SindyS10}--\ref{fig:SindyS20}). As previously discussed, it is observed that the learned model accurately captures the self-similar behavior (shown in grey shaded regions) of the normalized Reynolds stresses even in the testing cases which were not seen by the sparse regression method during training. 

\subsubsection{A note on non-inertial frames of reference} 
If a \emph{non-inertial} frame is to be considered \cite{Speziale1989, Jongen1998, Gatski1993}, one would need to modify the normalized, mean rotation rate tensor to include the rotation rate of the frame with respect to an inertial frame ($\boldsymbol{\Omega}$), i.e. $\hat{\mathbb{R}}_{ij} = \hat{R}_{ij} + \epsilon_{mji}\Omega_m$, where $\epsilon_{mji}$ denotes the permutation tensor. Additionally, Coriolis terms, $\left (\langle u_i u_k \rangle \epsilon_{mkj} \Omega_m + \langle u_j u_k \rangle \epsilon_{mki} \Omega_m \right )$, must be included in Eq.~\eqref{Eq:RSTransport_shear}.

\section{Turbulent flow through a periodically constricted channel \label{sec:WavyWall}}
\subsection{Problem statement} 
\begin{figure}
\centering
\includegraphics[width = 0.75\textwidth]{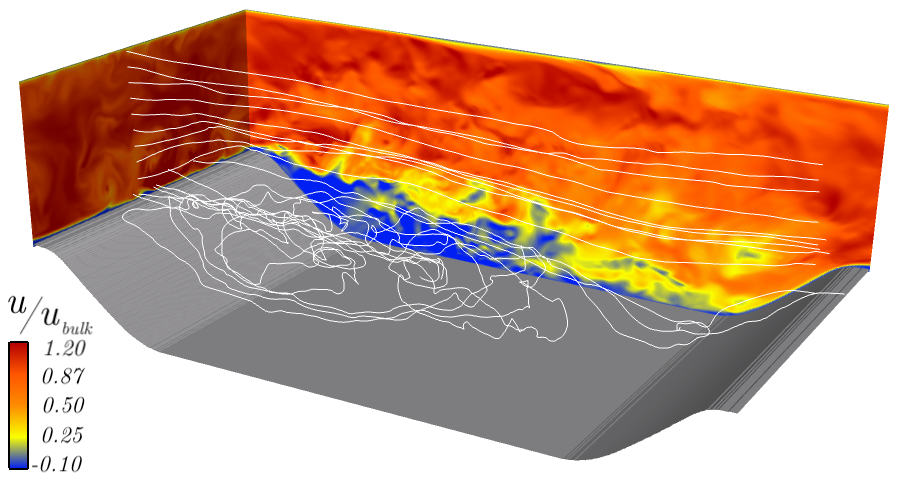}
\caption{Instantaneous streamwise velocity (color), with streamlines originating from $ = L_z/2$ (white lines).} \label{Fig:3Dwavy}
\end{figure}

In this section, we consider the classical case of turbulent flow through a periodically constricted channel as shown in Fig.~\ref{Fig:3Dwavy} and described in \citet{Breuer2009}. As discussed in Sec.~\ref{sec:CaseStudies}, two main approaches are typically taken when developing closures for the Reynolds stresses. In Sec.~\ref{sec:FreeShear}, the transport of the Reynolds stresses was addressed and in this section, algebraic closure of the Reynolds stresses will be developed. 

In this strategy, the algebraic closure for the Reynolds stresses depends on a model for the anisotropic stress tensor, such that $\langle u_i^{\prime} u_j^{\prime} \rangle = 2k(b_{ij} + \frac{1}{3} \delta_{ij})$. Further it has been well established that the model for $b_{ij}$ depends upon $\hat{S}_{ij}$ and $\hat{R}_{ij}$. Recalling from Section \ref{sec:FreeShear} that these quantities are normalized by TKE, $k$, and dissipation of TKE, $\varepsilon$, this method requires the transport of both $k$ and $\varepsilon$, which are given by  
\begin{align}
\frac{\partial k}{\partial t} + \frac{\partial  \left( k u_i \right)}{\partial x_i} &= \frac{\partial}{\partial x_j} \left \lbrack \left( \nu + \frac{\nu_t}{\sigma_k} \right) \frac{\partial k}{\partial x_j} \right \rbrack + \mathcal{P} - \varepsilon, \\
\frac{\partial \varepsilon }{\partial t} + \frac{\partial  \left( \varepsilon u_i \right) }{\partial x_i}&= \frac{\partial}{\partial x_j} \left \lbrack \left( \nu + \frac{\nu_t}{\sigma_{\varepsilon}} \right) \frac{\partial \varepsilon}{\partial x_j} \right \rbrack + C_{1\varepsilon}\frac{\varepsilon}{k}\mathcal{P} - C_{2\varepsilon} \frac{\varepsilon^2}{k}, 
\end{align}
where $\lbrack C_{\mu}, \sigma_{k}, \sigma_{\varepsilon},C_{1\varepsilon}, C_{2\varepsilon}\rbrack = \lbrack 0.09, 1.00, 1.30, 1.44, 1.92 \rbrack$. The turbulent viscosity, $\nu_t$, is given as $\nu_t = C_{\mu} k^2/\varepsilon$ where $C_{\mu} = 0.09$ \cite{pope2000turbulent}. 

Using these equations along with a model for the anisotropic stress tensor, the RANS equations (Eqs.~\eqref{Eq:RANScontinuity} and~\eqref{Eq:RANSmom}) are closed. The aim of this study is to use sparse regression to develop an improved algebraic closure for the anisotropic stress tensor. As is commonly used in the literature, the configuration under consideration here is turbulent flow through a periodically constricted channel (see Fig.~\ref{Fig:3Dwavy}). This flow configuration is particularly challenging because the quantities of interest are statistically 2-D (with dependence on the stream-wise and cross stream directions) and the presence of the constriction generates massive separation in the flow. 

The dataset used for training was simulated using NGA, described in Sec.~\ref{sec:FreeShear}. The top and bottom walls apply a no-slip boundary condition and the bottom, constricted wall is enforced using a cut-cell immersed boundary method \cite{capecelatro2013euler}. The geometry for the configuration under study matches the configuration described in \citet{Breuer2009} with uniform grid spacing discretized by $[N_x, N_y, N_z] = [512, 380, 380]$. A Reynolds number of 2800 is considered, where Re$=u_{\text{bulk}} h/\nu$. The bulk velocity $u_{\text{bulk}}$ is given by the mean velocity at the hill crest, and $h$ is the hill height. After reaching a statistically stationary point, the DNS data was averaged in the cross-stream ($z$--direction) and temporally for 44 flow through times.    

A Linear Eddy Viscosity Model (LEVM) is frequently used to close the Reynolds stresses that appear in the RANS equations \cite{pope2000turbulent}. This closure takes the form
\begin{equation}
b_{ij} = -C_{\mu}\hat{S}_{ij},
\end{equation}
which will serve for comparison purposes as the `existing' model. 

As discussed in Sec.~\ref{sec:Background}, NNs have been used in recent years to develop improved models for the anisotropic stress tensor and demonstrated marked improvement over traditional models. Here, sparse regression is applied to the same problem to seek improved closures for $b_{ij}$, while maintaining the benefits of interpretability and transportability. 

As outlined in Sec.~\ref{sec:Method}, the basis on which to train the model must first be identified. As previously derived \cite{ML_1975Pope}, a minimal integrity basis for the anisotropy tensor can be formulated using the normalized mean rotation and shear rate tensors, $\hat{R}_{ij}$ and $\hat{S}_{ij}$, respectively. Since the anisotropy stress tensor is symmetric and deviatoric, each of $\mathcal{T}_{ij}^{(n)}$ must also have these properties. After formulating combinations of $\hat{S}_{ij}$ and $\hat{R}_{ij}$ with these properties, and owing to the Cayley-Hamilton theory, all symmetric and deviatoric tensors that are combinations of $\hat{S}_{ij}$ and $\hat{R}_{ij}$ can be formed as a linear combination of the 10 basis tensors shown in Table \ref{tab:basis_b} \cite{ML_1975Pope}.  

\begin{table}[h!]
\begin{center}
\begin{tabular}{r l r l} 
\hline 
\hline 
$\mathcal{T}_{ij}^{(1)}$ &$ = \hat{{\textit S}}_{ij}$ & $\mathcal{T}_{ij}^{(6)}$ & $ =\hat{R}_{ik}\hat{R}_{kl}\hat{S}_{lj}+\hat{S}_{ik}\hat{R}_{kl}\hat{R}_{lj} - \frac{2}{3}\hat{S}_{pk}\hat{R}_{kl}\hat{R}_{lp} \delta_{ij} $\\
$\mathcal{T}_{ij}^{(2)}$ &$ = \hat{S}_{ik}\hat{R}_{kj} - \hat{R}_{ik}\hat{S}_{kj}$ & $\mathcal{T}_{ij}^{(7)} $ & $ = \hat{R}_{ik}\hat{S}_{kl} \hat{R}_{lp} \hat{R}_{pj} - \hat{R}_{ik}\hat{R}_{kl} \hat{S}_{lp} \hat{R}_{pj} $ \\
$\mathcal{T}_{ij}^{(3)}$ & $=\hat{S}_{ik}\hat{S}_{kj} - \frac{1}{3} \hat{S}_{lk}\hat{S}_{kl} \delta_{ij}$ & $\mathcal{T}_{ij}^{(8)}$ & $= \hat{S}_{ik}\hat{R}_{kl} \hat{S}_{lp} \hat{S}_{pj} - \hat{S}_{ik}\hat{S}_{kl} \hat{R}_{lp} \hat{S}_{pj} $ \\
$\mathcal{T}_{ij}^{(4)}$ & $ = \hat{R}_{ik}\hat{R}_{kj} - \frac{1}{3} \hat{R}_{lk}\hat{R}_{kl} \delta_{ij}$ & $\mathcal{T}_{ij}^{(9)}$ & $= \hat{R}_{ik}\hat{R}_{kl} \hat{S}_{lp} \hat{S}_{pj} + \hat{S}_{ik}\hat{S}_{kl} \hat{R}_{lp} \hat{R}_{pj}  - \frac{2}{3}\hat{S}_{qk}\hat{S}_{kl} \hat{R}_{lp} \hat{R}_{pq}$ \\
$\mathcal{T}_{ij}^{(5)}$ & $= \hat{R}_{ik}\hat{S}_{kl}\hat{S}_{lj} -  \hat{S}_{ik}\hat{S}_{kl}\hat{R}_{lj}$ & $\mathcal{T}_{ij}^{(10)}$ &$= \hat{R}_{ik}\hat{S}_{kl}\hat{S}_{lp}\hat{R}_{pq}\hat{R}_{qj} - \hat{R}_{ik}\hat{R}_{kl}\hat{S}_{lp}\hat{S}_{pq}\hat{R}_{qj}$\\
\hline 
\hline
\end{tabular}
\end{center}
\caption{The ten tensor bases that exactly describe the anisotropic stress tensor.}
\label{tab:basis_b}
\end{table}

Using this basis, the anisotropic stress tensor can be represented exactly as
\begin{equation}
b_{ij} = \sum_{n = 1}^{\infty} G^{(n)}\mathcal{T}^{(n)}_{ij}\left ( \hat{R}_{ij}, \hat{S}_{ij}\right). \label{eq:b_general}
\end{equation}

In the case of statistically two-dimensional flows, as is the case here, the basis simplifies to only three tensors and the coefficients depend on at most only two invariants as shown in Table \ref{tab:b_basis2} \cite{ML_1975Pope, Gatski1993}. 

\begin{table}
\centering 
\begin{tabular}{c l} 
\hline
\hline
$\mathcal{T}_{ij}^{(1)}$ & $\hat{S}_{ij}$ \\
$\mathcal{T}_{ij}^{(2)}$ & $\hat{S}_{ik}\hat{R}_{kj} - \hat{R}_{ij}\hat{S}_{kj}$ \\
$\mathcal{T}_{ij}^{(3)}$ & $\hat{S}_{ik}\hat{S}_{kj} - \frac{1}{3}\hat{S}_{lk}\hat{S}_{kl}\delta_{ij}$ \\
$\lambda_1$ & $\hat{S}_{lk}\hat{S}_{kl}$ \\
$\lambda_2$ & $\hat{R}_{lk}\hat{R}_{kl}$ \\
\hline
\hline
\end{tabular}
\caption{The reduced basis set for statistically two-dimensional flows.}
\label{tab:b_basis2}
\end{table}

Following the sparse regression methodology described in Sec.~\ref{sec:Method}, the DNS dataset is formulated into $\mathbb{D}$ and $\mathbb{T}$. However, instead of modeling $b_{ij}$ directly, the anisotropic stress tensor is split into linear and nonlinear portions, denoted by $b_{ij}^{\parallel}$ and $b_{ij}^{\perp}$, respectively. The linear portion will be taken as the standard LEVM and the nonlinear portion will be the subject of modeling efforts. 
\begin{align}
b_{ij} &= b_{ij}^{\perp} + b_{ij}^{\parallel} \\ 
         &= b_{ij}^{\perp}\left(k, \varepsilon, \hat{S}_{ij}, \hat{R}_{ij} \right) - C_{\mu} \hat{S}_{ij} \\
\mathcal{D}_{ij} &= b_{ij}^{\perp} = b_{ij} + C_{\mu}\hat{S}_{ij}. 
\end{align}

This strategy is employed based upon the recommendation of several works that have pointed out the ill-conditioning of the RANS equations \cite{Wu2019}. These works suggest that separating the model into a linear portion (solved implicitly with the viscous terms in the RANS solver) and a nonlinear portion (solved explicitly) improves stability of the integrated RANS solver \cite{Liu2019, Schmelzer2019}. Further, since the standard LEVM model is used as the starting point for modeling, the basis is formulated using data from a forward solution in OpenFOAM \cite{Weller1998} using the LEVM closure. Because the $k-\varepsilon$ equations contain models and are thereby a source of error in the `trusted' training data, this data must be used as a starting point for modeling. 

Using this formulation, sparse regression is employed to discover an improved model. This effort results in both an \emph{a priori} and an \emph{a posteriori} analysis of the model. In the former analysis, the training data is used to evaluate the accuracy of the learned model within the context of predicting the anisotropy tensor and compared against the NN performance of \citet{ML_2016Ling}. In the latter analysis, the learned model is implemented in OpenFOAM and the forward solution is compared against the trusted DNS data and the existing LEVM. Additionally, as an `upper end' metric, a look-up table was provided to the OpenFOAM RANS solver for the Reynolds stress terms that appear in both momentum and production in the $k-\varepsilon$ equations. This dataset serves as the performance of an ideal model that exactly captures the behavior of the Reynolds stresses while highlighting the model errors associated with the $k-\varepsilon$ model equations themselves.  

Two learned models are discovered using sparse regression, one with three terms ($\lambda = 0$, denoted Learned 1) and the second with two terms ($\lambda = 15$, denoted Learned 2). Both learned models take the form, 
 \begin{equation}
b_{ij}^{\perp} = \left \lbrack \frac{1}{1000 + \lambda_1^3} \right \rbrack \left(C_1 \mathcal{T}^{(1)} + C_2 \mathcal{T}^{(2)} + C_3 \mathcal{T}^{(3)} \right), 
\end{equation} 
and are detailed in Table~\ref{tab:coeff}. 

The \emph{a posteriori} analysis for the learned models includes an assessment of recirculation (Table~\ref{tab:Bubble}) and velocity predictions for the training case (Fig.~\ref{Fig:Posteriori}) and a test case at a higher Reynolds number (Fig.~\ref{Fig:VelProfile5600}). These results are discussed in detail in Sec.~\ref{Sec:posteriori}. 

\subsection{A priori analysis} 
Each model developed can be assessed using the data with which it was trained. This represents an \emph{a priori} assessment of the model, but has limitations as it does not take into account issues of stability or sensitivity that may be encountered within the context of a RANS solver. Further, since the forward solution is not computed here, all assessments of model accuracy are computed with respect to the anisotropic stress tensor. 

\begin{table}
\begin{tabular}{c @{\hskip 0.2in} c c c @{\hskip 0.2in} c c} 
\hline 
\hline
Model & $C_1$ & $C_2$ & $C_3$ & $\epsilon^{\bm{b}}$ & RMSE \\
\hline  
LEVM & - & - & - & 1.02 & 0.16 \\
Learned 1 & 63.12 & 51.42 & 10.98 & 0.64 & 0.10\\ [-1.75ex]
Learned 2 & 63.14 & 51.42 & 0 & 0.64 & 0.10 \\
Neural Network \cite{ML_2016Ling} & - &- &- & - & 0.08 \\ 
\hline
\hline
\end{tabular}
\caption{Summary of learned model coefficients and \emph{a priori} errors compared with the standard LEVM and an exemplary NN model developed in \citet{ML_2016Ling} for a periodically constricted channel.} 
\label{tab:coeff}
\end{table}

\begin{figure}
\centering
 \begin{tabular}{c c c c c} 
       & $b_{11}$ & $b_{12}$ & $b_{22}$ & $b_{33}$ \\
        &\includegraphics[height = 0.03\textwidth]{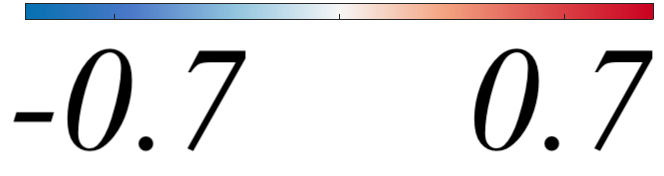}&
        \includegraphics[height = 0.03\textwidth]{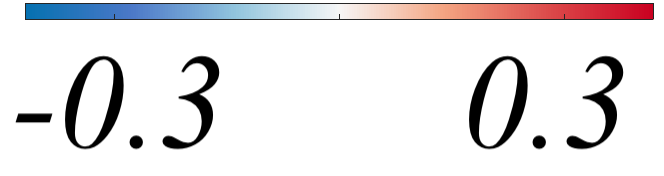}&
        \includegraphics[height = 0.03\textwidth]{legend1}&
        \includegraphics[height = 0.03\textwidth]{legend1}\\
       \includegraphics[height = 0.07\textwidth]{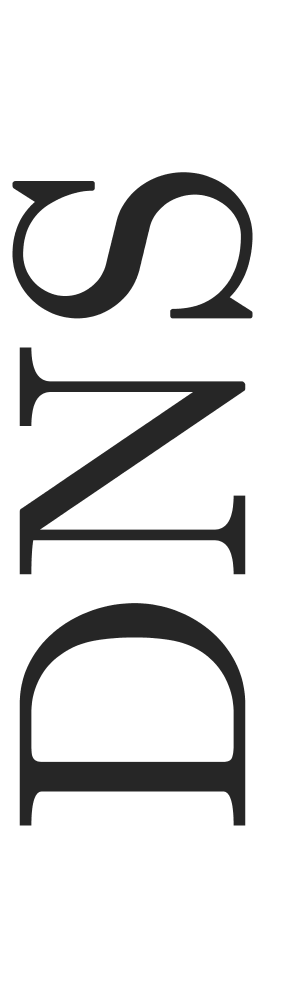} &
        \includegraphics[height = 0.08\textwidth]{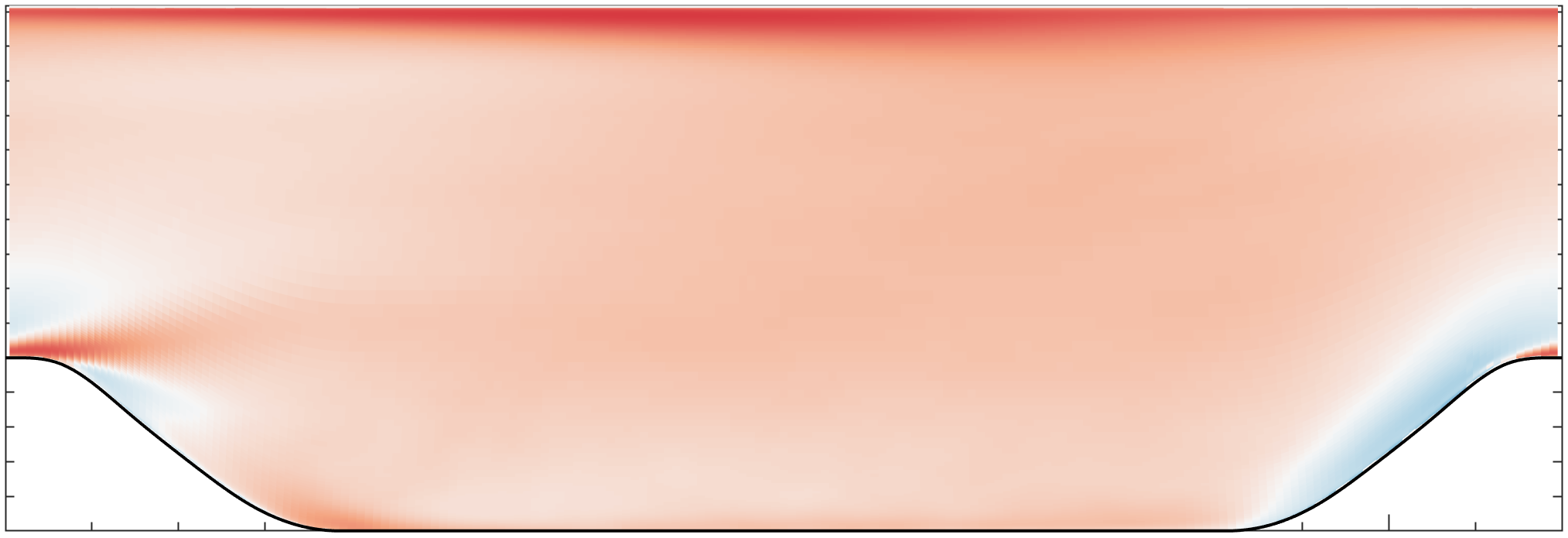} &
        \includegraphics[height = 0.08\textwidth]{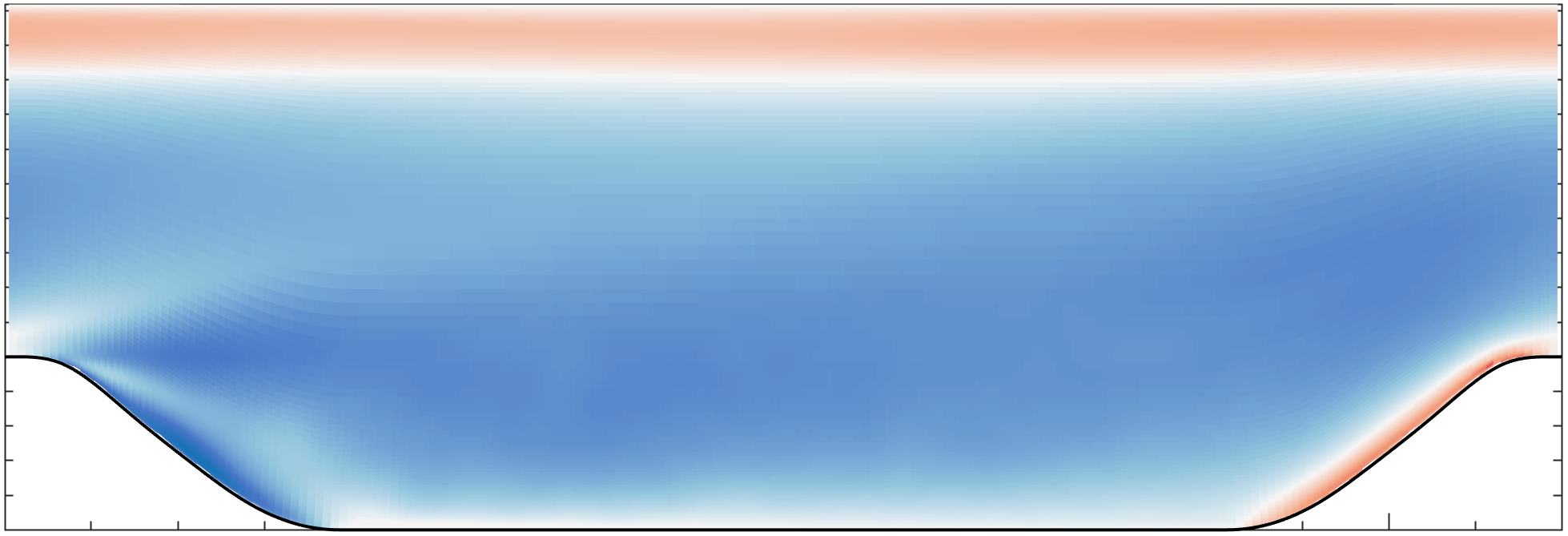} &
         \includegraphics[height = 0.08\textwidth]{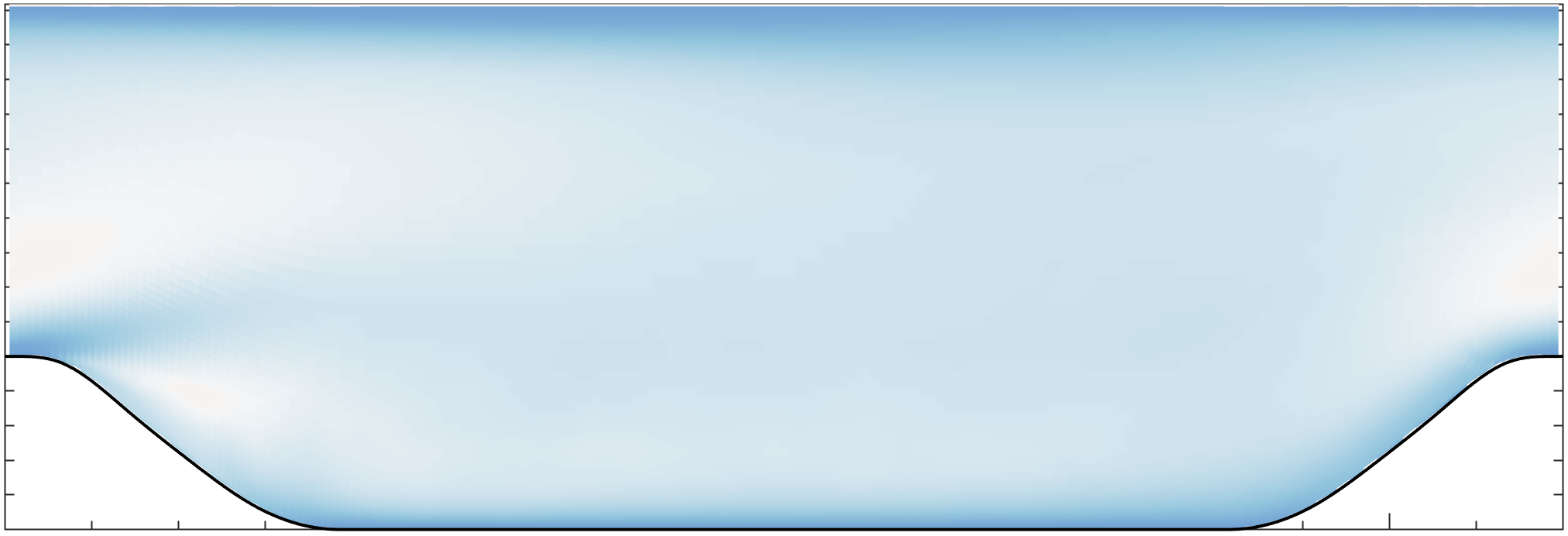} &
          \includegraphics[height = 0.08\textwidth]{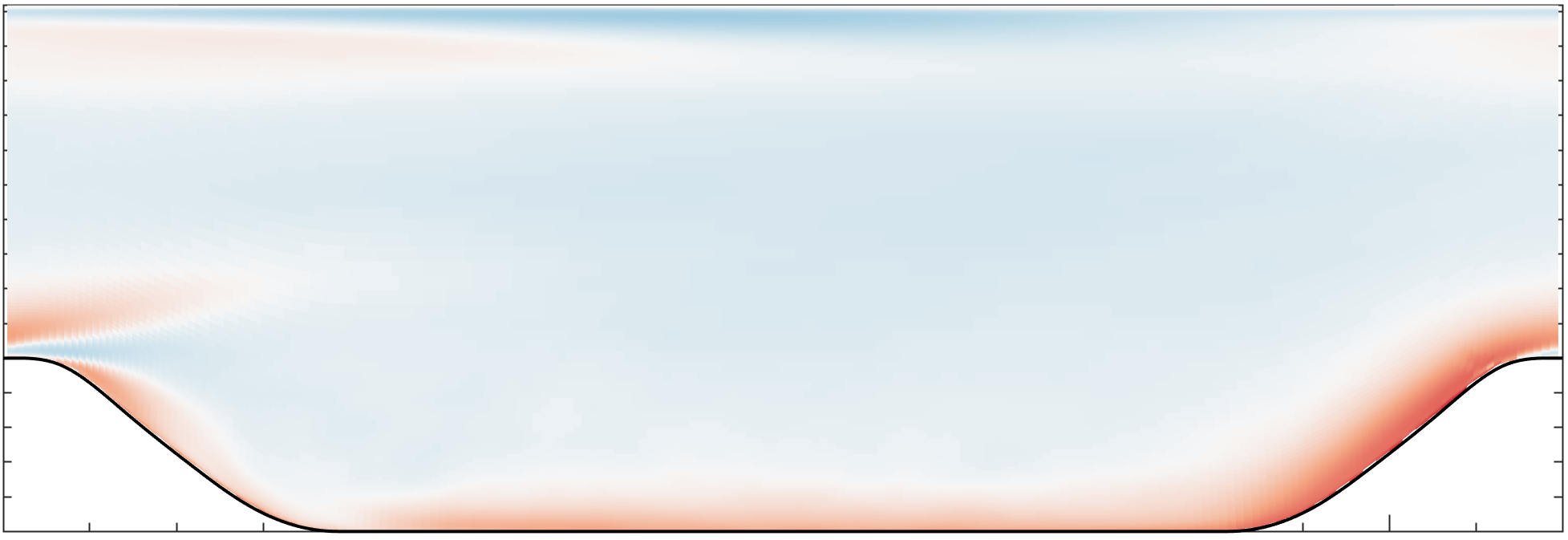} \\[0.5 ex]
           \includegraphics[height = 0.07\textwidth]{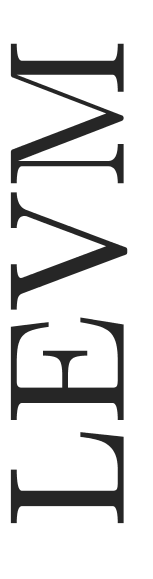} &
        \includegraphics[height = 0.08\textwidth]{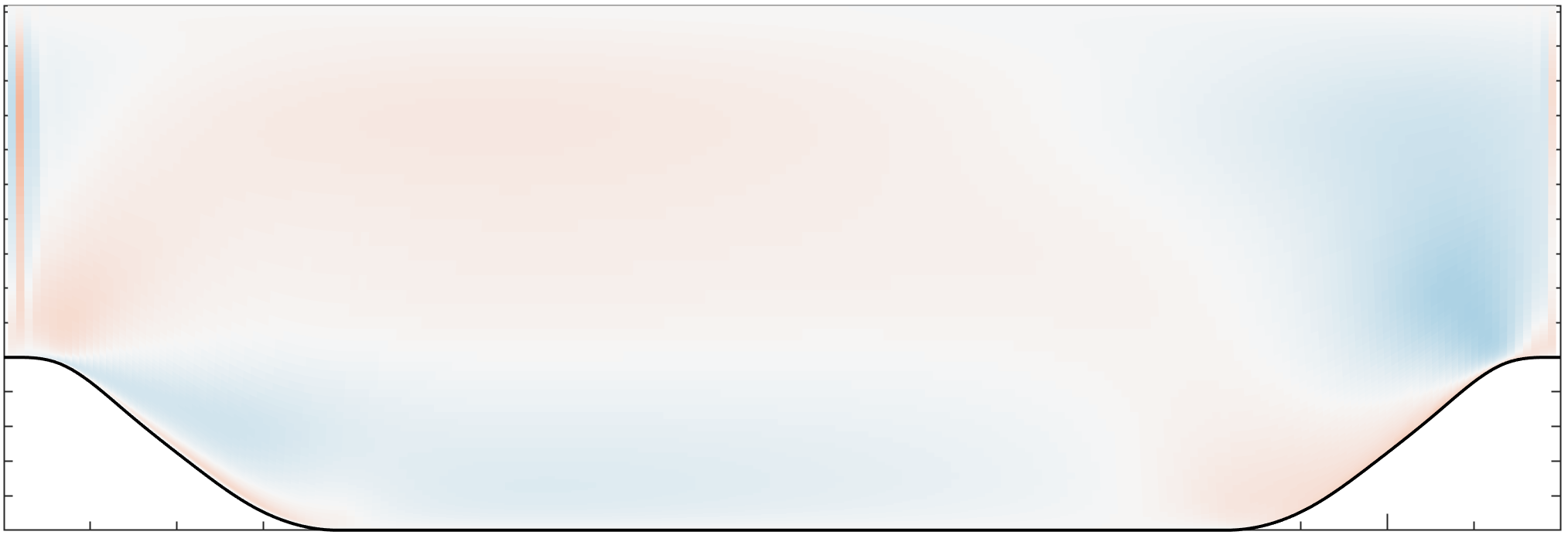} &
        \includegraphics[height = 0.08\textwidth]{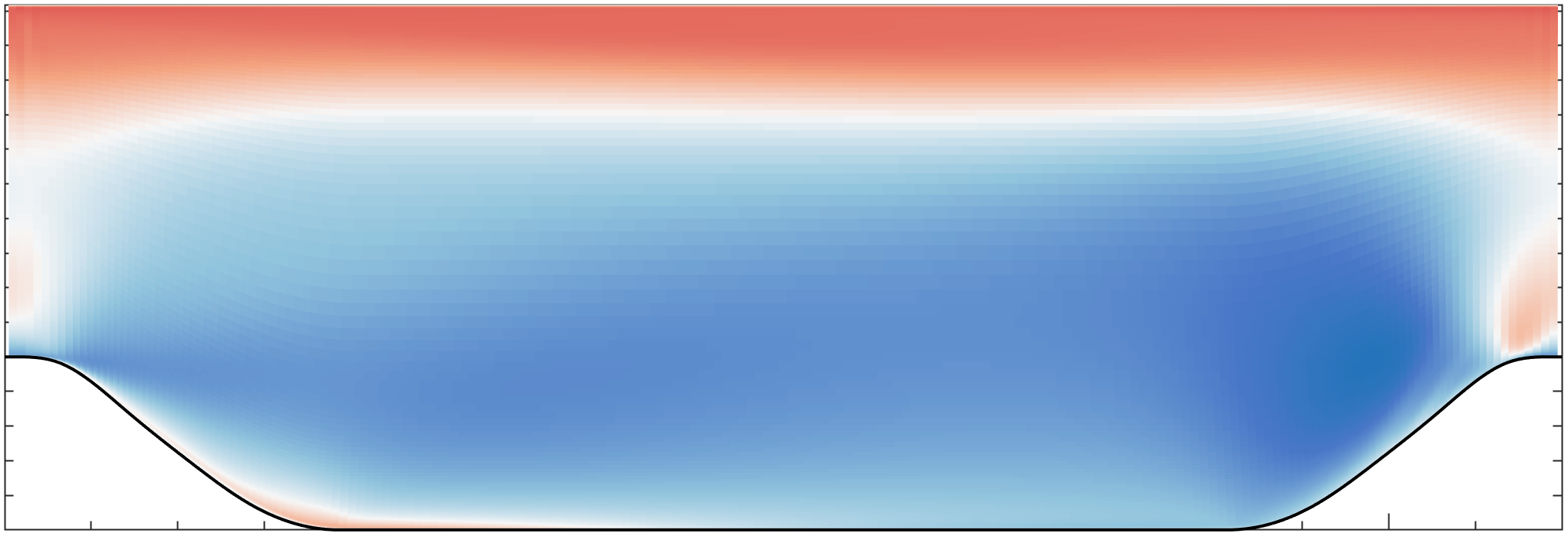} &
        \includegraphics[height = 0.08\textwidth]{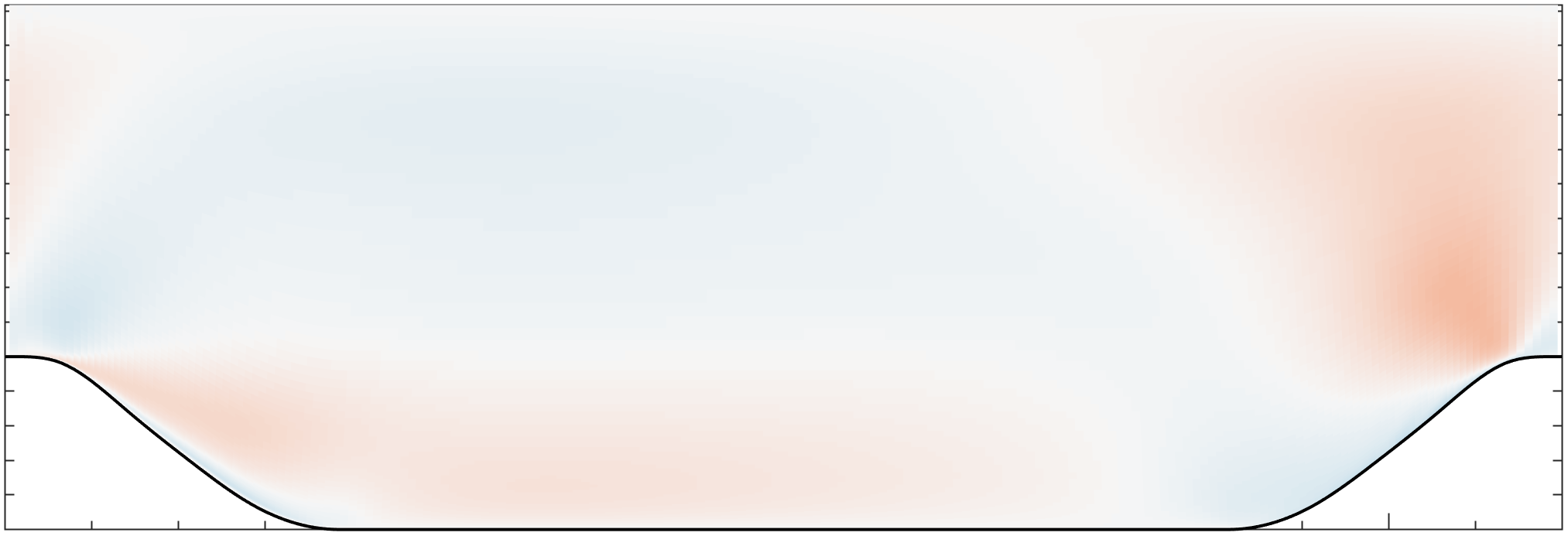} &
        \includegraphics[height = 0.08\textwidth]{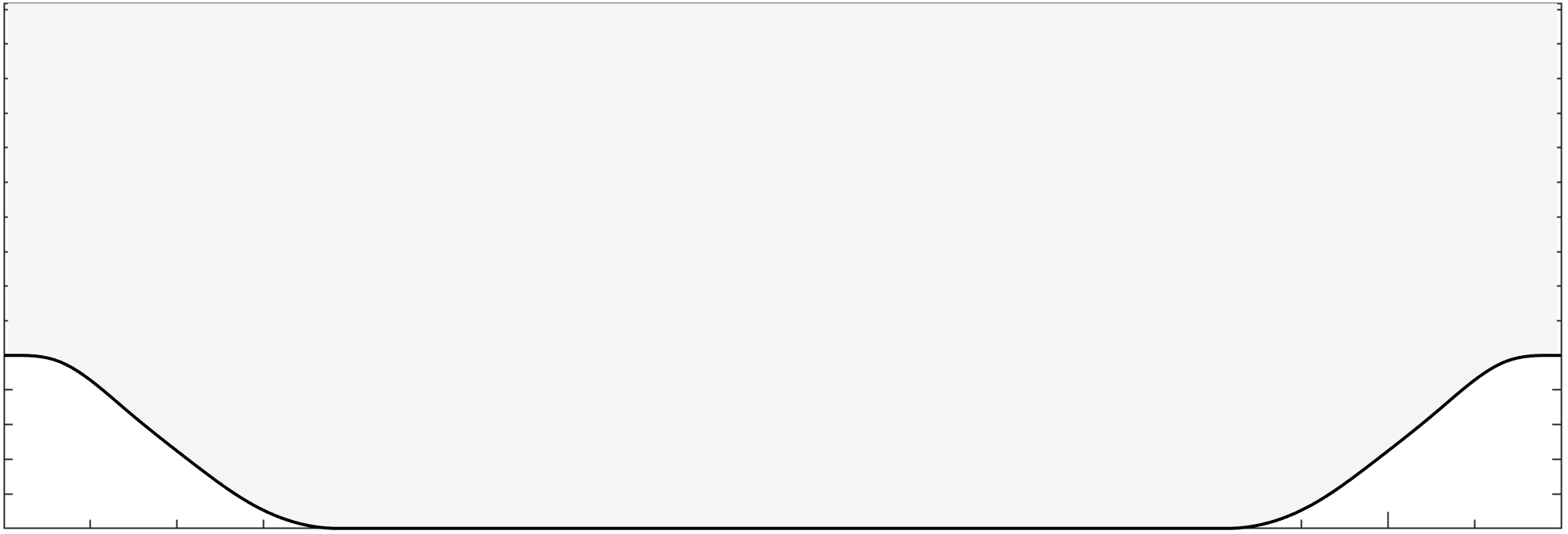} \\  
         \includegraphics[height = 0.09\textwidth]{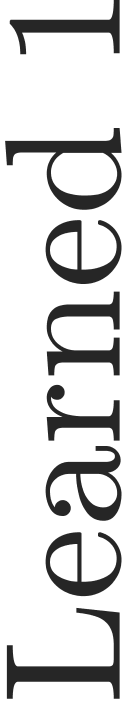} &
        \includegraphics[height = 0.08\textwidth]{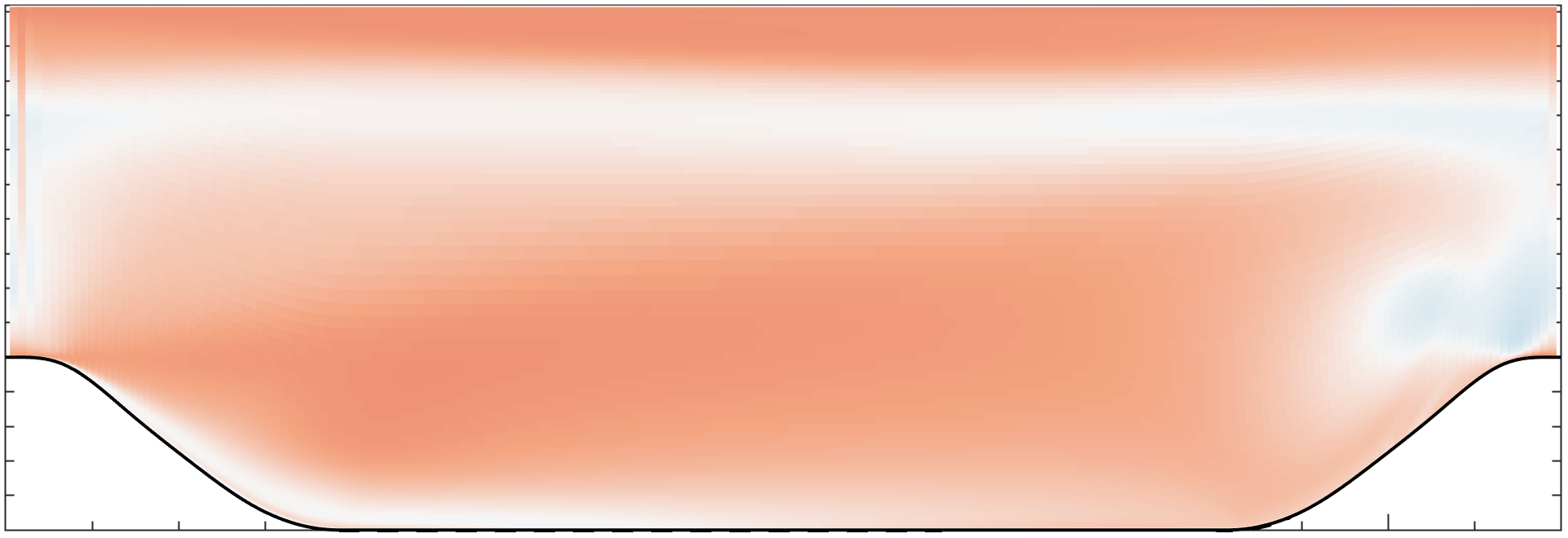} &
        \includegraphics[height = 0.08\textwidth]{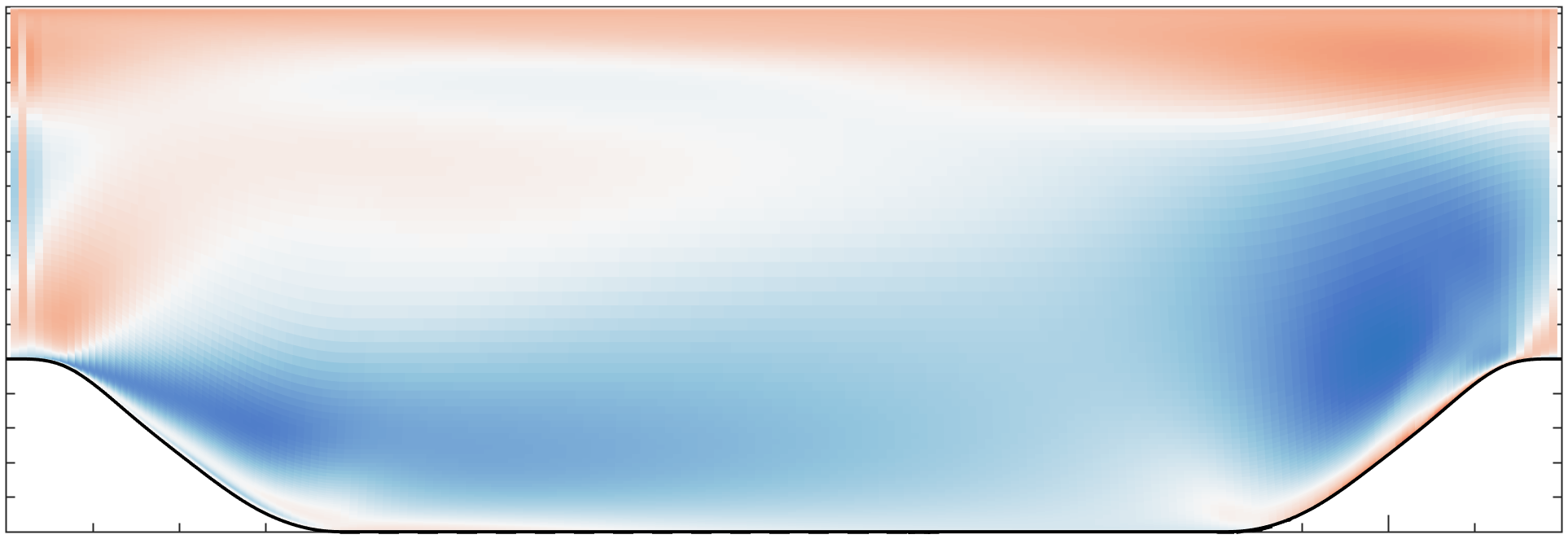} &
        \includegraphics[height = 0.08\textwidth]{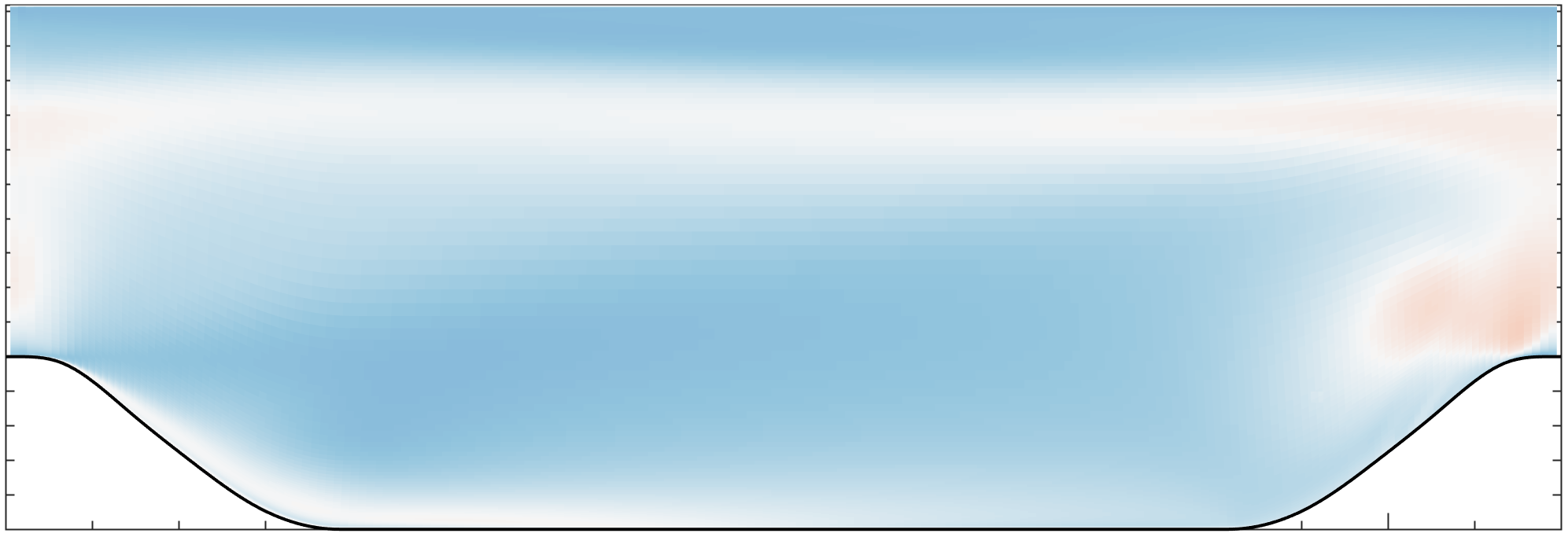} &
        \includegraphics[height = 0.08\textwidth]{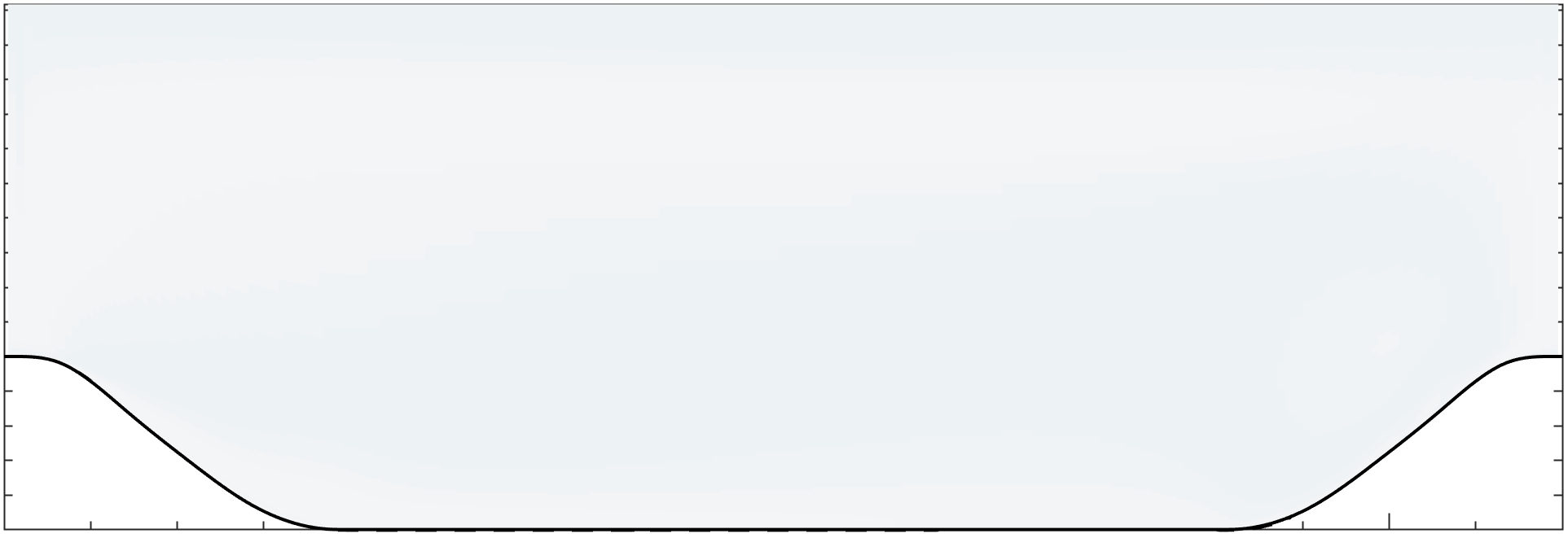} \\ 
        \includegraphics[height = 0.09\textwidth]{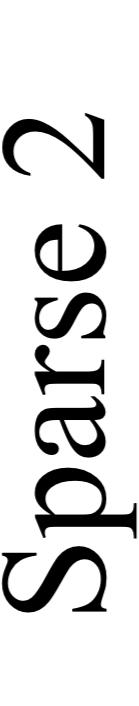} &
        \includegraphics[height = 0.08\textwidth]{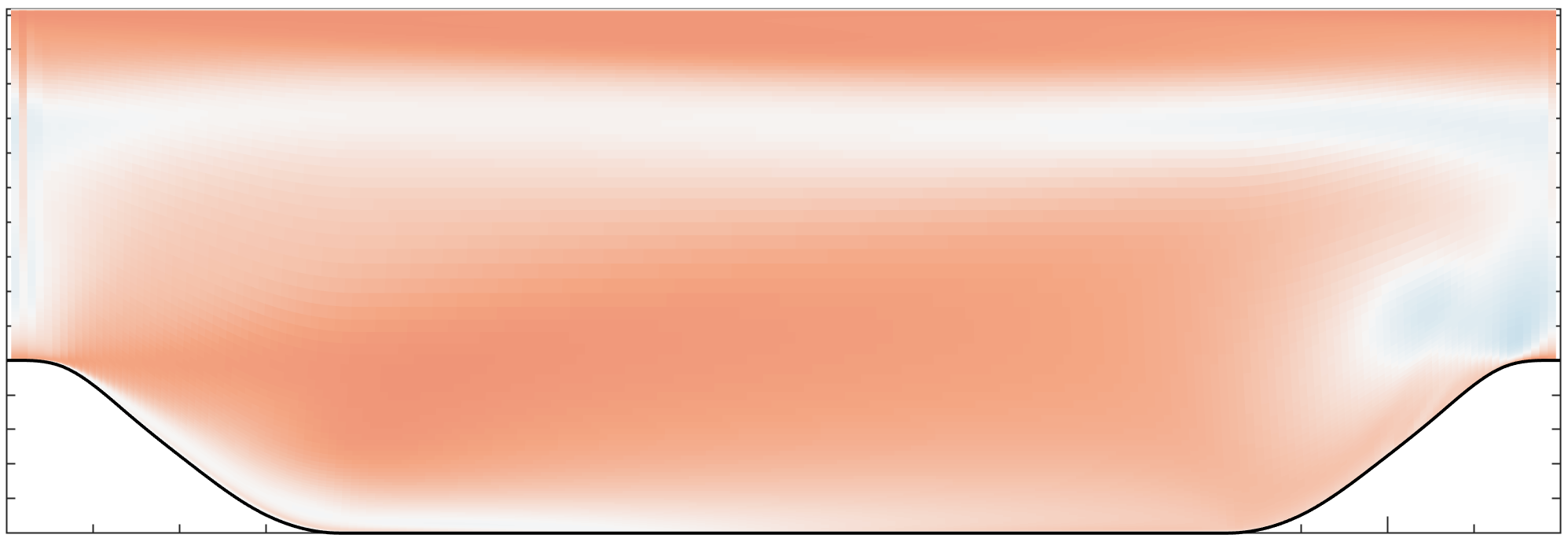} &
        \includegraphics[height = 0.08\textwidth]{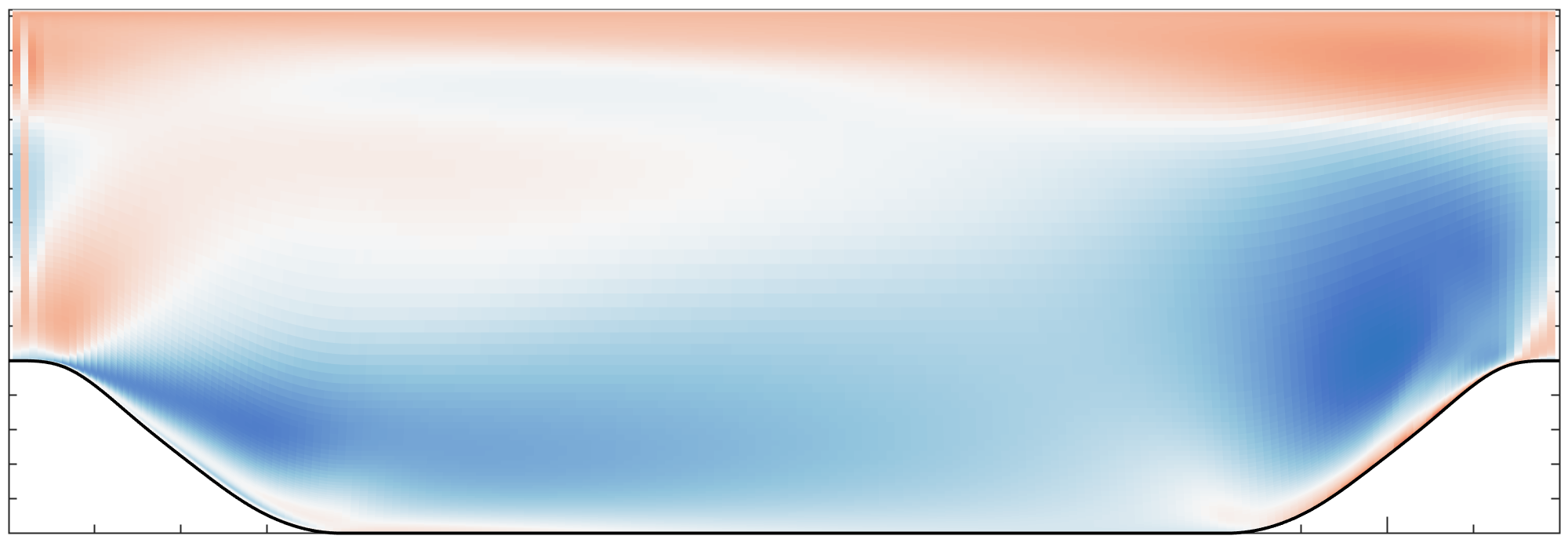} &
        \includegraphics[height = 0.08\textwidth]{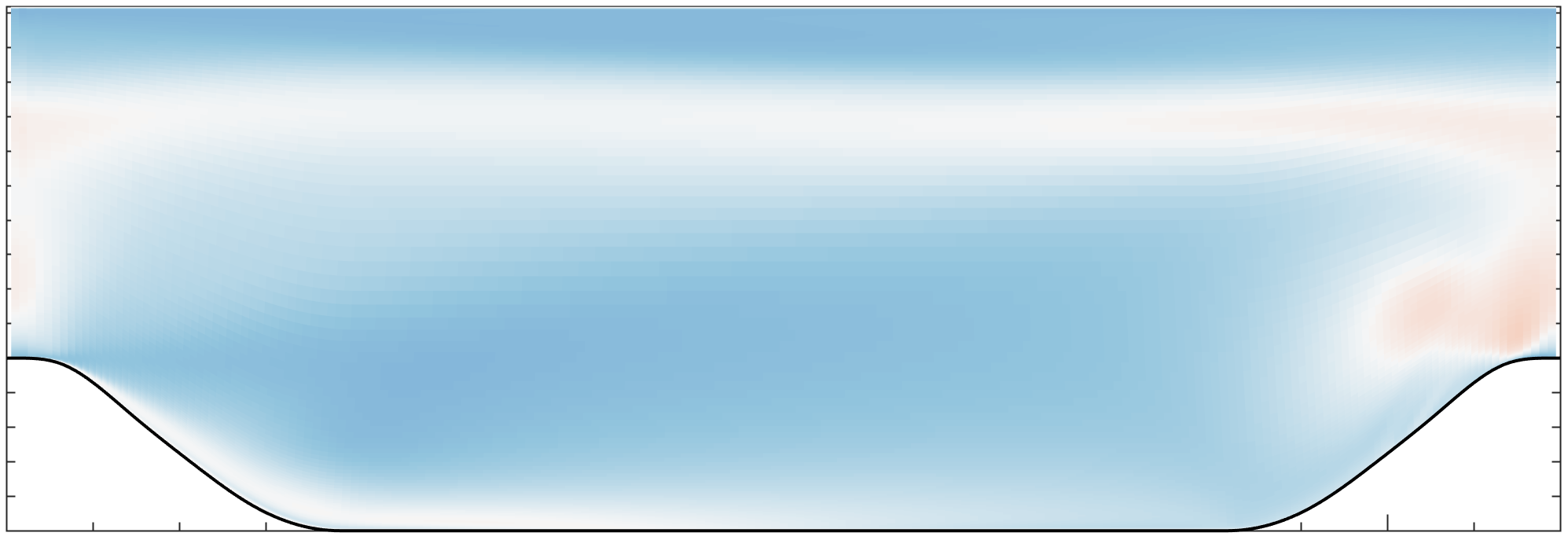} &
        \includegraphics[height = 0.08\textwidth]{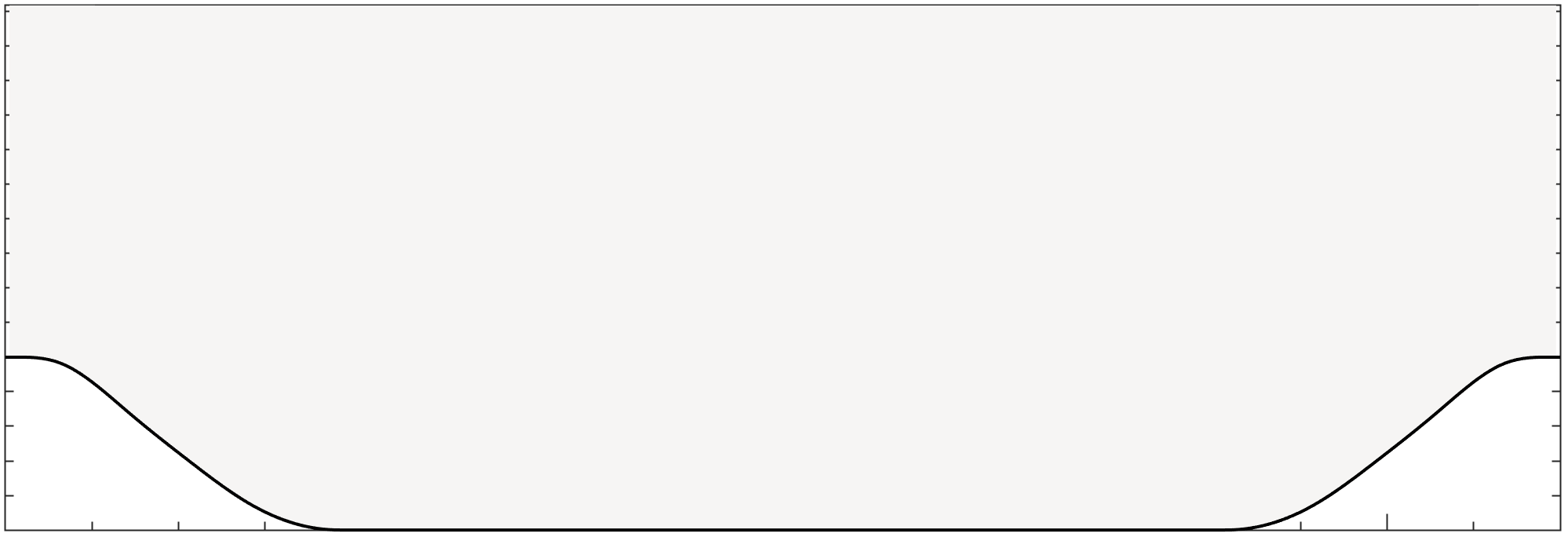} \\[-2ex]
        &\includegraphics[height = 0.023\textwidth]{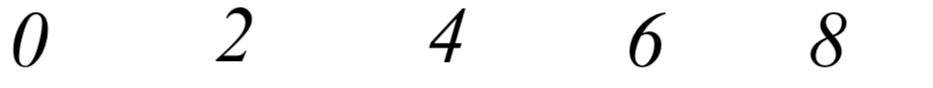} &
        \includegraphics[height = 0.023\textwidth]{axis} &
        \includegraphics[height = 0.023\textwidth]{axis} &
        \includegraphics[height = 0.023\textwidth]{axis} \\[-2ex]
        &$x/h$ & $x/h$ & $x/h$ & $x/h$ 
         \end{tabular}
        \caption{Comparisons of the components of the Reynolds stress tensor computed using DNS, LEVM, and the two learned models.}
\label{Fig:Apriori1}
       \end{figure}

Shown in Fig.~\ref{Fig:Apriori1}, the standard LEVM does a reasonable job predicting the $b_{12}$ component of the anisotropy tensor, but it struggles for the diagonal components. In all three cases, both the sign and magnitude are incorrect. The learned models, in contrast, capture the correct sign for the diagonal components and improve the magnitude inaccuracies present in the standard LEVM for the $b_{12}$ component. However, for Learned 2, with the elimination of the third basis term, the prediction for $b_{33}$ is also lost.

Using the \textit{L}-2 norm as a metric for error, the learned models reduce model error in the anisotropic stress tensor by 41\% with respect to LEVM. The root-mean-square-error (RMSE) of the learned models is 0.10, which is comparable to the performance reported by \citet{ML_2016Ling}, who used a NN to model the anisotropic stress tensor. However, unlike models determined using NNs, sparse regression returns an algebraic closure, which sheds light on which terms are most important to capturing critical flow features and reducing error.   

\subsection{A posteriori analysis \label{Sec:posteriori}}

The true test of any model is its performance in the context of a forward solver. It is in this sense that model shortcomings become apparent, e.g., sensitivity or stability issues. Further, while the aim of Reynolds stress modeling is to improve accuracy in describing the stresses, the ultimate goal is that these models will improve predictions in the velocity field.  

\begin{figure}[h!]
      \centering
               \subcaptionbox{DNS \label{fig:Velpost_DNS}}
        {\includegraphics[width = 0.47\textwidth]{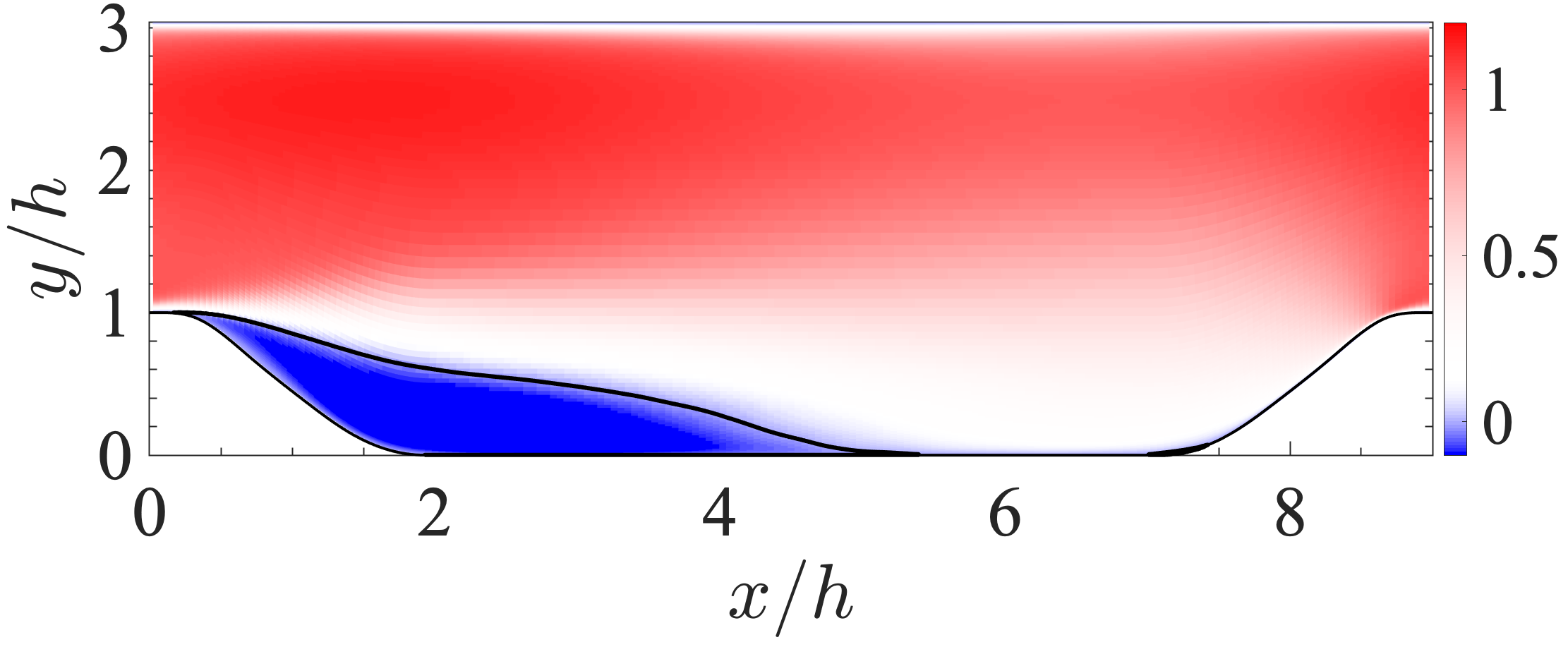}} \\
      \subcaptionbox{LEVM \label{fig:Velpost_LEVM}}
        {\includegraphics[width = 0.47\textwidth]{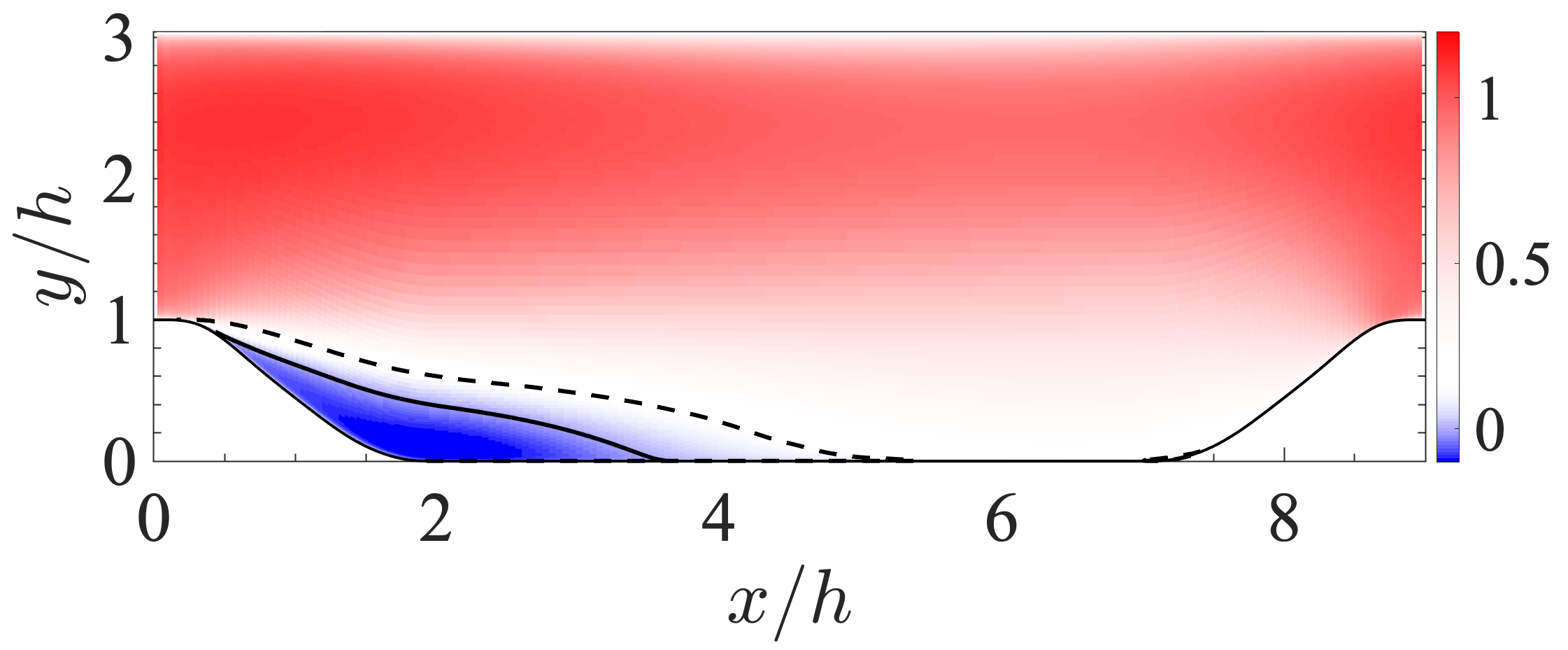}}
          \subcaptionbox{Lookup table for $b_{ij}$ \label{fig:Velpost_Lookup}}
        {\includegraphics[width = 0.47\textwidth]{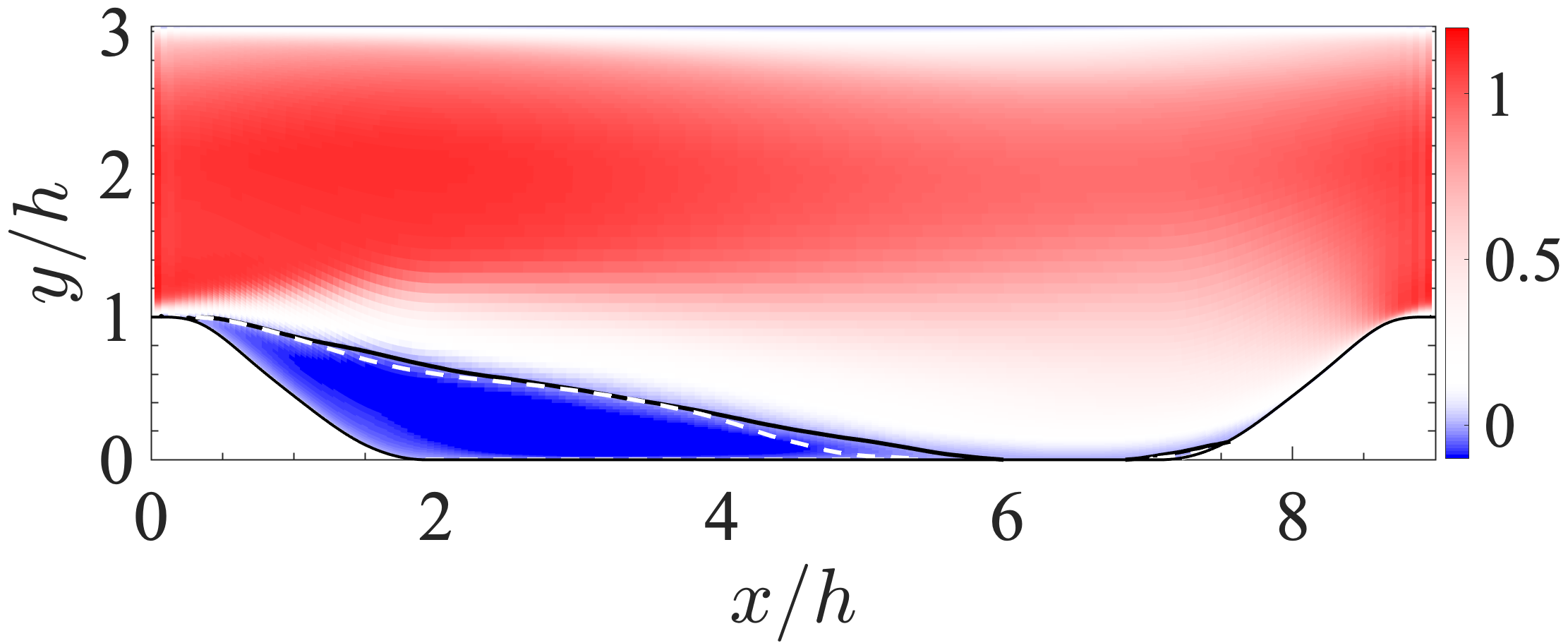}} \\
         \subcaptionbox{Learned 1\label{fig:Velpost_LearnedOne}}
        {\includegraphics[width = 0.47\textwidth]{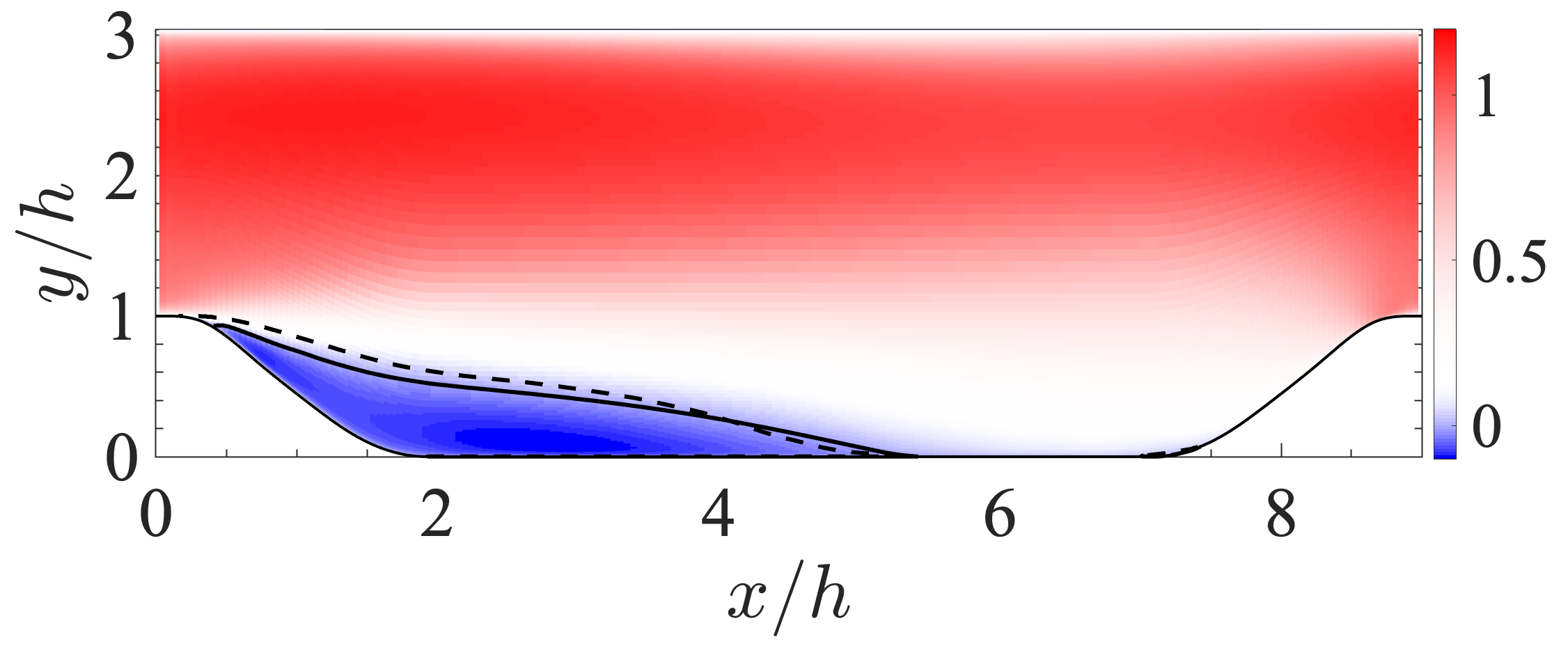}}
       \subcaptionbox{Learned 2 \label{fig:Velpost_SparseTwo}}
        {\includegraphics[width = 0.47\textwidth]{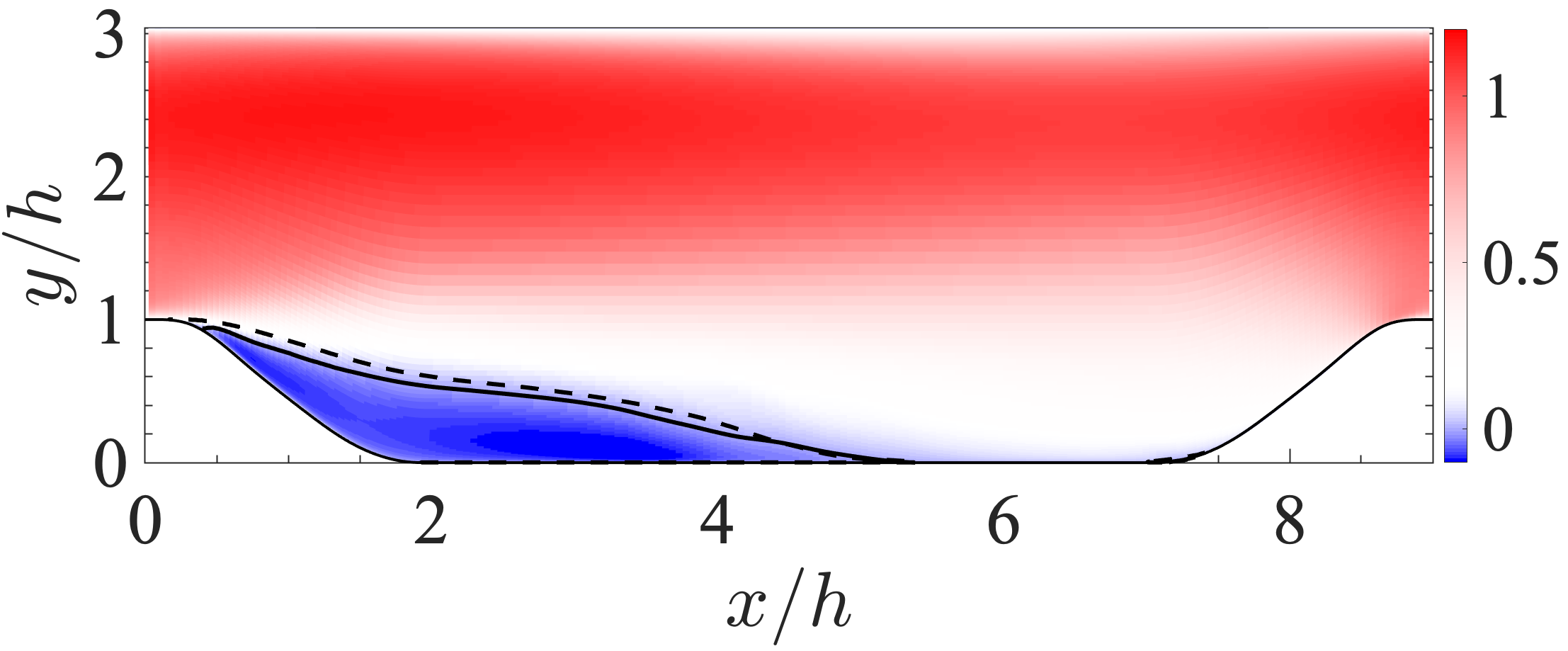}}

             \caption{Forward solutions of the mean, normalized velocity, $\langle u\rangle/u_{\text{bulk}}$, for the standard LEVM model, the two learned models, the lookup table for DNS values of $b_{ij}$ and the DNS results. The solid line represents the region of recirculation and the dashed line overlays where this region exists in the DNS data.} \label{Fig:Posteriori}
\end{figure}

In order to assess the improvement of the learned model over the LEVM, the learned models were integrated into OpenFOAM, solved in conjunction with the $k-\varepsilon$ equations, and compared with the LEVM model and the $k-\varepsilon$ equations with a look-up table containing the DNS values for the anisotropy tensor. In each case, the RANS equations were solved on a two-dimensional grid of resolution $(N_x, N_y) = (200,160)$ with the same physical dimensions as described in \citet{Breuer2009} (and used for the DNS computations). Periodic conditions were imposed at the left and right faces, and `patch' conditions were imposed on the front and back faces to enforce a 2-D solution. The bottom and top walls were treated as no slip and a forcing term was added such that the velocity at the top of the hill crest enforced the desired Reynolds number. 

\begin{table}
\begin{tabular}{c c @{\hskip 0.2in} c c @{\hskip 0.2in} c c } 
\hline
\hline
\multirow{2}{*}{Reynolds number} & \multirow{2}{*}{Model} & \multicolumn{2}{c}{Primary} & \multicolumn{2}{c}{Secondary} \\[-2ex]
 & & Separation & Reattachment  & Separation & Reattachment \\
\hline
 & LEVM & 0.43 (1.62) & 3.64 (0.32) &-- &-- \\[-1ex]
Re $=2800$ &Learned 1 & 0.40 (1.53) & 5.38 (0.005) & 7.14 (0.04) & 7.31 (0.008) \\ [-2ex]
&Learned 2 & 0.40 (1.48) & 5.38 (0.005) & 7.14 (0.04) & 7.31 (0.01) \\[-1ex]
&DNS & 0.16 (--)& 5.35 (--)& 6.87 (--)& 7.25 (--)\\[2 ex]
 & LEVM & -- & -- & -- & -- \\ [-1 ex]
{Re = 5600} & Learned 1 & 0.40 & 5.38 & 7.14 & 7.30  \\ [-1 ex]
 & DNS \cite{Breuer2009} & 0.18 & 5.14 & -- & -- \\ [-2ex]
 & DNS \cite{Krank2018} & 0.17 & 5.04$\pm$0.09 & 7.04 & 7.31 \\ 
\hline 
\hline
\end{tabular}
\caption{Summary of separation and reattachment locations for all models compared with DNS. Relative error with respect to the DNS values are shown in parentheses.} 
\label{tab:Bubble} 
\end{table}

The mean velocity normalized by the bulk velocity, $u_{\text{bulk}}$, is shown in Fig.~\ref{Fig:Posteriori} and the detached regions are delineated by a black line. It is observed that LEVM under predicts recirculation compared with the DNS results (Fig.~\ref{fig:Velpost_LEVM}), while both learned models demonstrate marked qualitative improvement in velocity prediction. Quantitative measurements of separation and reattachment locations for both the primary and secondary recirculation regions are detailed in Table \ref{tab:Bubble}. The learned models predict both primary and secondary reattachment points within 1\% of the DNS values, with exception of the primary separation point. In comparison, LEVM under predicts the primary reattachment point by 32\% compared with DNS and fails to predict existence of the secondary recirculation. 

\begin{figure}
\centering
 \begin{tabular}{c c c c c} 
       & $b_{11}$ & $b_{12}$ & $b_{22}$ &$b_{33}$ \\
       &\includegraphics[height = 0.03\textwidth]{legend1}&
        \includegraphics[height = 0.03\textwidth]{legend2}&
        \includegraphics[height = 0.03\textwidth]{legend1}&
        \includegraphics[height = 0.03\textwidth]{legend1}\\ 
                  \includegraphics[height = 0.07\textwidth]{DNS} &
        \includegraphics[height = 0.08\textwidth]{DNS_b11} &
        \includegraphics[height = 0.08\textwidth]{DNS_b12} &
         \includegraphics[height = 0.08\textwidth]{DNS_b22} &
          \includegraphics[height = 0.08\textwidth]{DNS_b33} \\ [0.5ex]
         \includegraphics[height = 0.09\textwidth]{Model1_Learned} &
        \includegraphics[height = 0.08\textwidth]{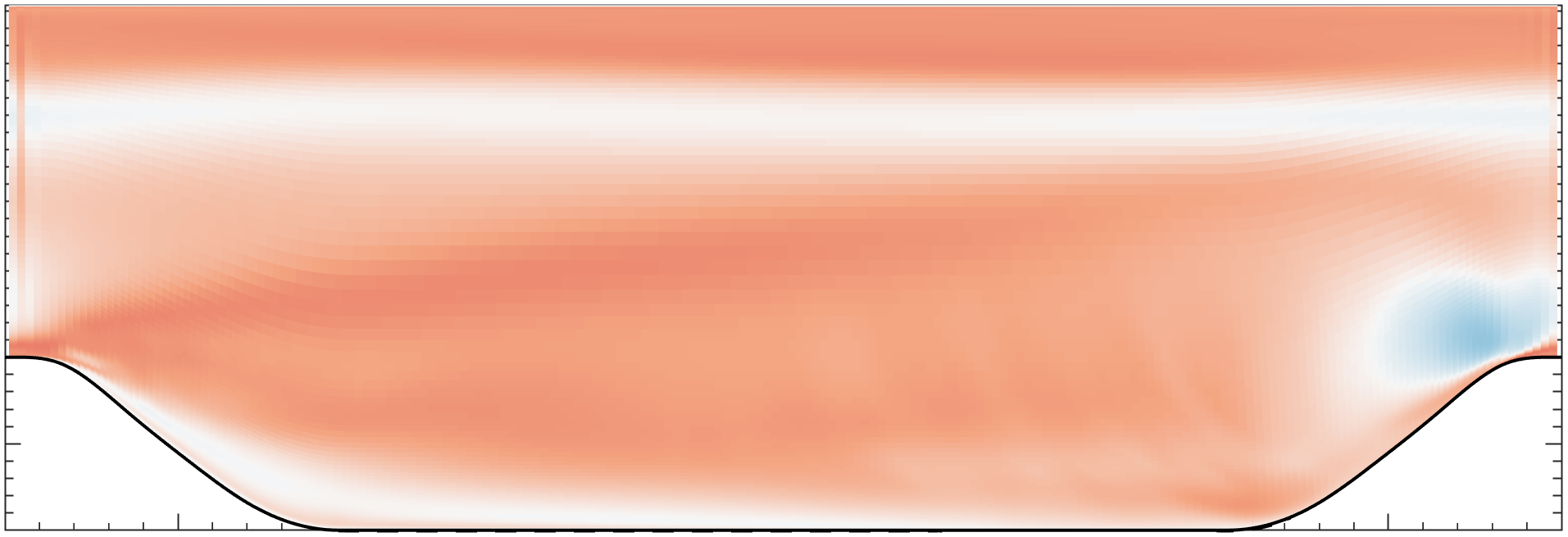} &
        \includegraphics[height = 0.08\textwidth]{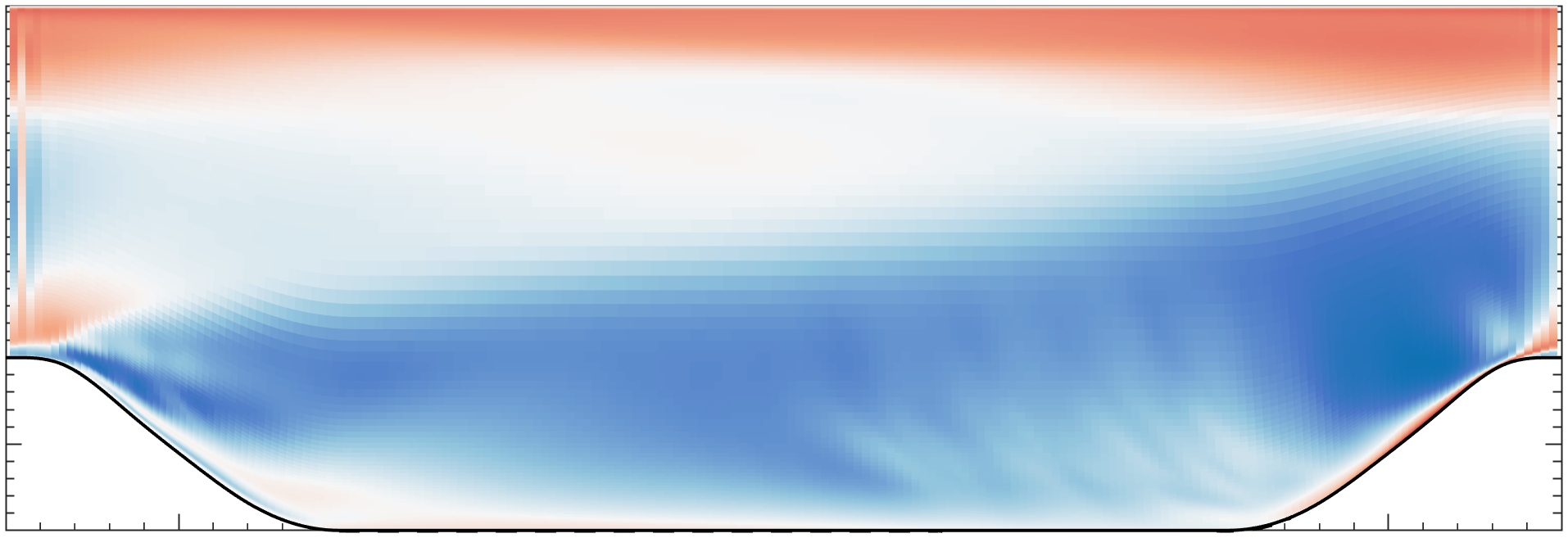} &
        \includegraphics[height = 0.08\textwidth]{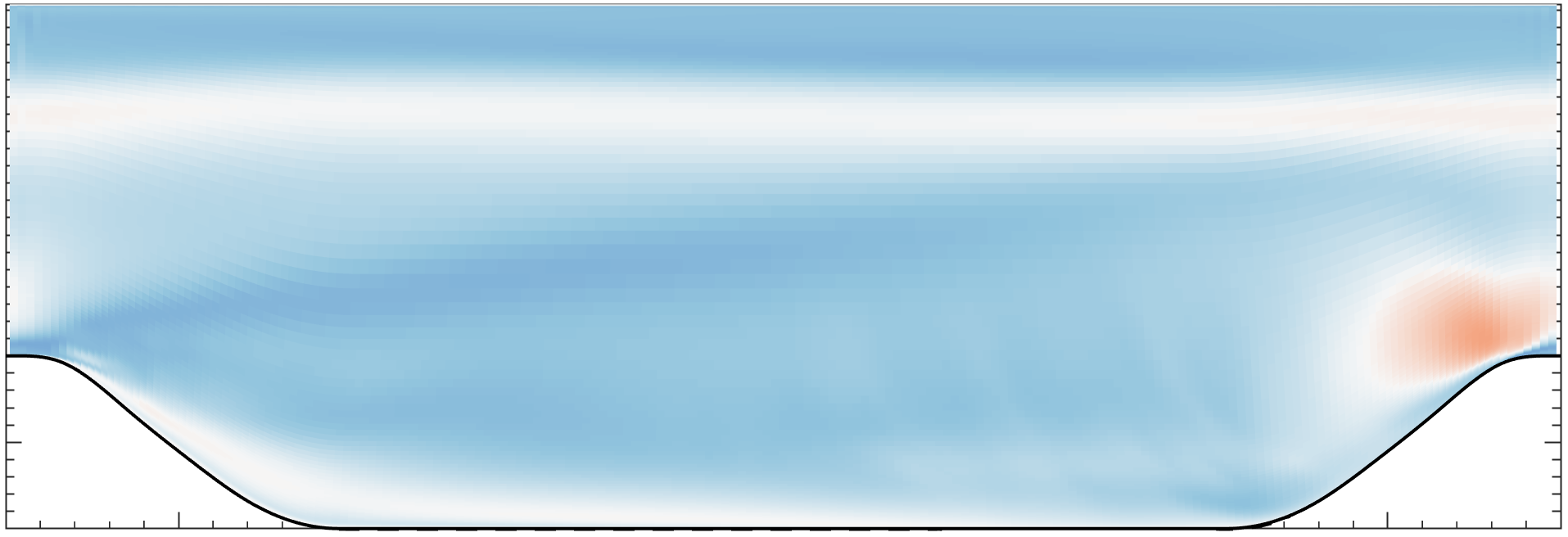} &
        \includegraphics[height = 0.08\textwidth]{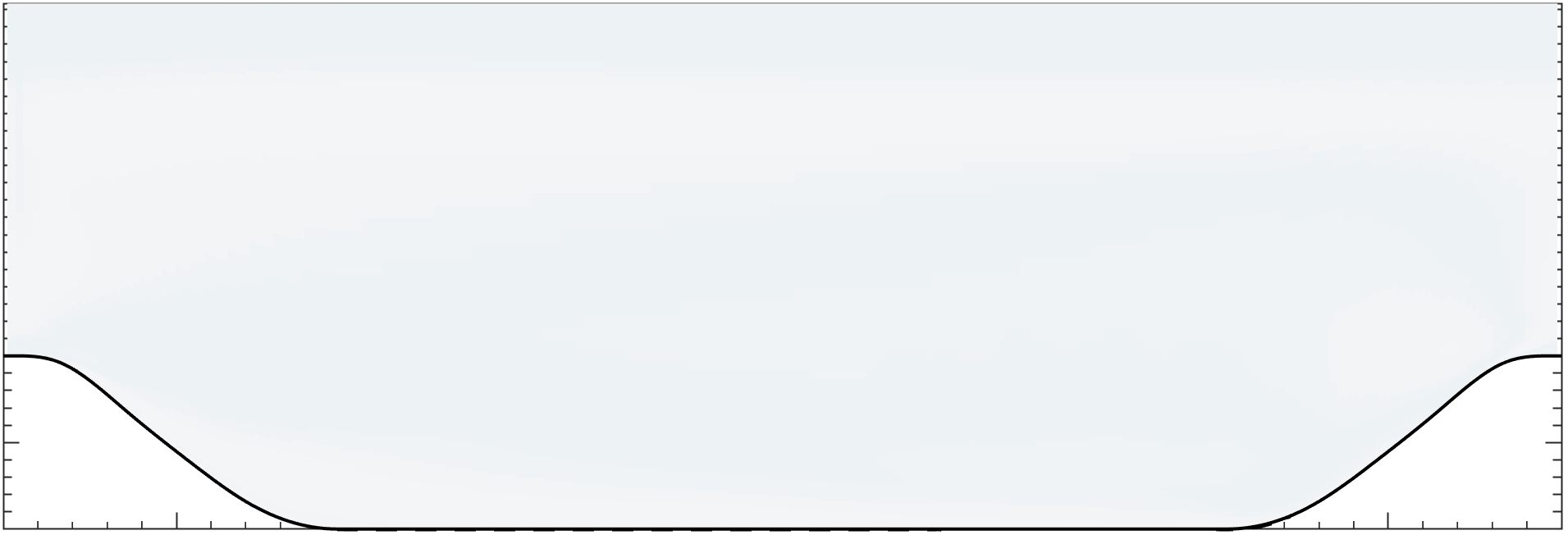} \\[0.5ex]
& $C_i \mathcal{T}^{(i)}_{11}$ & $C_i \mathcal{T}^{(i)}_{12}$ & $C_i \mathcal{T}^{(i)}_{22}$ &$C_i \mathcal{T}^{(i)}_{33}$ \\
       &\includegraphics[height = 0.03\textwidth]{legend1}&
        \includegraphics[height = 0.03\textwidth]{legend2}&
        \includegraphics[height = 0.03\textwidth]{legend1}&
        \includegraphics[height = 0.03\textwidth]{legend1}\\ 
         \includegraphics[height = 0.09\textwidth]{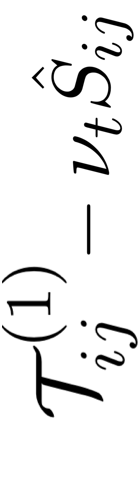} &
        \includegraphics[height = 0.08\textwidth]{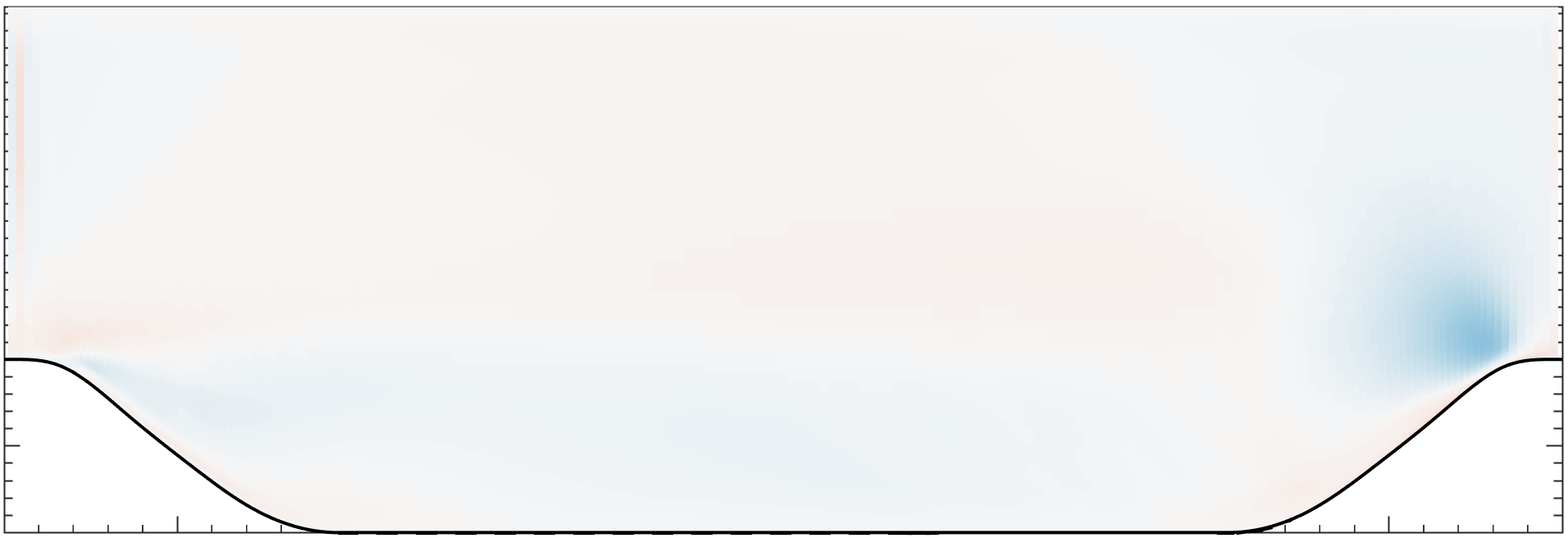} &
        \includegraphics[height = 0.08\textwidth]{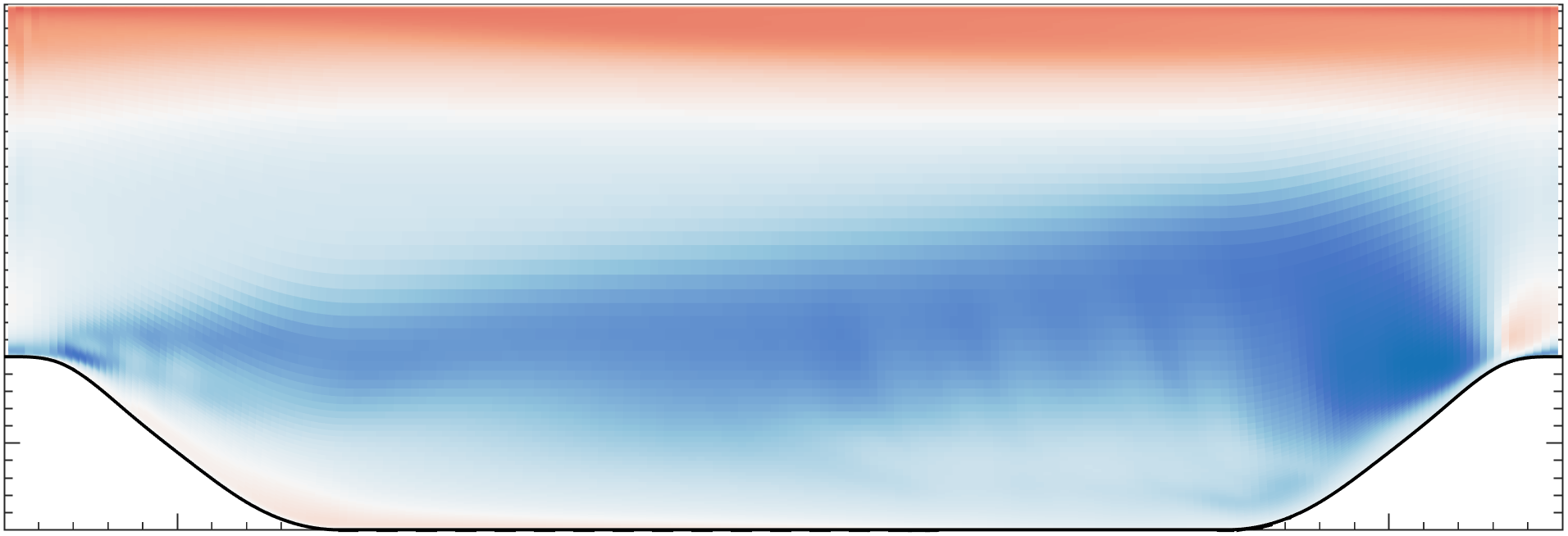} &
        \includegraphics[height = 0.08\textwidth]{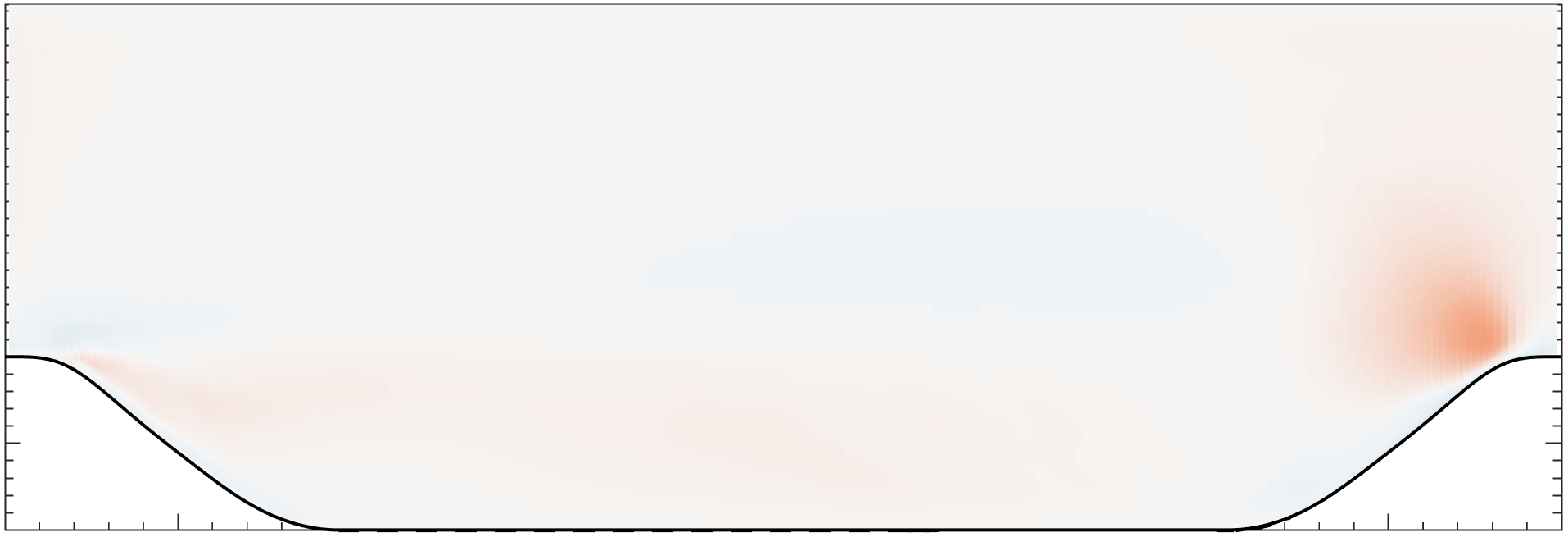} &
        \includegraphics[height = 0.08\textwidth]{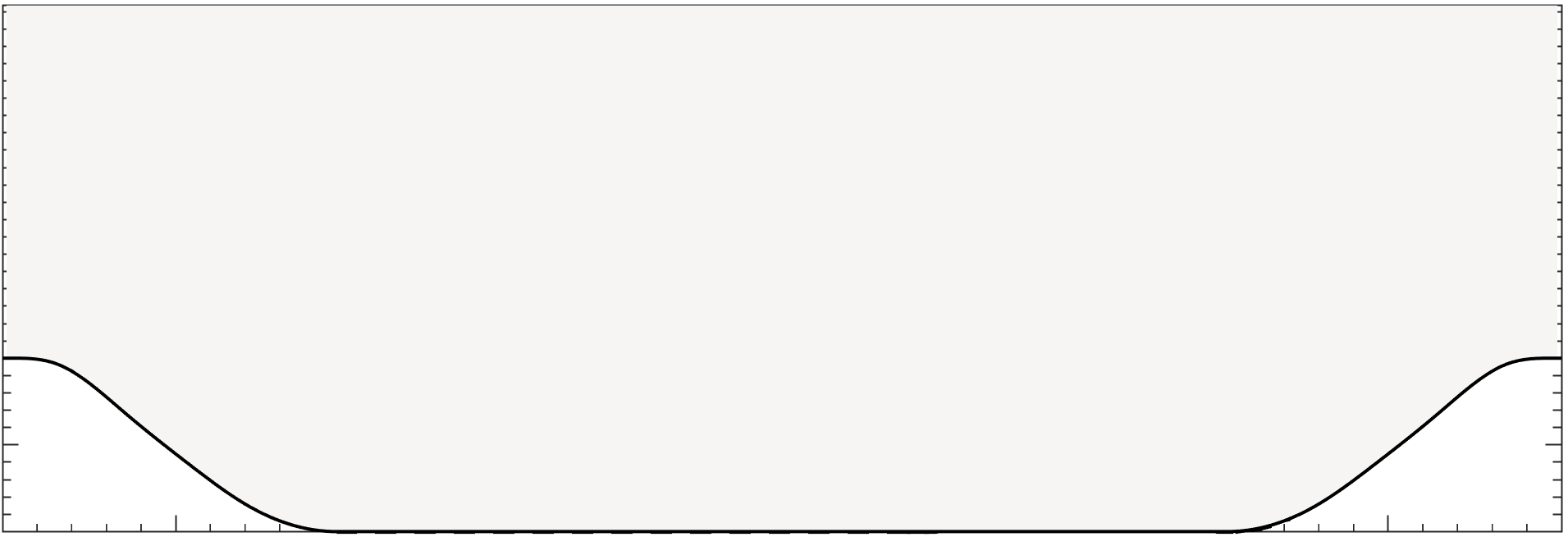} \\[0.5ex]
         \includegraphics[height = 0.09\textwidth]{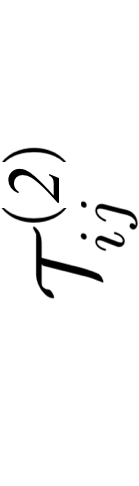} &
        \includegraphics[height = 0.08\textwidth]{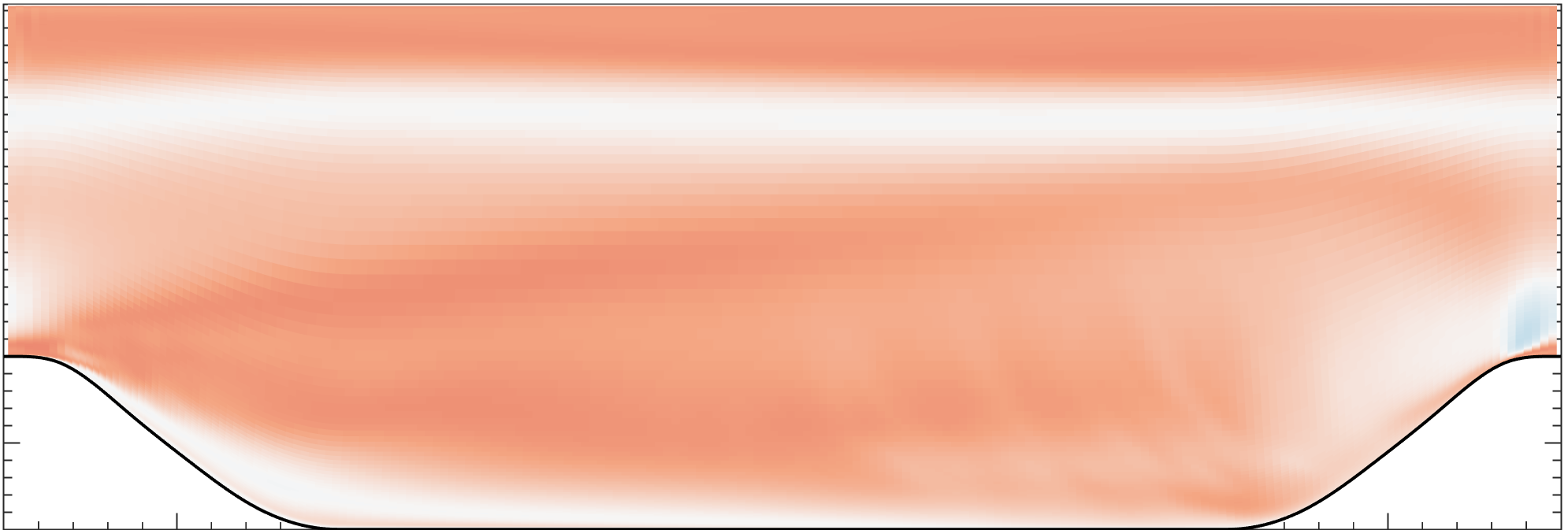} &
        \includegraphics[height = 0.08\textwidth]{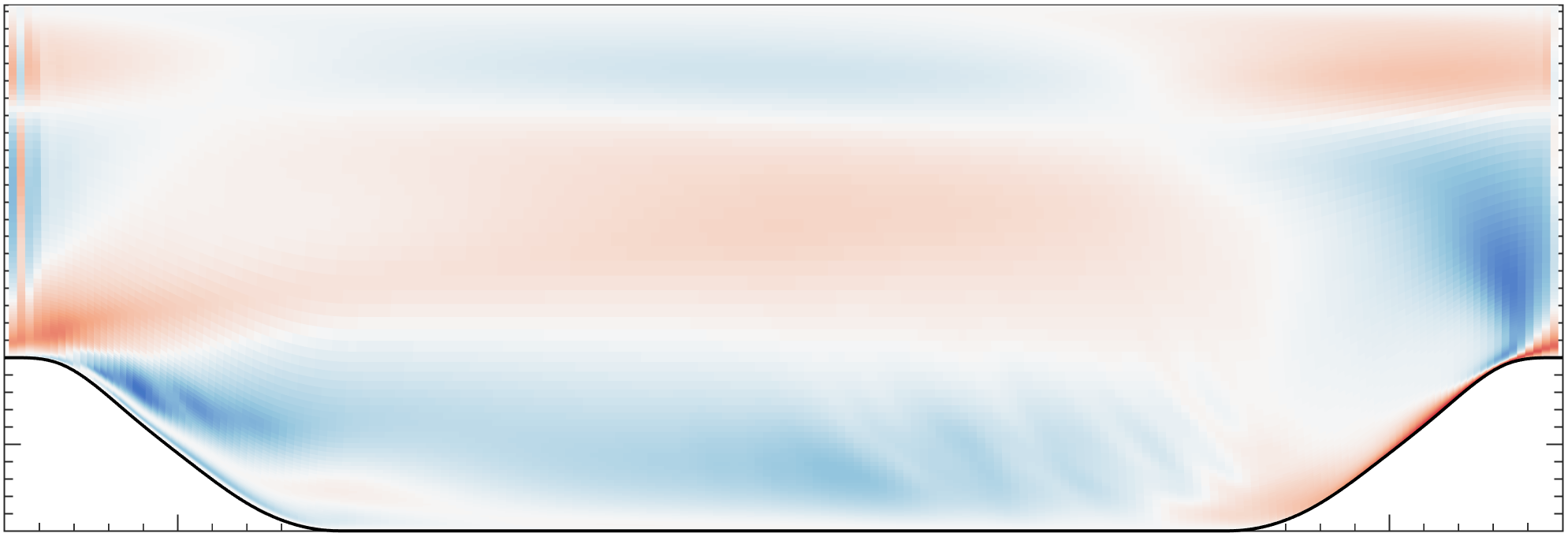} &
        \includegraphics[height = 0.08\textwidth]{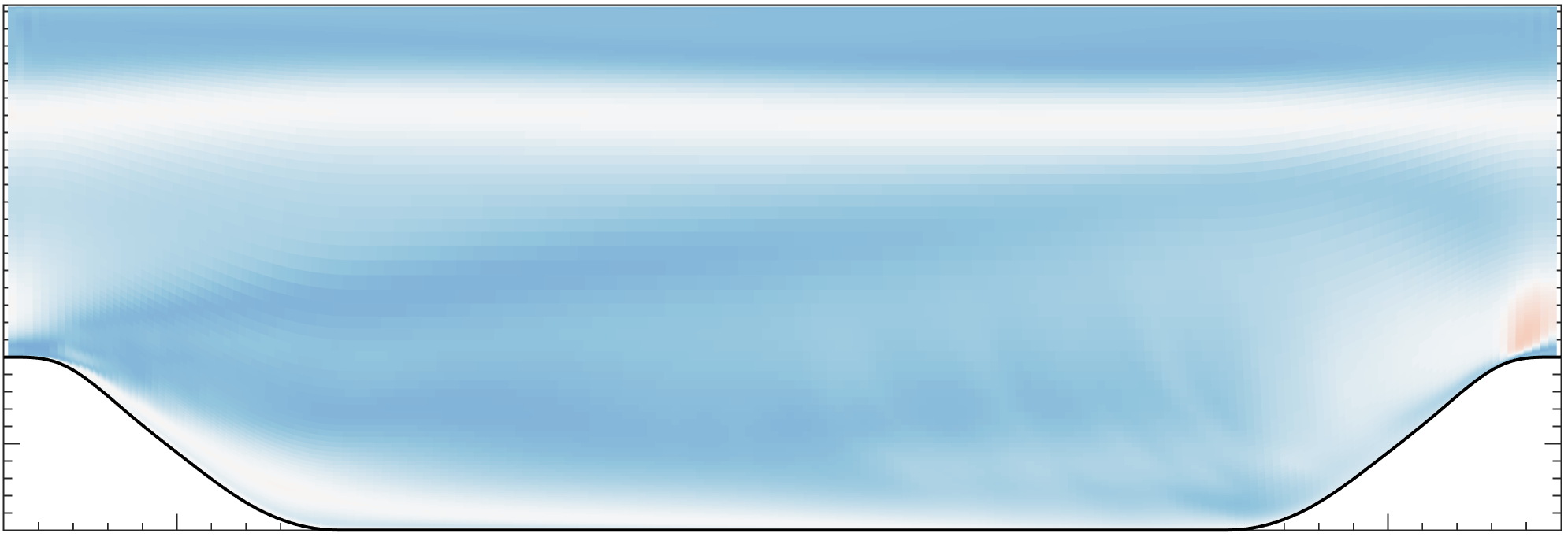} &
        \includegraphics[height = 0.08\textwidth]{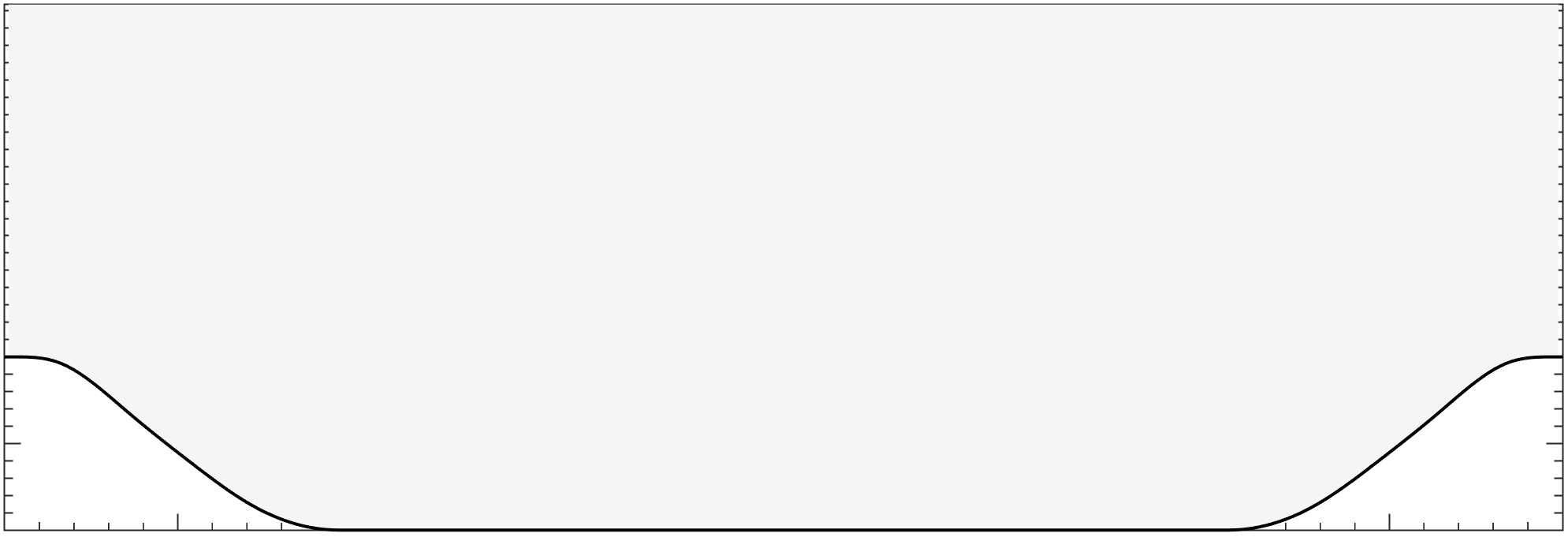} \\[0.5ex]
        \includegraphics[height = 0.09\textwidth]{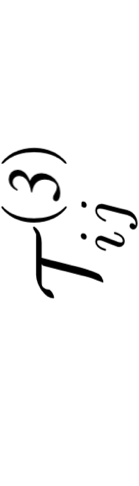} &
        \includegraphics[height = 0.08\textwidth]{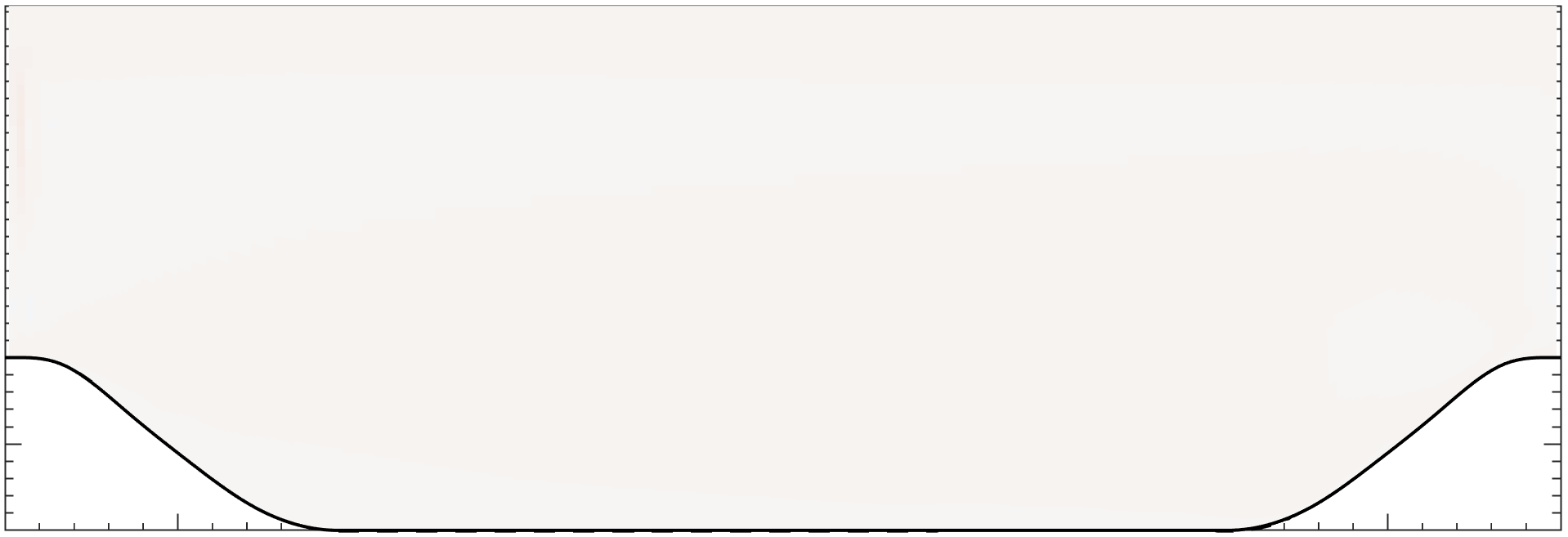} &
        \includegraphics[height = 0.08\textwidth]{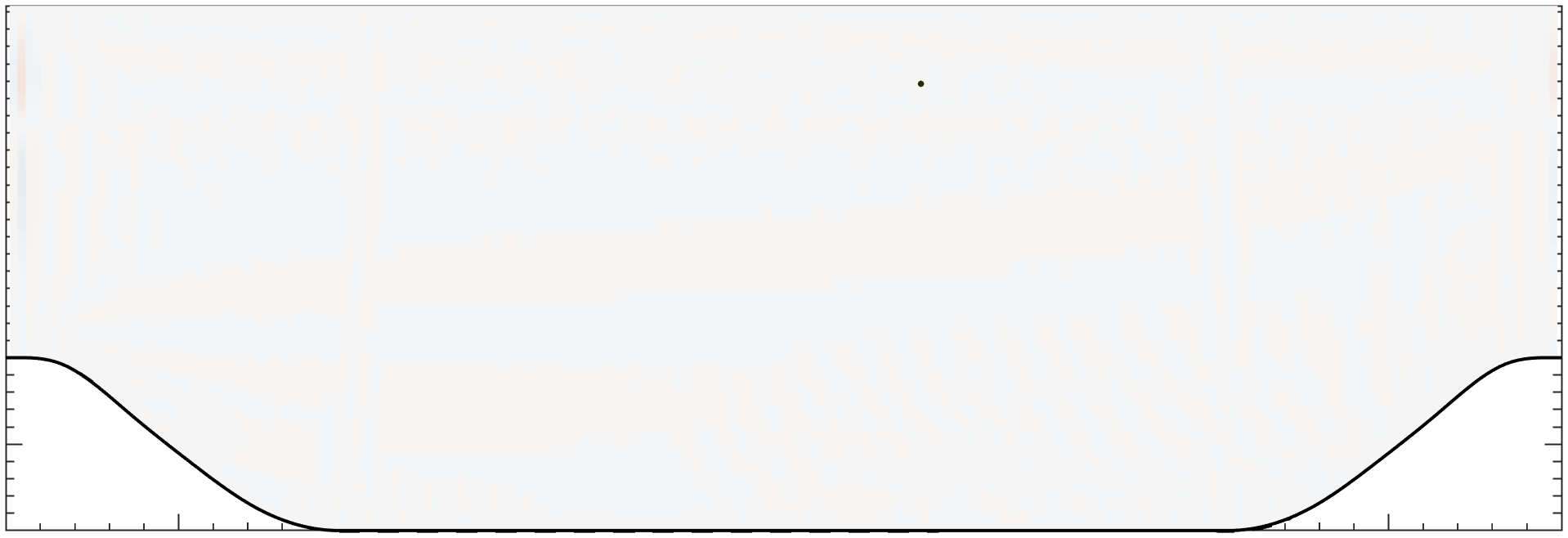} &
        \includegraphics[height = 0.08\textwidth]{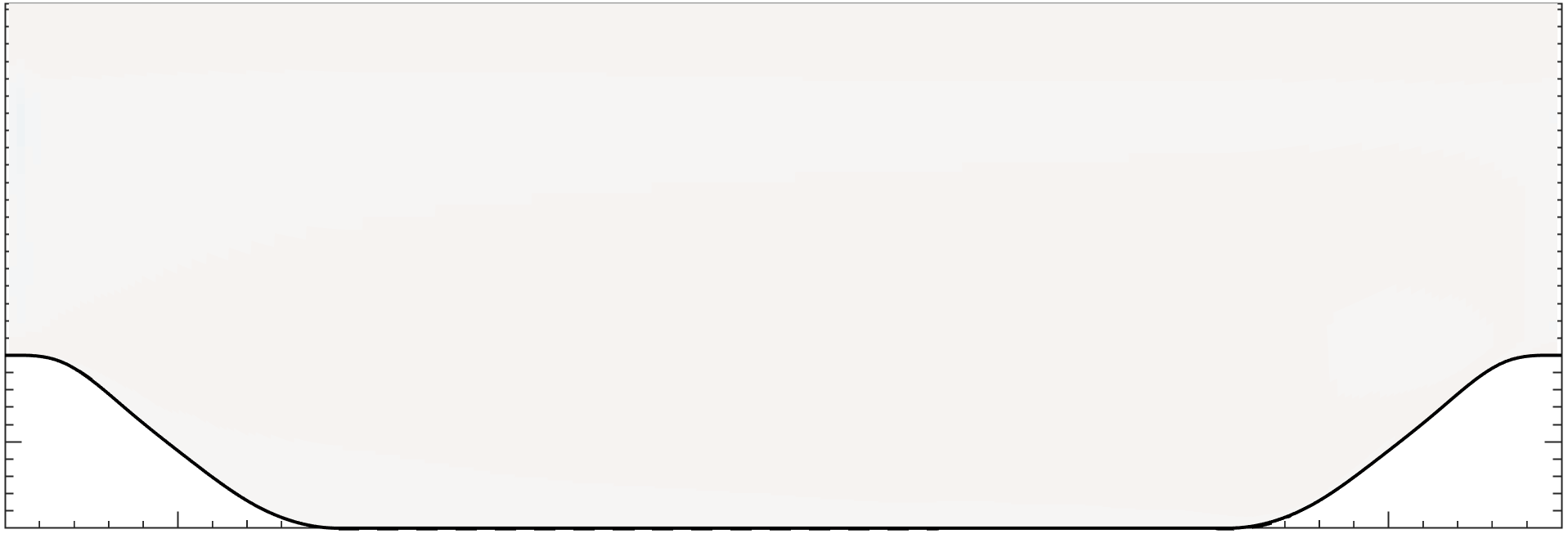} &
        \includegraphics[height = 0.08\textwidth]{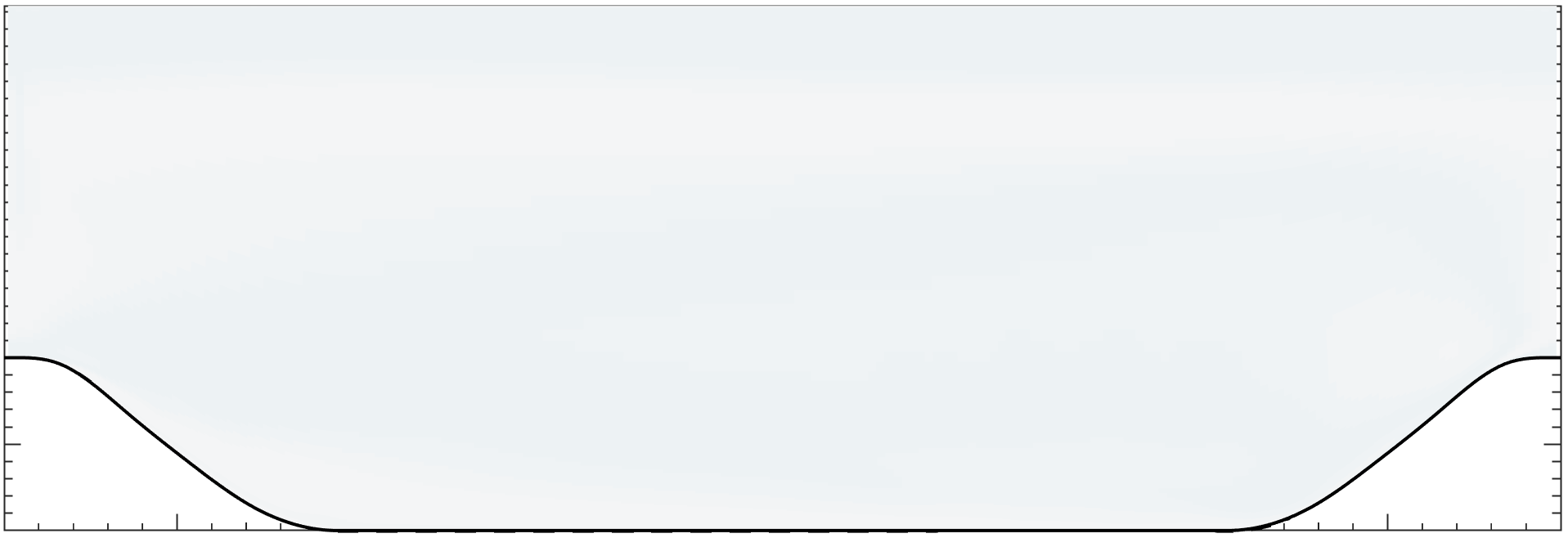} \\[-2ex]
      &\includegraphics[height = 0.023\textwidth]{axis} &
        \includegraphics[height = 0.023\textwidth]{axis} &
        \includegraphics[height = 0.023\textwidth]{axis} &
        \includegraphics[height = 0.023\textwidth]{axis} \\[-2ex]
        &$x/h$ & $x/h$ & $x/h$ & $x/h$ 
         \end{tabular}
        \caption{Contributions to each component of the anisotropy tensor from each of the bases for the model Learned 1, compared with the DNS field}
\label{Fig:Aposteriori2}
\end{figure}

Examination of the momentum RANS equation (Eq.~\eqref{Eq:RANSmom}) makes clear that $\langle u^{\prime} u^{\prime} \rangle$ and $\langle u^{\prime} v^{\prime} \rangle$ are the only Reynolds stress components that contribute to $\langle u \rangle$ and therefore to the prediction of recirculation. By examining $b_{11}$ and $b_{12}$ in Fig.\ref{Fig:Aposteriori2}, it can be seen that both components of anisotropy contribute to predicting the location of separation, however the $b_{12}$ component is most important for the prediction of reattachment. Further, as shown in Fig.~\ref{Fig:Aposteriori2}, the second basis tensor, $\mathcal{T}^{(2)}_{ij}$, is the most important contribution for accurately describing $b_{11}$ and $b_{22}$ and the first basis tensor, $\mathcal{T}^{(1)}_{ij}$, is the most dominant contribution for modeling $b_{12}$. The third basis is critical for accurately describing the $b_{33}$ component, though for this particular configuration (since $z$ is a homogeneous direction), accuracy in this component is not required for predicting the statistically two-dimensional mean flow field.

\begin{figure}[h!]
\centering 
\includegraphics[width = \textwidth]{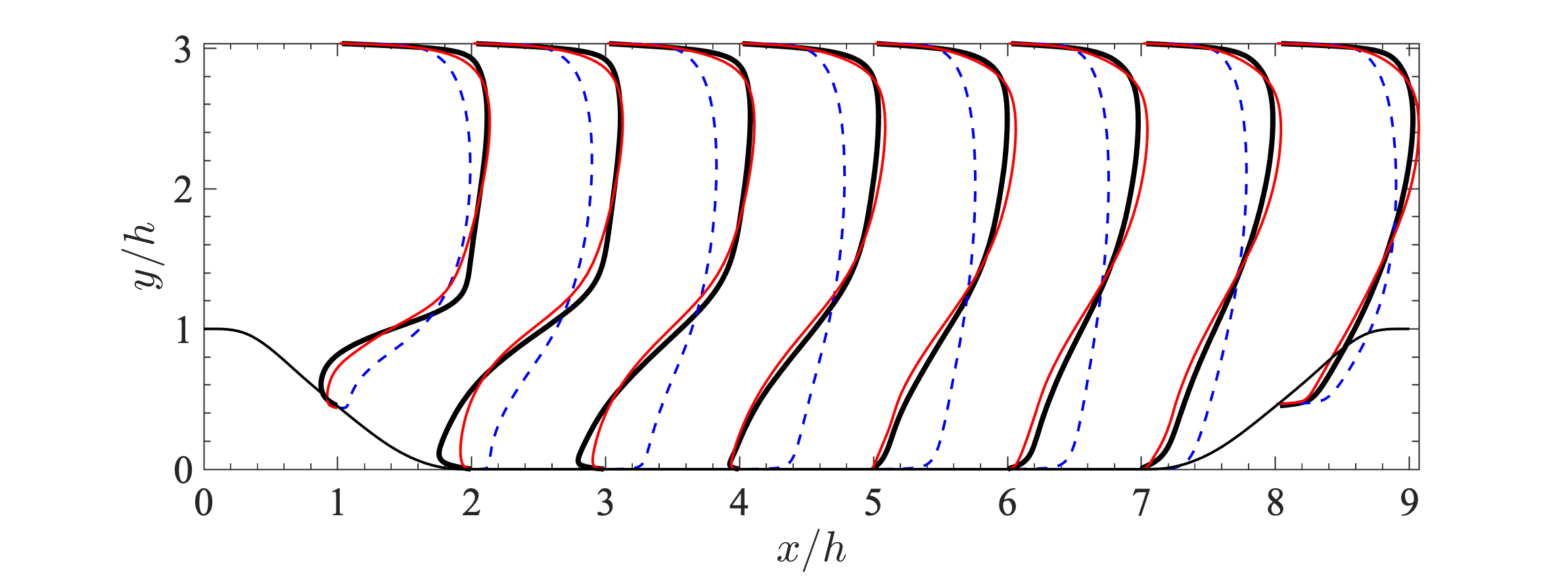}
\caption{Velocity profiles for Re $=5600$. Learned 1 (\protect \Learned), LEVM (\protect \LEVM) and DNS (\protect \DNS) \cite{Breuer2009} }
\label{Fig:VelProfile5600}
\end{figure}

As a final assessment of the learned model (Learned 1), the model was again implemented in OpenFOAM, but for a higher Reynolds number and compared against the openly available dataset provided by \citet{Breuer2009}, in which only a primary recirculation region is observed. Dramatic improvement over LEVM is observed in this case, as shown in Fig.~\ref{Fig:VelProfile5600} and Table \ref{tab:Bubble}. The LEVM solution fails to predict any recirculation while the learned model predicts the reattachment location within 5\% of the DNS value reported in \citet{Breuer2009} and within 5-9\% of the DNS value reported in \citet{Krank2018}. The primary separation location is slightly over two times further in the stream-wise direction as compared to both DNS results \cite{Breuer2009, Krank2018}. The learned model also predicts the small secondary recirculation region that is reported in \citet{Krank2018}. \citet{Breuer2009} does not observe this secondary recirculation, however this appears to be due to differences in numerical schemes and order of accuracy as compared with \citet{Krank2018}. In this secondary region, the learned model predicts the separation and reattachment points within 1\% and 0.1\%, respectively, as compared with the DNS reported in \citet{Krank2018}. 


\subsection{Modeling with sparse data} 
Because the sparse regression methodology seeks to uncover underlying physics, far less data is required to achieve reasonable learned models. This is demonstrated in two contexts. First, a model is learned using only the $y-$dependent data along eight stream-wise locations (see Table \ref{Tab:sparse}). Due to the grid spacing of the RANS simulation, only 160 data points were used for training of each case (compared with 32,000 when using the full dataset in the previous section). As seen in Table \ref{Tab:sparse}, similar model performance is observed for models trained using data located at $x/h = (4-8)$ when compared with the model learned using the full dataset. Secondary recirculation is predicted in three of these training sets. Interestingly, the model trained at $x/h = 8$, where recirculation is not present, is able to predict recirculation in both regions of flow separation. Additionally, it is notable that the model is insensitive to variation in coefficients, especially for the first and third terms.  

\begin{table}
\begin{tabular}{c @{\hskip 0.2in} c c c @{\hskip 0.2in} c @{\hskip 0.2in}  c c @{\hskip 0.2in}  c c} 
\hline
\hline
Training & \multicolumn{3}{c}{Learned coefficients} & \multicolumn{1}{c}{Error} & \multicolumn{2}{c}{Primary} & \multicolumn{2}{c}{Secondary}\\ [-2ex]
($x/h$)& $C_1$ & $C_2$ & $C_3$ &  $\epsilon^{\langle u \rangle}$ & separation & reattachment & separation & reattachment\\
\hline 
1 & 32.35 & 45.42 & 19.69 & 0.17 & 0.36 (1.24) & 4.15 (0.22) & -- & -- \\ [-1ex]
2 & 36.72 & 46.85 & 36.43 & 0.16 & 0.35 (1.23) &  4.27 (0.20) & -- & --\\[-1ex]
3 & 51.81 & 48.75 & 37.59 & 0.15 & 0.34 (1.17) & 4.78 (0.11) & -- & -- \\[-1ex]
4 & 51.38 & 53.05 & 40.99 & 0.12 & 0.40 (1.52) & 5.32 (0.01) & 7.22 (0.05) & 7.35 (0.01) \\[-1ex]
5 & 48.63 & 55.83 & 35.85 & 0.13 & 0.41 (1.56) & 5.10 (0.04) & 7.17 (0.04) & 7.18 (0.01) \\[-1ex]
6 & 49.85 & 55.77 & 23.34 & 0.13 & 0.41 (1.56) & 5.11 (0.04) &  -- & -- \\[-1ex]
7 & 58.34 & 54.19 & -4.04 & 0.13 & 0.41 (1.56) & 5.14 (0.04) & -- & -- \\[-1ex]
8 & 112.56 & 51.42 & -50.00 & 0.12 & 0.37 (1.31) & 4.35 (0.18) & 7.10 (0.03) & 7.27 (0.002) \\ [1ex]
 & \multicolumn{3}{r}{Learned 1 \quad \quad} & 0.12 & 0.40 (1.53) & 5.38 (0.005) & 7.14 (0.04) & 7.31 (0.008)  \\ 
 & \multicolumn{3}{r}{LEVM \quad \quad} & 0.17 & 0.43 (1.61)& 3.64 (0.32)& -- & -- \\
 & \multicolumn{3}{r}{DNS \quad \quad} & -- & 0.16 (--) & 5.35 (--) & 6.87 (--) & 7.25 (--) \\
\hline
\hline 
\end{tabular}
\caption{Summary of learned coefficients using sparse data, i.e. only the $y$-dependent data at the specified $x/h$ location. Model error is reported for the \emph{a posteriori} velocity, and separation and reattachment points are compared with DNS, LEVM, and the `Learned 1' model for Re $= 2800$.}
\label{Tab:sparse}
\end{table}

To further assess the performance of sparse regression in using sparse data, subsets of data are randomly chosen throughout the domain and used as training data. Datasets ranging from 30,000 to 50 training points were assessed (see Table~\ref{Tab:sparseRand}). While the learned coefficients change as the dataset is reduced, the \emph{a priori} model error in the anisotropic stress tensor only increases by 8\%. This suggests that sparse regression would make an excellent modeling construct for extremely sparse datasets, such as those available from experiments where obtaining a high level of resolution is challenging. 

\begin{table}
\centering
\begin{tabular}{c @{\hskip 0.2in} c c c  @{\hskip 0.2in} c} 
\hline 
\hline
  & \multicolumn{3}{c}{Coefficients} & \\ [-2ex]
$n^{train}$ & $C_1$ & $C_2$ & $C_3$ &  $\epsilon^{\bm{b}} $ \\ 
  \hline 
$nx \times ny$ & 63.12 &51.42 &10.98 &0.64 \\[-1ex]
30,000 & 62.50 &51.52&11.24&0.64\\[-1ex]
20,000& 52.50 &45.77&12.19&0.65\\[-1ex]
10,000 & 42.03 &38.84&16.92&0.67 \\[-1ex]
5,000 & 37.35 &36.37&19.69&0.69 \\[-1ex]
1,000 & 33.32 &35.41&21.76 &0.69 \\[-1ex]
500 & 33.91&35.07&20.33&0.69 \\[-1ex]
100 & 33.00 &32.5&23.34&0.71\\[-1ex]
50 & 31.14 &38.28&20.66&0.68 \\
 \hline
 \hline
 \end{tabular}
 \caption{Summary of the learned coefficients for sparse, randomly sampled data using $n^{train}$ training points. The error reported is the \emph{a priori} error in the anisotropic stress tensor.} 
 \label{Tab:sparseRand}
\end{table}

\begin{figure}
\centering
        {\includegraphics[width = 0.7 \textwidth]{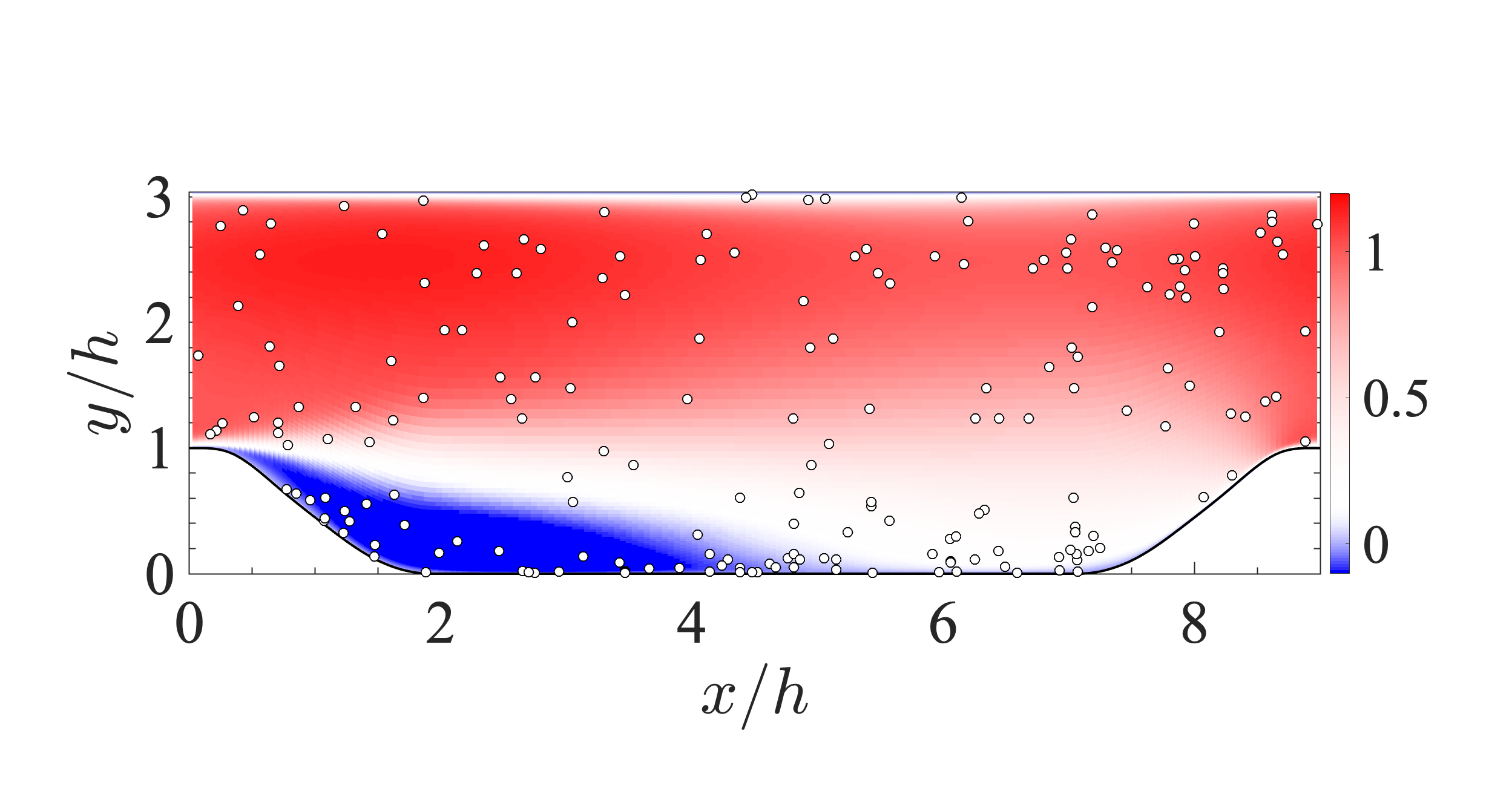}}   
\caption{Example of the random points used for training set corresponding to $n^{train} = 200$. Location of points used for training ($\circ$), $\langle u \rangle$ from DNS (color).} 
        \label{Fig:SparseRandom}
\end{figure}

\section{Conclusion}
In this work, a turbulence closure modeling methodology has been proposed as an alternative to other machine learning techniques, such as NNs. This method is based upon sparse regression which uses an L-2 norm with an L-1 norm penalty cost functional to produce a compact, algebraic model. Further, the inputs to the optimization algorithm are specifically tailored in order to ensure form invariance. This is specifically accomplished by arranging the trusted and basis tensorial data into column vectors, thereby constraining coefficients to be invariant with respect to direction. By generating a model in this form, several important modeling properties can be achieved: form (or Galilean) invariance, interpretability, and ease of dissemination. Using two canonical cases, it was demonstrated that this technique produces results with model accuracies similar to that of modern NN methodologies, even when using a drastically reduced training dataset.

Using homogeneous free shear turbulence as a preliminary example, sparse regression was able to return the LRR-IP model used to generate a synthetic dataset, even when large amounts of noise were applied. Next, using DNS data for homogeneous free shear turbulence, sparse regression learned a model that reduces model error by 70\% as compared to the existing LRR-IP and LRR-QI models. This performance was also observed in the testing data evaluated. 

In the case of turbulent flow through a periodically constricted channel, sparse regression uncovered a model that has comparable performance to a modern NN considering the same flow, however this performance can be achieved using a drastically minimal dataset and the resultant model form is available in a compact, algebraic form. Additionally, the learned model demonstrated significant improvements in performance as compared with LEVM for a much higher Reynolds number, and outside the scope of its training. Further, due to the ability of sparse regression to learn predictive models using minimal datasets and noisy data (as demonstrated in Sec.~\ref{sec:FreeShear}), it is an ideal candidate for translating experimental data, which may be both noisy and sparse, into accurate models.

Finally, sparse regression assumes complete generality and thus does not strictly require an existing model upon which to augment. This is an important property for other open areas of research, e.g., modeling multiphase turbulence \cite{fox2014, capecelatro2014cit, capecelatro2015on, capecelatro2016channel2, beetham2019}, for which existing models are either unavailable or too inaccurate to reliably use as a baseline model upon which to build.  Such an approach can also be applied to turbulent combustion, in which heat release due to chemical reactions give rise to `back scatter' and existing models based on an energy cascade fail to be predictive \cite{Veynante2002, Pitsch2006}. 

\section*{Acknowledgements} 
This material is based upon work supported by the National Science Foundation Graduate Research Fellowship. We would also like to acknowledge the National Science Foundation for partial support from award CBET 1846054. The computing resources and assistance provided by the staff of the Advanced Research Computing at the University of Michigan, Ann Arbor are greatly appreciated. Finally, the authors gratefully acknowledge Prof. M. Houssem Kasbaoui for the code used to generate homogeneous sheared turbulence data and Prof. C. Petty for his discussions on non-inertial reference frames.  

\section*{References}

%
\newpage
\bibliographystyle{plainnat}
\bibliography{text}

\end{document}